\documentclass[aps,pre,twocolumn,preprintnumbers,amsmath,amssymb,floatfix,superscriptaddress,nofootinbib]{revtex4-1}
\usepackage{graphicx}
\usepackage{bm}
\usepackage{color}
\usepackage[normalem]{ulem} 

\def\nablab{{\bm \nabla}}

\def\O{\mathcal{O}}

\def\eq{{\rm ref}}

\def\figures{.}

\definecolor{gray}{rgb}{0.5,0.5,0.5}
\definecolor{lgray}{rgb}{0.8,0.8,0.8}
\definecolor{dgray}{rgb}{0.6,0.6,0.6}
\definecolor{dred}{rgb}{0.5,0.0,0.0}
\definecolor{dgreen}{rgb}{0.0,0.5,0.0}
\definecolor{dblue}{rgb}{0.0,0.0,0.5}
\definecolor{violet}{rgb}{0.7,0.0,0.5}

\definecolor{lred}{rgb}{1.0,0.5,0.5}
\definecolor{lgreen}{rgb}{0.5,1.0,0.5}
\definecolor{lblue}{rgb}{0.5,0.5,1.0}

\newcommand{\mcb}{\color{dgreen} }

\def\mcbtxt{green}

\def\L{{\mathcal L}}
\def\H{{\mathcal H}}
\def\O{{\mathcal O}}

\def\P{{\mathcal P}}
\def\E{{\mathcal E}}

\makeatother
\begin{document}


\title{Testing the conservative character of particle simulations: \\ II. Spurious heating of guiding centers and full orbits subject to fluctuations expressed in terms of ${\bm E}$ and ${\bm B}$}

\author{A.~Bierwage}
\email[]{bierwage.andreas@qst.go.jp}
\affiliation{QST, Rokkasho Fusion Institute, Aomori 039-3212, Japan}
\affiliation{QST, Naka Fusion Institute, Ibaraki 311-0193, Japan}
\author{K.~Shinohara}
\affiliation{\mbox{Department of Complexity Science and Engineering, The University of Tokyo, Kashiwa, Chiba 277-8561, Japan}}
\affiliation{QST, Naka Fusion Institute, Ibaraki 311-0193, Japan}

\date{\today}

\begin{abstract}
For an axisymmetric tokamak plasma, Hamiltonian theory predicts that the orbits of charged particles must stay on invariant tori of conserved energy in the moving frame of reference of a wave that propagates along the torus with a fixed angular phase velocity. In principle, this is true for arbitrary mode structures in the poloidal plane, but only if the fluctuations are expressed in terms of potentials $\Phi$ and ${\bm A}$, which satisfy Faraday's law by definition. Here, we use the physical fields ${\bm E}$ and ${\bm B}$, where Faraday's law may be violated by errors introduced during the process of computing or designing the wave field through numerical inaccuracies, approximations, or gross negligence. Numerical heating caused by noise-like artifacts on the grid scale can to some extent be reduced via shorter time steps. In contrast, coherent inconsistencies between ${\bm E}$ and ${\bm B}$ cause spurious acceleration that is independent of time steps or numerical methods, but can be sensitive to geometry. In particular, we show that secular acceleration is enhanced when one imposes nonnormal modes that possess strong up-down asymmetry instead of the usual in-out asymmetry of normal toroidal (eigen)modes. Our tests are performed for full gyroorbit and guiding center (GC) models, which give similar results. In addition, we show that $N$-point gyroaveraging is not a recommendable method to enhance the realism of GC simulations. Besides breaking conservation laws, $N$-point gyroaveraging in our example makes the GC results deviate further from the full orbit results, showing that this method can even give the wrong trend.{\looseness=-1}
\end{abstract}

\maketitle

\thispagestyle{empty}
\everypar{\looseness=-1} 

\tableofcontents

\section{Introduction}
\label{sec:intro}

The motion of charged particles can be simulated using the physical electromagnetic fields ${\bm E}$ and ${\bm B}$ or their scalar and vector potentials $\Phi$ and ${\bm A}$. Both approaches have advantages and disadvantages with respect to the complexity of the equations, ease of model integration, numerical efficiency, accuracy and physical integrity. In this work, we are concerned with the conservative character of wave-particle interactions in the absence of collisional and radiative processes, simulated using ${\bm E}$ and ${\bm B}$ in realistic tokamak geometry, and subject to practical constraints with respect to numerical accuracy.

We begin with classical gyroorbits whose equations of motion, namely the Newton-Lorentz equations, are embedded in the Lagrangian $\L$ and Hamiltonian $\H$ given by
\begin{subequations}
\begin{align}
\L({\bm x},{\bm v},\dot{\bm x},t) =& (M{\bm v} + Ze{\bm A})\cdot\dot{\bm x} - \H,
\label{eq:lh_l}
\\
\H({\bm x},{\bm v},t) =& Mv^2/2 + Ze\Phi,
\label{eq:lh_h}
\end{align}
\label{eq:lh}\vspace{-0.4cm}
\end{subequations}

\noindent where $Ze$ and $M$ are the electric charge and mass of the particle, ${\bm v}$ and ${\bm x}$ its velocity and position vectors, with magnitude $v = |{\bm v}|$ and total time derivative $\dot{\bm x} \equiv {\rm d}{\bm x}/{\rm d}t$. Equation~(\ref{eq:lh}) acquires canonical form when substituting ${\bm v}$ with the canonical momentum
\begin{equation}
{\bm P}_{\rm c} = M{\bm v} + Ze{\bm A}.
\end{equation}

\noindent Note that $\H$ and ${\bm P}_{\rm c}$ are expressed in terms of the potentials $\Phi$ and ${\bm A}$. This implies that not only the Newton-Lorentz equations, but also Faraday's law
\begin{equation}
\partial_t{\bm B} = -\nablab\times{\bm E}
\label{eq:faraday}
\end{equation}

\noindent is built into Eq.~(\ref{eq:lh}) via the definitions ${\bm E} = -\nablab\Phi - \partial_t{\bm A}$ and ${\bm B} = \nablab\times{\bm A}$. However, this implicit constraint is lost when the resulting equations of motion are expressed in terms of ${\bm E}$ and ${\bm B}$. In such cases, it is important to ensure explicitly that ${\bm E}$ and ${\bm B}$ are consistent with each other in order to preserve the Hamiltonian character of the system.

In practice, the task of ensuring the physical consistency between ${\bm E}$ and ${\bm B}$ may pose numerical challenges and requires attention when the fluctuating fields have been subject to some form of design. For instance, the fields may be designed using simple parametric models, which can be useful for systematic analyses of wave-particle interactions. Even self-consistently computed wave fields from a Maxwell-Vlasov or magnetohydrodynamic (MHD) simulation are often subject to some form of post-processing, such as Fourier analysis that usually involves truncation, or discretized coordinate transformations. Such designed or processed modes may enter an integrated modeling workflow and serve as input to orbit-following codes that compute particle distribution functions in the presence of radio-frequency (RF) waves or MHD modes. Modifications on the fluctuating fields may even be performed on-the-fly during a particle simulation. Two prominent examples are the interpolations that are performed as part of the widely-used particle-in-cell (PIC)\cite{BirdsallLangdon} method, and $N$-point gyroaveraging during a guiding center (GC) simulation.

In the present work, we wish to demonstrate how flaws in the design of the fluctuations expressed in terms of ${\bm E}$ and ${\bm B}$ can influence the accuracy and physical integrity of the results of an orbit-following simulation code that solves equations of motion derived on the basis of Hamiltonian theory. For this purpose, we examine the conservative character of the motion of deuterons in a JT-60U tokamak plasma using full orbit and GC models (see the Appendices for details). The simulation particles are subject to a prescribed wave field with a fixed amplitude function $\tilde\Phi_{n,\omega}(R,z)$ and fixed angular phase velocity $\omega/n$, where $\omega$ is the angular frequency and $n$ the toroidal mode number. That is, we design the fluctuating fields ${\bm E}(t)$ and $\delta{\bm B}(t)$ from a model of the electric potential with the form
\begin{equation}
\Phi_{\rm mdl}(R,\varphi,z,t) = \Re\{\widetilde{\Phi}_{n,\omega}(R,z) \exp(-i\omega t - in\varphi)\}.
\label{eq:epot}
\end{equation}

\noindent The tilde indicates a complex quantity and we use right-handed cylinder coordinates $(R,\varphi,z)$ with major radius $R$, geometric toroidal angle $\varphi$ and height $z$. When Eq.~(\ref{eq:epot}) is satisfied, Hamiltonian theory predicts that the particle's energy $\E' \equiv \E + \frac{\omega}{n}\P_\varphi$ in the wave's rotating frame of reference is conserved (see Appendix~\ref{apdx:fo_erot} for details and generalization):
\begin{equation}
n\dot{\E}' = n\dot{\E} + \omega\dot{\P}_\varphi = 0.
\label{eq:derot_dt_0}
\end{equation}

\noindent Here, $\E = \H = Mv^2/2 + Ze\Phi$ is the total energy of a particle and $\P_\varphi = {\bm P}_{\rm c}\cdot\partial_\varphi{\bm x}$ is the covariant toroidal component of the canonical momentum ${\bm P}_{\rm c}$.

Evidently, the condition $\E' = {\rm const}$.\ --- which defines invariant tori of the Kolmogorov-Arnold-Moser (KAM) theorem --- holds for arbitrary potential structures $\widetilde{\Phi}_{n,\omega}(R,z)$ and $\widetilde{\bm A}_{n,\omega}(R,z)$ in the poloidal plane. Thus, when the fluctuations are prescribed using $\Phi$ and ${\bm A}$ --- as is done in Hamiltonian GC codes like {\tt ORBIT} \cite{White84} --- the conservation of rotating frame energy (\ref{eq:derot_dt_0}) and the conservation of phase space density (Liouville's theorem) are guaranteed down to the level of the numerical scheme's accuracy.

This is not true for arbitrary $\widetilde{\bm E}_{n,\omega}(R,z)$ and $\widetilde{\bm B}_{n,\omega}(R,z)$. For instance, when one designs a model potential $\Phi_{\rm mdl}$ and then prescribes the electric field as ${\bm E} = {\bm E}_\perp = -\nablab_\perp\Phi_{\rm mdl}$ with ${\bm E}_\perp \equiv \hat{\bm b}\times({\bm E}\times\hat{\bm b})$ and $\hat{\bm b} \equiv {\bm B}/B$, the simulation must explicitly include a consistent fluctuating magnetic field $\delta{\bm B}(t)$ satisfying Faraday's law in the form $\partial_t\delta{\bm B} = -\nablab\times\nablab_\perp\Phi_{\rm mdl}$. This model is not foolproof, since it is possible to run such simulations without the magnetic fluctuation (as is, for instance, done in some RF heating codes). We will present some illustrative examples showing how this produces unphysical secular particle acceleration that violates conservation of energy and phase space density in both full orbit and GC models.

The spurious heating caused by such violations of the laws of electromagnetism is numerically robust; i.e., independent of resolution in space and time. Interestingly, however, the unphysical secular dynamics caused by this sort of errors can remain relatively small if the poloidal structure of the applied perturbation resembles a normal (eigen)mode of the toroidal plasma with predominant in-out asymmetry along $R$. In contrast, the secular dynamics are strongly enhanced, and thus more easily seen, when one perturbs the system using `nonnormal' modes that posses a strong up-down asymmetry along $z$, so that their intensity distribution is `unnatural' for a toroidal plasma. This observation means that geometric effects can influence how strongly the breaking of the laws of electromagnetism affects the simulation results. This mechanism can be utilized when testing new codes as suggested in the companion paper,\cite{Bierwage22b} where the same effect was seen in tests involving variations of the Hamiltonian GC code {\tt ORBIT}.

Significant spurious heating may not only occur when Faraday's law (\ref{eq:faraday}) is grossly violated by the neglect of $\delta{\bm B}(t)$ as in the above example. The conservative character of the system can be broken also as the result of more subtle errors in the numerical accuracy of the poloidal mode structure, or by smoothing operations, such as gyroaveraging.

In the examples that we will discuss in this paper, interpolation artifacts are introduced through a coordinate transformation operation. We design $\widetilde\Phi_{n,\omega}(R,z)$ appearing in Eq.~(\ref{eq:epot}) from a set of poloidal Fourier harmonics $\exp(- im\vartheta)$ that are defined in toroidal magnetic coordinate system $(\Psi_{\rm P},\vartheta,\zeta)$, where $\Psi_{\rm P}$ is the poloidal magnetic flux serving as a minor radial coordinate, $\vartheta$ is a poloidal angle, and $\zeta$ another toroidal angle (here, $\zeta = -\varphi$). Numerical artifacts are introduced when mapping the perturbation to cylinder coordinates $(R,\varphi,z)$ because the magnetic flux coordinate mesh tends to be highly nonuniform, so that there is usually a region where the flux coordinate mesh density is lower than the cylinder coordinate grid. The noise-like artifacts are then enhanced with each differential operation $\nablab$ needed to compute first ${\bm E}$ and then $\delta{\bm B}$. GC equations contain additional derivatives of $\delta{\bm B}$ that enter the mirror force and drift terms. The `noisy' $\delta{\bm B}$ field satisfies Faraday's law and the solenoidal condition $\nablab\cdot\delta{\bm B} = 0$ on the discrete mesh, so it is not unphysical; it only has an unrealistic fine structure. Via the Courant-Friedrichs-Lewy (CFL) condition, the noise imposes a (soft) upper bound $\dot{\bm x}\Delta t_{\rm CFL} = \Delta{\bm x}_{\rm grid}$ on the time step for pushing simulation particles with speed $\dot{\bm x}$ across the mesh whose spacing $\Delta{\bm x}_{\rm grid}$ is much shorter than that of the physical waves of interest. In our working example, where we follow $400\,{\rm keV}$ deuterons in a tokamak plasma perturbed with long-wavelength modes, the consequence is that the GC simulation has to be run with shorter time steps that lie within a factor 10 of an efficient full orbit solver. This reduces the GC model's advantage in terms of computational speed, unless grid-scale noise is efficiently suppressed (or an implicit solver is used).

We also test the influence of $N$-point gyroaveraging that is used in some codes as an attempt to improve the quantitative accuracy of the drift-kinetic GC model. We use the procedure described previously in Ref.~\onlinecite{Bierwage16c}, where $N$-point gyroaveraging is applied to the fluctuating fields ${\bm E}$ and $\delta{\bm B}$ in the poloidal $(R,z)$ plane. This procedure is not based on Hamiltonian theory, so a violation of conservation laws was to be expected and is indeed observed here. This is one reason for why we think it is preferable to follow full orbits instead of the GCs in cases where gyroaveraging effects are quantitatively important. Further strengths and weaknesses of the full orbit model are discussed in Appendix~\ref{apdx:reasons}.

Overall, our impression is that the quality of the perturbed field's structure can be more important than the choice of coordinates or the numerical scheme used to push the particles. In the main part of the paper, we employ the explicit 4th-order Runge-Kutta (RK4) scheme. For the same time step size $\Delta t$, the modified leap-frog (MLF)\cite{Hirvioki14} solver is faster than RK4, but measurements of $\P_\varphi$ and $\E'$ are less accurate with MLF since positions and velocities are computed at staggered times that lie $\Delta t/2$ apart. The $\varphi$-dependence of the fields is evaluated using the particle-in-Fourier (PIF)\cite{Evstatiev13, AmeresPhdThesis, Mitchell19} method due to its higher accuracy and efficiency compared to PIC (see Appendix~\ref{apdx:pif_pic}).

The perturbation model is described in Section~\ref{sec:model}. Results of our simulations are presented in Section~\ref{sec:results}, followed by a discussion in Section~\ref{sec:summary}. The Appendices~\ref{apdx:orbtop}--\ref{apdx:gc} contain a detailed description of the physical models and their numerical implementation in the current version of the new orbit-following simulation code {\tt ORBTOP} that we employ here. Appendices~\ref{apdx:pif_pic} and \ref{apdx:reasons} contain additional discussions pertaining to accuracy and performance.

\section{Perturbation model}
\label{sec:model}

\subsection{Polarization: electrostatic and electromagnetic scenarios}
\label{sec:model_scenarios}

The electric and magnetic field vectors ${\bm E}$ and ${\bm B}$ can be written in terms of a fluctuating electric potential $\Phi(t)$ and vector potential ${\bm A}(t) = {\bm A}_\eq + \delta{\bm A}(t)$ as
\begin{subequations}
\begin{align}
{\bm E} &= -\nablab\Phi - \partial_t\delta{\bm A}, \\
{\bm B} &= {\bm B}_\eq + \underbrace{\nablab\times\delta{\bm A}}\limits_{\delta{\bm B}(t)}, \quad \partial_t\delta{\bm B} = -\nablab\times{\bm E}.
\end{align}\vspace{-0.4cm}
\end{subequations}

\noindent In our case, the stationary reference field
\begin{equation}
{\bm B}_\eq = \nablab\times{\bm A}_\eq = \nablab\Psi_{\rm P}\times\nablab\varphi + B_{\eq,\varphi}\nablab\varphi
\end{equation}

\noindent is obtained from an ideal MHD equilibrium solver in terms of the poloidal flux function $\Psi_{\rm P} = A_{{\rm ref},\varphi}$ and the covariant toroidal field component $B_{\eq,\varphi} = {\bm B}_\eq\cdot\partial_\varphi{\bm x}$. In the present work, we use prescribed perturbations ${\bm E}(t)$ and $\delta{\bm B}(t)$ that are constructed from a parametric model for the electric potential $\Phi$, which will be introduced in Section~\ref{sec:model_mstruc} below. The following two scenarios were constructed:
\begin{enumerate}
\item[(i)]  Purely electrostatic mode:
\begin{subequations}
\begin{align}
{\bm E} &= -\nablab\Phi \;\; \Rightarrow \;\; \partial_t\delta{\bm B} = \nablab\times\nablab\Phi = 0, \\
{\bm B} &= {\bm B}_\eq \quad \text{(unperturbed reference field)}.
\end{align}
\label{eq:scenario_1}\vspace{-0.6cm}
\end{subequations}

\item[(ii)]  Electromagnetic mode with transverse polarization relative to the reference field ${\bm B}_\eq$:
\begin{subequations}
\begin{align}
{\bm E} &= -\nablab_\perp^\eq\Phi \;\; \Rightarrow \;\; \partial_t\delta{\bm A} = -\nablab_\parallel^\eq\Phi,
\\
{\bm B} &= {\bm B}_\eq + \underbrace{(-\partial_t^{-1}\nablab\times{\bm E})}\limits_{\delta{\bm B}(t) = \nablab\times\delta{\bm A}},
\end{align}
\label{eq:scenario_2}\vspace{-0.6cm}
\end{subequations}
\end{enumerate}

\noindent where the following definitions were used:
\begin{gather}
\nablab_\perp^\eq\Phi \equiv \hat{\bm b}_\eq\times(\nablab\Phi\times\hat{\bm b}_\eq), \quad
\nablab_\parallel^\eq \equiv \hat{\bm b}_\eq \hat{\bm b}_\eq \cdot\nablab, \nonumber
\\
\partial_t^{-1}X \rightarrow \Re\{X/(-i\omega)\}, \quad \hat{\bm b}_\eq \equiv {\bm B}_\eq/B_\eq.
\end{gather}

\noindent In addition, we will demonstrate what happens if one ignores the {\it time-dependent} magnetic perturbation in scenario (ii) or, equivalently, if one ignores the parallel component of the electric field in scenario (i), giving
\begin{enumerate}
\item[(iii)]  Unphysical mode:
\begin{subequations}
\begin{align}
{\bm E} &= -\nablab_\perp^\eq\Phi \quad (E_\parallel^\eq \equiv {\bm E}\cdot\hat{\bm b}_\eq \rightarrow 0),
\\
{\bm B} &= {\bm B}_\eq \quad (\delta{\bm B}(t) \rightarrow 0).
\end{align}
\label{eq:scenario_3}\vspace{-0.6cm}
\end{subequations}
\end{enumerate}

\noindent While scenario (i) may be unrealistic for a tokamak plasma and for the frequency $\omega$ that we will use, it is not unphysical as it satisfies the laws of electromagnetism. In contrast, scenario (iii) with $E_\parallel^\eq,\delta{\bm B}(t) \rightarrow 0$ violates Faraday's law (\ref{eq:faraday}), which dictates that $E_\parallel^\eq$ and $\delta{\bm B}(t)$ must not be zero at the same time.

\begin{figure}[tbp]
\centering\vspace{-0.25cm}
\includegraphics[width=0.48\textwidth]{\figures/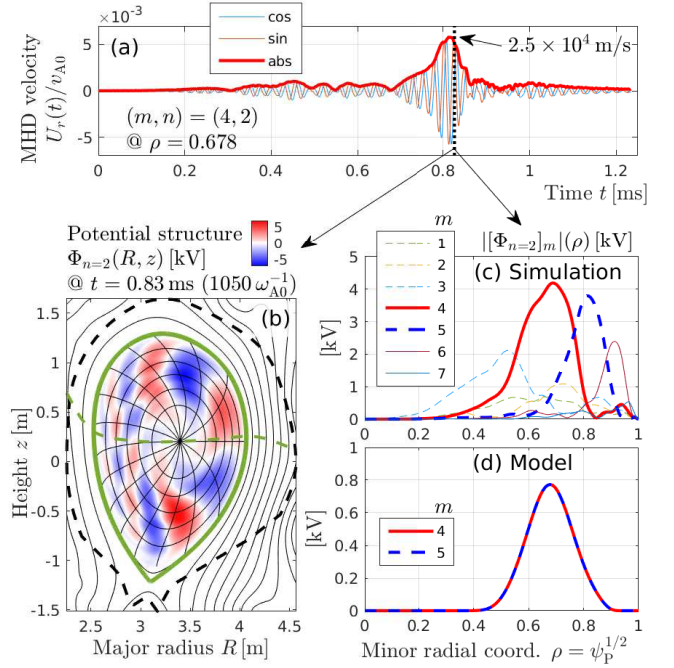}
\caption{$n = 2$ harmonic of the fluctuating field during an Abrupt Large-amplitude Event (ALE) in JT-60U, simulated by the kinetic-MHD hybrid code {\tt MEGA} as reported in Ref.~\protect\onlinecite{Bierwage18}. This mode served as a reference for the perturbation model used in this work. Panel (a) shows the temporal evolution of the $n=2$ harmonic of the radial component $U_r(t) = {\bm U}\cdot\hat{\bm e}_r$ of the MHD velocity ${\bm U} \approx {\bm v}_{\rm E}$ in the {\tt MEGA} simulation. Plotted here are the cosine and sine components and the absolute value of the $(m,n) = (4,2)$ harmonic of $U_r(t)$ at $\rho = 0.678$ ($r/a = 0.587$). Panel (b) shows the poloidal mode structure of the approximate electric potential $\Phi_{n=2}(R,z) \approx B_0 \sum_m \frac{r}{m}\left(U_r^{\rm c}(r,m)\cos(m\vartheta) - U_r^{\rm s}(r,m)\sin(m\vartheta)\right)$, where $r/a$ is another radial coordinate approximately proportional to the square root of the toroidal flux and $a \approx 1\,{\rm m}$ is the mean minor radius of the plasma. The dashed black line is the wall, the dashed green line is the midplane (where ${\bm B}_\eq\cdot\nablab B_\eq = 0$), and the bold green line is the plasma boundary (edge). Panel (c) shows the radial profiles of the mode's poloidal Fourier harmonics $|[\Phi_{n=2}]_m|(\rho)$. Panel (d) shows the model used in this work, which consists of two poloidal harmonics, $m_{\rm a} = 4$ (red) and $m_{\bm b} = 5$ (blue dashed), with identical radial profiles.}\vspace{-0.15cm}
\label{fig:01_ale_mdl}%
\end{figure}

\subsection{Mode structure and amplitude}
\label{sec:model_mstruc}

As in Ref.~\onlinecite{Bierwage22b}, we construct a simple parametric model for $\Phi$ from Fourier harmonics in toroidal flux coordinates,
\begin{equation}
\widetilde{\Phi}(\rho,\varphi,\vartheta,t) = e^{-i\omega t - in\varphi} \sum_m e^{i\Theta_{0,m} - i m\vartheta} \Phi_m(\rho),
\label{eq:mdl}
\end{equation}

\noindent with a given toroidal mode number $n$ and oscillation frequency $\omega = 2\pi\nu$. The triplet $(\rho,\varphi,\vartheta)$ forms a right-handed set of toroidal flux coordinates with geometric toroidal angle $\varphi$, and the poloidal angle $\vartheta$ defined such that magnetic field lines ${\bm B}_\eq$ are straight in the $(\vartheta,\varphi)$ plane. The flux label $\rho = \psi_{\rm P}^{1/2} \in [0,1]$ serves as a minor radial coordinate, where $\psi_{\rm P} = (\Psi_{\rm P} - \Psi_{\rm P,0})/(\Psi_{\rm P,edge} - \Psi_{\rm P,0})$ is the normalized poloidal flux function. $\Psi_{\rm P,0}$ and $\Psi_{\rm P,edge}$ are the values of $\Psi_{\rm P}$ at the magnetic axis and plasma edge. In Eq.~(\ref{eq:mdl}) and hereafter, the tilde indicates a complex quantity and the physical electric and magnetic field perturbations are taken to be
\begin{subequations}
\begin{align}
{\bm E} =& \Re\{\widetilde{\bm E} \} = \Re\{-\nablab\widetilde\Phi - \partial_t\delta\widetilde{\bm A}\},
\\
\delta{\bm B} =& \Re\{\nablab\times\widetilde{\bm E}/(i\omega)\} = \Im\{\nablab\times\widetilde{\bm E}/\omega\}.
\end{align}\vspace{-0.3cm}
\end{subequations}

\noindent The radial profile function $\Phi_m(\rho)$ for each poloidal Fourier harmonic $m$ in Eq.~(\ref{eq:mdl}) is
\begin{subequations}
\begin{align}
\Phi_m(\rho) =& \Phi_{0,m} \exp\left(-\frac{(\rho^2 - 0.5)^8}{0.3^8}\right) \int_0^\rho{\rm d}\rho' \hat{E}_m(\rho'),
\label{eq:mdl_prof_phi}
\\
\hat{E}_m(\rho) =& \exp\left(-\frac{(\rho - \rho_m)^2}{2 d_m^2}\right) \tanh\left(\frac{\rho - \rho_m}{d_m}\right).
\label{eq:mdl_prof_e}
\end{align}
\label{eq:mdl_prof}\vspace{-0.0cm}
\end{subequations}

\noindent The exponential factor in Eq.~(\ref{eq:mdl_prof_phi}) is a super-Gaussian cut-off function, which ensures that the profile rapidly drops to zero in the regions $\rho < 0.2$ and $\rho > 0.8$. The parameter values are summarized in Tables~\ref{tab:parm_prof} and \ref{tab:parm_pol}.

\begin{table}[tbp]\vspace{-0.35cm}
\caption{Model parameters for Eqs.~(\protect\ref{eq:mdl}) and (\protect\ref{eq:mdl_prof}). All poloidal harmonics $m$ listed in Table~\protect\ref{tab:parm_pol} have the same radial profile $\Phi_m(\rho)$, whose peak value is ${\rm max}\{\Phi_m(\rho)\} \approx 0.8\,{\rm kV}$.}
\begin{ruledtabular}
\begin{tabular}{c|c|c|c|c}
$\nu = \omega/(2\pi)$ & $n$ & $\Phi_{0,m}$ & $\rho_m$ & $d_m$ \\
\hline $47.4\,{\rm kHz}$ & $2$ & $3\times 7.3915\,{\rm kV}$ & $\sqrt{0.46}$ & $0.1$ \\
\end{tabular}
\label{tab:parm_prof}
\end{ruledtabular}
\end{table}

\begin{table}[tbp]\begin{ruledtabular}
\caption{Poloidal Fourier harmonics $m$ and their phases $\Theta_{0,m}$ characterizing the $\vartheta$-dependence of the perturbation. Parameters are given for all five cases that are shown in Fig.~\protect\ref{fig:02_modes}. Among these, only cases (A)--(D) are used in the main part of the paper. The absolute values of the phases are chosen such that the dominant resonance in scenario (ii) defined by Eq.~(\protect\ref{eq:scenario_2}) has an elliptic point at $\vartheta = 0$.}
\begin{tabular}{c|c|c|c|c|c}
Case & $m_{\rm a}$ & $m_{\rm b}$ & $\Theta_{0,{\rm a}}$ & $\Theta_{0,{\rm b}}$ & $\Delta\Theta_0 = \Theta_{0,{\rm b}} - \Theta_{0,{\rm a}}$ \\
\hline (O) & $4$ & $4$ & $0$ & $0$ & $0$ \\
(A) & $4$ & $5$ & $0$ & $0$ & $0$ \\ 
(B) & $4$ & $5$ & $-\pi$ & $\pi - \pi$ & $\pi$ \\ 
(C) & $4$ & $5$ & $-\pi/4$ & $\pi/2 - \pi/4$ & $\pi/2$ \\ 
(D) & $4$ & $5$ & $\pi/3$ & $3\pi/2 + \pi/3$ & $3\pi/2$ \\ 
\end{tabular}
\label{tab:parm_pol}
\end{ruledtabular}
\end{table}

Our setup is motivated by kinetic-MHD hybrid simulations of Abrupt Large-amplitude Events (ALE) as reported in Ref.~\onlinecite{Bierwage18}. We also use the same MHD equilibrium field ${\bm B}_\eq$. The magnetic axis is located at major radius $R_0 = 3.4\,{\rm m}$ and height $z_0 = 0.2\,{\rm m}$. The central field strength is $B_0 = |{\bm B}_\eq(R_0,z_0)| = 1.16\,{\rm T}$ and the plasma current is $I_{\rm p} = 0.57\,{\rm MA}$. Field and current both flow in the $-\varphi$ direction, so that $\Psi_{\rm P}$ increases monotonically from the center to the edge of the plasma. As a reference for our perturbation model, we have chosen the $n=2$ harmonic of the fluctuating field, whose temporal evolution during the ALE simulation is shown in Fig.~\ref{fig:01_ale_mdl}(a). The peak around time $t \approx 0.8\,{\rm ms}$ is the ALE.\footnote{The ALE also has large $n=1$ and $n=3$ components\cite{Bierwage18} that are ignored here.}
The poloidal structure of the $n=2$ component of the electric potential at that time is shown in Fig.~\ref{fig:01_ale_mdl}(b), and panel (c) shows the radial profiles of its main poloidal Fourier components with $m = 1...7$.

\begin{figure}[tbp]
\centering\vspace{-0.1cm}
\includegraphics[width=0.48\textwidth]{\figures/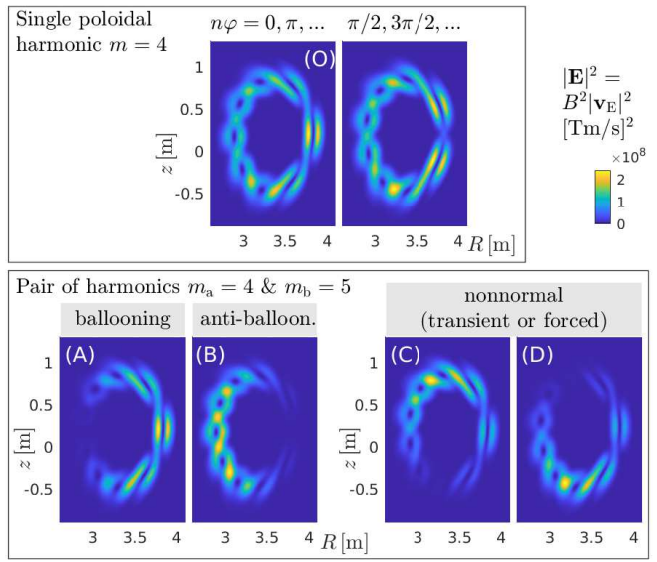}
\caption{Examples of modeled electric field intensity distributions $|{\bm E}|^2 = B^2|{\bm v}_{\rm E}|^2$ in the poloidal $(R,z)$ plane. The fields are designed to resemble normal and nonnormal modes. Normal (eigen)modes of a toroidal plasma possess primarily in-out asymmetry as in cases (O), (A) and (B). Cases (C) and (D) show extreme examples of nonnormal modes that are entirely out-of-phase with the magnetic field nonuniformity $B \propto 1/R$ and peak at the top and bottom of the plasma.}
\label{fig:02_modes}%
\end{figure}

Our simplified model is shown in Fig.~\ref{fig:01_ale_mdl}(d). We include only two poloidal harmonics, $m_{\rm a} = 4$ and $m_{\rm b} = 5$, and let them have identical radial profiles $\Phi_m(\rho)$. Their amplitude parameter in Eq.~(\ref{eq:mdl_prof}) is fixed at $\Phi_{0,m} \approx 22.2\,{\rm kV}$, which yields profiles with a peak value of ${\rm max}\{\Phi_m(\rho)\} \approx 0.8\,{\rm kV}$ that is approximately 5 times smaller than what is seen in Fig.~\ref{fig:01_ale_mdl}(c). Our model for $\Phi$ gives rise to an electric drift velocity field whose values reach ${\rm max}|{\bm v}_{\rm E}| \approx 1.5\times 10^4\,{\rm m/s}$. This corresponds to $0.35\%$ of the on-axis Alfv\'{e}n velocity $v_{\rm A0} \approx 4.3\times 10^6\,{\rm m/s}$ and amounts to about $60\%$ of the peak velocity seen during the ALE in Fig.~\ref{fig:01_ale_mdl}(a).

The poloidal mode structures of the five model cases (O), (A), (B), (C) and (D) listed in Table~\ref{tab:parm_pol} differ primarily in the poloidal structure of the electric field intensity $|{\bm E}|^2(R,z)$, as shown in Fig.~\ref{fig:02_modes}. Cases (A)--(D) are similar but not identical to those used in the companion paper,\cite{Bierwage22b} one difference being the choice of magnetic coordinates. As was explained in Ref.~\onlinecite{Bierwage22b}, the beat between the wave components $\Phi_m(\rho)e^{i\Theta_{0,m} - im\vartheta}$ in Eq.~(\ref{eq:mdl}) determines the poloidal location of the intensity peak, which is independent of the toroidal angle. The two parameters controlling the beat wave are the difference $\Delta m = m_{\rm b} - m_{\rm a}$ between the poloidal mode numbers, and their relative phase $\Delta\Theta_0 = \Theta_{0,{\rm b}} - \Theta_{0,{\rm a}}$ (Table~\ref{tab:parm_pol}).

\begin{figure*}[tbp]
\centering\vspace{-0.35cm}
\includegraphics[width=0.96\textwidth]{\figures/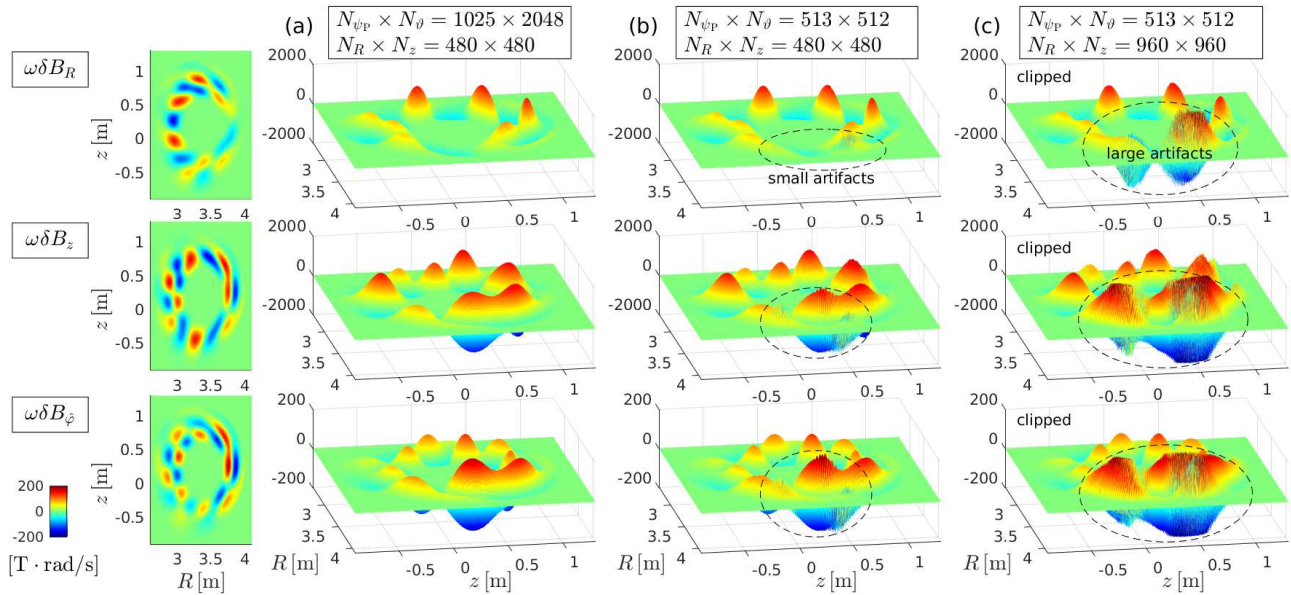}
\caption{Poloidal structure of the magntic perturbation $\omega\delta{\bm B} = \Im\{\nablab\times\widetilde{\bm E}\}$ in scenario (ii) for the nonnormal mode in case (C). The three rows show the $R$, $z$ and $\hat{\varphi}$ component, respectively, where $\delta B_{\hat{\varphi}} = \delta{\bm B}\cdot\partial_\varphi{\bm x}/|\partial_\varphi{\bm x}| = \delta B_\varphi/R$. Column (a) shows 2-D and 3-D contour plots of the default case. The number of grid points in the original flux coordinate mesh, where $\widetilde{\Phi}_{\rm mdl}$ was modeled, is $N_{\psi_{\rm P}}\times N_\vartheta = 1025\times 2048$. Using scattered data interpolation as described in the text, $\widetilde{\Phi}_{\rm mdl}$ was then mapped to a cylindrical mesh with $N_R\times N_z = 480\times 480$ grid points. Then we computed $\widetilde{\bm E} = -\nabla_\perp^\eq\widetilde{\Phi}$ and finally $\nablab\times\widetilde{\bm E}$ using 4th-order finite differences. Columns (b) and (c) show the situation with reduced resolution in $(\psi_{\rm P},\vartheta)$. Dashed ellipses indicate coordinate transformation artifacts that attained readily visible amplitudes on the low-field side of the plasma, where the $(\psi_{\rm P},\vartheta)$ grid is sparsest (cf.\ Fig.~\protect\ref{fig:01_ale_mdl}(b)). In column (c), we have increased the resolution in $(R,z)$ without changing the $(\psi_{\rm P},\vartheta)$ grid, which enhanced the artifacts. The contour plots in (c) were clipped to have the same vertical axis limits as (a) and (b).}\vspace{-0.15cm}
\label{fig:03_rotE_converg}%
\end{figure*}

\subsection{Normal and nonnormal modes}
\label{sec:model_cases}

Cases (O), (A) and (B) in Fig.~\ref{fig:02_modes} are meant to represent the electric field intensity distribution $|{\bm E}|^2(R,z)$ of normal (eigen)modes of a toroidal plasma. Normal modes are fluctuation patterns whose overall structure in the ideal MHD domain (outside resonant layers) remains unchanged for infinitely long times, even if the amplitude is damped. In a toroidal plasma, normal modes have electric field intensity distributions $|{\bm E}|^2 = B^2|{\bm v}_{\rm E}|^2$ that possess an in-out asymmetry in the poloidal $(R,z)$ plane, in accordance with the variation of the magnetic field strength $B \propto 1/R$. Case (O) resembles the structure of a beta-induced Alfv\'{e}n eigenmode (BAE)\cite{Heidbrink93} that is dominated by a single poloidal harmonic (here $m=4$). One can see in Fig.~\ref{fig:02_modes}(O) that the toroidal geometry of the magnetic field produces a certain in-out asymmetry in $|{\bm E}|^2$ along the $R$-axis even for a single-$m$ mode. That in-out asymmetry is enhanced in multi-$m$ modes, such as cases (A) and (B) in Fig.~\ref{fig:02_modes} that consist of two adjacent poloidal harmonics, $m_{\rm a} = 4$ and $m_{\rm b} = 5$, and resemble the structure of a toroidicity-induced Alfv\'{e}n eigenmode (TAE).\cite{Cheng85} While the poloidal harmonics of true TAEs peak at different radii (near rational surfaces of the field helicty $q$), the two harmonics in our model were chosen to have identical profiles in order to emphasize the poloidal symmetry breaking of the resulting beat wave. Modes peaking on the low-field side (larger $R \propto B^{-1}$) are said to have a `ballooning' structure (A) and those peaking on the high-field side (smaller $R$) are said to be `anti-ballooning' (B). Note that the poloidal phase of the beat is independent of the toroidal angle $\varphi$. Only the locations of local minima and maxima vary, as one can see by comparing the two panels for case (O) in the upper box of Fig.~\ref{fig:02_modes}, which show the structure of $|{\bm E}|^2(R,z)$ at $n\varphi = 0,\pi,...$ and $n\varphi = \pi/2,3\pi/2,...$, respectively.

Cases (C) and (D) in Fig.~\ref{fig:02_modes} are two arbitrary examples of nonnormal modes, whose intensity peaks are located at the top and bottom of the plasma, respectively, so they are $\pm 90$ degrees out-of-phase with the variation of the field strength $B$. Note that the nonnormal modes are not unphysical. They are merely unlikely to survive for long periods of time without suitable external drive, so they tend to be transient or change their shape rapidly when forming spontaneously during the nonlinear evolution of a plasma.

\subsection{Coordinate transformation, differentiation and numerical artifacts}
\label{sec:model_transform}

After preparing the poloidal mode structure of $\widetilde{\Phi}$, we mapped it from uniformly meshed polar coordinates $(\psi_{\rm P},\vartheta)$ to uniformly meshed rectangular coordinates $(R,z)$ using {\tt MATLAB}'s {\tt scatteredInterpolant} and {\tt interp1} functions with method `{\tt linear}'.\footnote{{\tt MATLAB} release 2020a was used in this work.}
Finally, we took derivatives of $\widetilde{\Phi}$ to compute $\widetilde{\bm E}$ and $\delta\widetilde{\bm B}$. This was done using 4th-order finite differencing routines adopted from the hybrid code {\tt MEGA}.\cite{Todo98, Todo05, Todo06}

The mapping $(\psi_{\rm P},\vartheta) \rightarrow (R,z)$ is a critical step. It is desirable to have at least one grid point of $(\psi_{\rm P},\vartheta)$ in each cell of the $(R,z)$ mesh, but this is obviously difficult to realize throughout the plasma, because the grid points of the magnetic coordinate mesh are distributed in a highly nonuniform fashion in the $(R,z)$ plane, as one can infer from the thin black lines in Fig.~\ref{fig:01_ale_mdl}(b). This produces interpolation artifacts that tend to be largest in regions where the flux coordinate grid is relatively sparse, especially around the outer midplane on the low-field side of the plasma ($R \sim 3.5...4.1\,{\rm m}$).

As an example, Fig.~\ref{fig:03_rotE_converg} shows the structure of the magnetic perturbation $\Im\{\nablab\times\widetilde{\bm E}\} = \Im\{i\omega\delta\widetilde{\bm B}\} = \Re\{\omega\delta\widetilde{\bm B}\}$ in case (C) for scenario (ii) defined in Eq.~(\ref{eq:scenario_2}), where $\widetilde{\bm E} = -\nablab_\perp^\eq\widetilde{\Phi}$. There are no readily visible artifacts in the default case in column (a) of Fig.~\ref{fig:03_rotE_converg}. The artifacts become visible in column (b), where the resolution in polar coordinates $(\psi_{\rm P},\vartheta)$ was reduced by a factor 2 in both dimensions. These noise-like artifacts grow further to exceed the magnitude of the signal in column (c), where the resolution in $(R,z)$ was increased by a factor 2 while keeping the reduced resolution in $(\psi_{\rm P},\vartheta)$.

Note that, in spite of these interpolation artifacts, the magnetic field vector satisfies the solenoidal condition $\nabla\cdot\delta\widetilde{\bm B} = 0$ when the divergence is computed using a 4th-order finite-difference scheme that is consistent with the scheme used to evaluate the rotation operator, $\nablab\times\widetilde{\bm E}$. For all data in Fig.~\ref{fig:03_rotE_converg}, we find ${\rm max}|\nabla\cdot\delta\widetilde{\bm B}| \sim \O(10^{-14}\,{\rm T/m})$, which is not far from the theoretical limit of 64-bit floating-point precision.

Most results reported in the following Section~\ref{sec:results} were obtained using the $\delta\widetilde{\bm B}$ field in column (a) of Fig.~\ref{fig:03_rotE_converg}, where the interpolation artifacts are sufficiently small to be not readily visible at a glance. The only exception will be Fig.~\ref{fig:11_cfo-gco_low-res_C}, where we examine the effect of the noise seen in Fig.~\ref{fig:03_rotE_converg}(b).

\begin{figure}[tbp]
\centering\vspace{-0.1cm}
\includegraphics[width=0.48\textwidth]{\figures/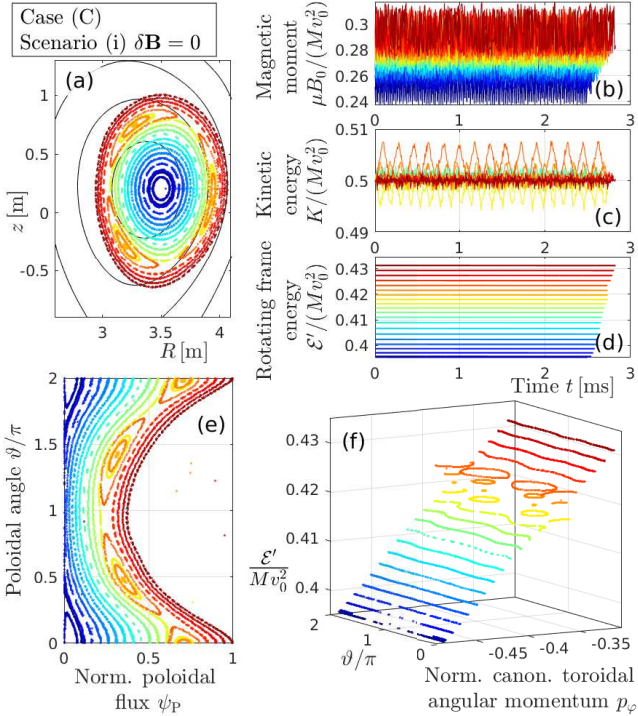}
\caption{Results for the nonnormal mode case (C) in scenario (i) with $\delta{\bm B} = 0$, simulated with the full orbit model. Each color represents a different initial position of a tracer particle along the $\psi_{\rm P}$-axis at $\vartheta = 0$. Panels (a), (e) and (f) show Poincar\'{e} plots in configuration space, flux space and angular momentum space. Panels (b--d) show time traces of the normalized magnetic moment $\hat{\mu}(t)$, kinetic energy $\hat{K}(t)$ and rotating frame energy $\hat{\E}'(t)$. The latter is conserved with high accuracy and used also for the vertical axis in panel (f).}
\label{fig:04_cfo_overview_C}%
\end{figure}

\section{Results}
\label{sec:results}

For the purpose of introducing the analysis and visualization methods used in this study, Fig.~\ref{fig:04_cfo_overview_C} shows an overview of the results for the nonnormal mode case (C) simulated with the full orbit model that is described in Appendix~\protect\ref{apdx:fo}. We have chosen scenario (i) defined in Eq.~(\ref{eq:scenario_1}), where the effect of the interpolation artifacts is weakest, since $\delta{\bm B} = 0$, so that conservation laws are satisfied with high accuracy.

Figure~\ref{fig:04_cfo_overview_C}(a) shows a Poincar\'{e} plot of the phase space topology in the poloidal plane that moves with the toroidally propagating wave field as $\varphi + \omega t/n = 0$. For full orbit simulations like the one whose results are shown in Fig.~\ref{fig:04_cfo_overview_C}, our Poincar\'{e} maps do not show the actual positions ${\bm x}$ of simulation particles. Instead, we recorded the points in the $(R,z)$-plane where their GC positions ${\bm x}^{\rm gc} = {\bm x} + {\bm v}\times{\bm B}/\Omega_{\rm g}$ satisfy $\varphi^{\rm gc}(t) = -\omega t/n$, where $\Omega_g = ZeB/M$ is the instantaneous value of the gyrofrequency. Also recorded at these times were the samples of the magnetic moment $\mu = M v_\perp^2/(2B)$, kinetic energy $K = M v^2/2$ and rotating frame energy $\E'$ of the simulation particles. Their time traces are shown in panels (b)--(d) of Fig.~\ref{fig:04_cfo_overview_C}, using the normalizations $\hat{\mu} = \mu B_0/(M v_0^2)$, $\hat{K} = K/(M v_0^2)$ and $\hat{\E}' = \E'/(Mv_0^2)$, where velocities are normalized by their initial value $v_0 = \sqrt{2K_0/M}$. The normalized rotating frame energy
\begin{equation}
\hat{\E}' = \hat{K} + \frac{\hat{\Phi} + \omega\hat{\P}_\varphi/n}{\rho_0},
\label{eq:erot_nrm}
\end{equation}

\noindent with characteristic gyroradius $\rho_0 = v_0/\Omega_{\rm g0}$ and on-axis gyrofrequency $\Omega_{\rm g0} = ZeB_0/M \approx 2\pi\times 9\,{\rm MHz}$, is evaluated with
\begin{subequations}
\begin{gather}
\hat{\Phi} = \frac{\Phi}{v_0 B_0}, \quad
\delta\hat{A}_\varphi = R\frac{\delta A_{\hat{\varphi}}}{B_0} = \Re\left\{\frac{in\widetilde{\Phi} - R \widetilde{E}_{\hat{\varphi}}}{-i\omega v_0 B_0}\right\},
\\
\hat{\P}_\varphi = \frac{\P_\varphi}{Ze B_0} = \frac{\Psi_{\rm P} + \delta A_\varphi}{B_0} + \rho_0 \hat{v}_\varphi.
\label{eq:pphi_nrm}
\end{gather}
\label{eq:epot_pphi_nrm}
\end{subequations}

In most cases we use deuterons with $K_0 \equiv K(t=0) = 400\,{\rm keV}$ ($\rho_0 \approx 0.11\,{\rm m}$) and initial pitch angle $\alpha(t=0) = \sin^{-1}(v_\parallel(t=0)/v_0) = 0.25\pi$ in the domain of co-passing particles. The only exception is Fig.~\ref{fig:08_cfo_1keV_ACD}, where $K_0 = 1\,{\rm keV}$ ($\rho_0 = 6\,{\rm mm}$). Deuterons gyrate clockwise in the $(R,z)$ plane and the initial direction of their perpendicular velocity ${\bm v}_\perp$ is chosen to be $\hat{\bm e}_z\times \hat{\bm b}$ (pointing approximately in the positive-$R$ direction). The colors in Fig.~\ref{fig:04_cfo_overview_C} identify the initial position of a simulation particle along the $\psi_{\rm P}$-axis at $\vartheta = 0$.

The same data as in Fig.~\ref{fig:04_cfo_overview_C}(a) are plotted again in Fig.~\ref{fig:04_cfo_overview_C}(e), now mapped into polar coordinates $(\psi_{\rm P},\vartheta)$. Here one can clearly see the radial extent of the magnetic drifts. Finally, Fig.~\ref{fig:04_cfo_overview_C}(f) shows a 3-D plot of the Poincar\'{e} data with the rotating frame energy $\hat{\E}'$ along the vertical axis and the normalized canonical toroidal angular momentum
\begin{equation}
p_\varphi = \frac{\P_\varphi/(Ze)}{\Psi_{\rm P,edge} - \Psi_{\rm P,0}} = \frac{\Psi_{\rm P} + B_0 v_\varphi/\Omega_{\rm g0}}{\Psi_{\rm P,edge} - \Psi_{\rm P,0}}
\end{equation}

\noindent serving as a radial coordinate. Note that $v_\varphi < 0$ in our setup. In the limit $v_\varphi/\Omega_{\rm g0} = \hat{v}_\varphi\rho_0 \rightarrow 0$, we have $p_\varphi \rightarrow \psi_{\rm P}$. Similarly to $\psi_{\rm P}$, which measures radial distance in flux space, $p_\varphi$ is a radial coordinate in drift orbit space.

Figure~\ref{fig:04_cfo_overview_C}(f) shows that all Poincar\'{e} contours are distinct in energy even when they appear to overlap in the 2-D projection of panels (a) and (e). Such overlaps can be avoided by initializing simulation particles with the same rotating frame energy $\E'(t=0)$ rather than $K_0$, but this was not done here for various reasons, including mere simplicity.

Our main point of interest here is the long-time evolution of the system on the multi-millisecond time scale. Each Poincar\'{e} contour in Fig.~\ref{fig:04_cfo_overview_C} consists of 450 points, and it took about $3\,{\rm ms}$ of physical time for our deuterons to complete their 450 toroidal transits in the wave's moving frame of reference.

The most efficient wave-particle interactions occur near resonances. Here, the dominant resonance is located around $p_\varphi \approx -0.37$ ($\psi_{\rm P} = 0.3...0.7$) and has three elliptic (O-)points in the poloidal direction. Note that it is the same island that appears three times in $0\leq \vartheta< 2\pi$. Particles inside the resonant island exhibit the largest oscillations in the kinetic energy $\hat{K}(t)$ in Fig.~\ref{fig:04_cfo_overview_C}(c), with $|\Delta\hat{K}|/\hat{K} \lesssim 2\%$ ($\approx 8\,{\rm keV}$) starting from $\hat{K}_0 = 0.5$. The oscillations of $\hat{K}(t)$ in Fig.~\ref{fig:04_cfo_overview_C}(c) occur between constant bounds, and $\E'(t)$ in Fig.~\ref{fig:04_cfo_overview_C}(d) is effectively constant, showing the conservative character of the system at hand.

The magnetic moment $\hat{\mu}(t)$ in Fig.~\ref{fig:04_cfo_overview_C}(b) is also conserved on average. Its oscillation has a magnitude of about $|\Delta\hat{\mu}|/\hat{\mu} \sim 10\%$ that is similar for all particles, inside and outside the resonant island. This is because the oscillations in $\hat{\mu}(t)$ are mainly due to the nonuniformity of the magnetic field across the gyroorbit, whose diameter is about $2\rho_{\rm g} \approx 2\rho_0 v_\perp/v_0 \approx 0.16\,{\rm m}$ for our energetic deuterons with $K_0 = 400\,{\rm keV}$.

\subsection{Poloidal mode structure matters}
\label{sec:results_mode}

In this section, we demonstrate how the poloidal mode structure of the fluctuating fields affects the results of an orbit-following simulation, in particular, with respect to its conservative character. We begin with the full orbit simulations that solve the equations given in Appendix~\ref{apdx:fo}.

Figure~\ref{fig:05_cfo_overview_ABCD} contains an overview of Poincar\'{e} plots showing the phase space topology. Results for cases (A)--(D) are arranged column-wise, with the normal mode cases (A) and (B) on the left and the nonnormal mode cases (C) and (D) on the right. That is, we vary the effect of `mode geometry' between each column. For convenience, the upper row shows the poloidal structures $\Phi(R,z)$ of the four modes and the locations of the Poincar\'{e} contours of our co-passing $400\,{\rm keV}$ deuterons in the $(R,z)$ plane of our JT-60U plasma.

Results for scenarios (i)--(iii) defined by Eqs.~(\ref{eq:scenario_1})--(\ref{eq:scenario_3}) are arranged row-wise in Fig.~\ref{fig:05_cfo_overview_ABCD}. The electrostatic scenario (i) is at the top, followed by the transversely polarized electromagnetic scenario (ii), and the unphysical scenario (iii) in the bottom row, which differs from the self-consistent scenarios (i) and (ii) by the simultaneous neglect of both the parallel electric field component $E_\parallel^\eq$ and the time-dependent magnetic fluctuation $\delta{\bm B}(t)$. In other words, we vary the `mode electromagnetism' between rows (i)--(iii).

In the physical scenarios (i) and (ii), the main difference between the four cases (A), (B), (C) and (D) is the width of the primary resonant island. Cases (A) and (B) have the largest and smallest width, respectively, because our co-passing deuterium orbits are shifted outward in $R$, where the modes in cases (A) and (B) have high and low intensity, respectively.

Concerning the conservative character of the dynamics, scenario (i) exhibits good invariant surfaces in all cases shown in Fig.~\ref{fig:05_cfo_overview_ABCD}. The same is true for scenario (ii), although a small amount of broadening can be seen on Poincar\'{e} contours around $p_\varphi > -0.35$, which are associated with orbits that pass near the plasma boundary on the outboard side ($R \approx 4.1\,{\rm m}$), where numerical errors in $\delta{\bm B}$ are expected to be largest (according to Fig.~\ref{fig:03_rotE_converg}). In both physical scenarios (i) and (ii), the magnetic moment $\hat{\mu}(t)$ and kinetic energy $\hat{K}(t)$ are well-conserved on average. Their time traces are similar to those in Fig.~\ref{fig:04_cfo_overview_C}(b,c) for case (C), so we have not plotted them again for the other cases. The same counts for the rotating frame energy $\E'(t)$ that was shown in Fig.~\ref{fig:04_cfo_overview_C}(d) and is effectively constant.

A dramatically different picture is seen in the scenario (iii) with an unphysical mode structure due to the neglect of both $E_\parallel^\eq$ and $\delta{\bm B}(t)$. Panels (C-iii) and (D-iii) of Fig.~\ref{fig:05_cfo_overview_ABCD} show that the Poincar\'{e} contours for the nonnormal mode cases (C) and (D) are smeared in the $p_\varphi$ direction, especially near the resonance. The corresponding time traces of $\hat{\E}'(t)$ in Fig.~\ref{fig:06_cfo_k_ABCD}(C,D) show a significant amount of secular acceleration that violates the conservation laws of our setup. Very similar behavior was reported in the companion paper,\cite{Bierwage22b} where we had systematically violated the conservation of energy and phase space density by omitting certain terms in the equations of motion. Here, the same effect was achieved by corrupting the self-consistency of the wave's poloidal mode structure in scenario (iii) by letting both $E_\parallel^\eq$ and $\delta{\bm B}(t)$ be zero simultaneously.

\begin{figure*}[tbp]
\centering\vspace{-0.3cm}
\includegraphics[width=0.96\textwidth]{\figures/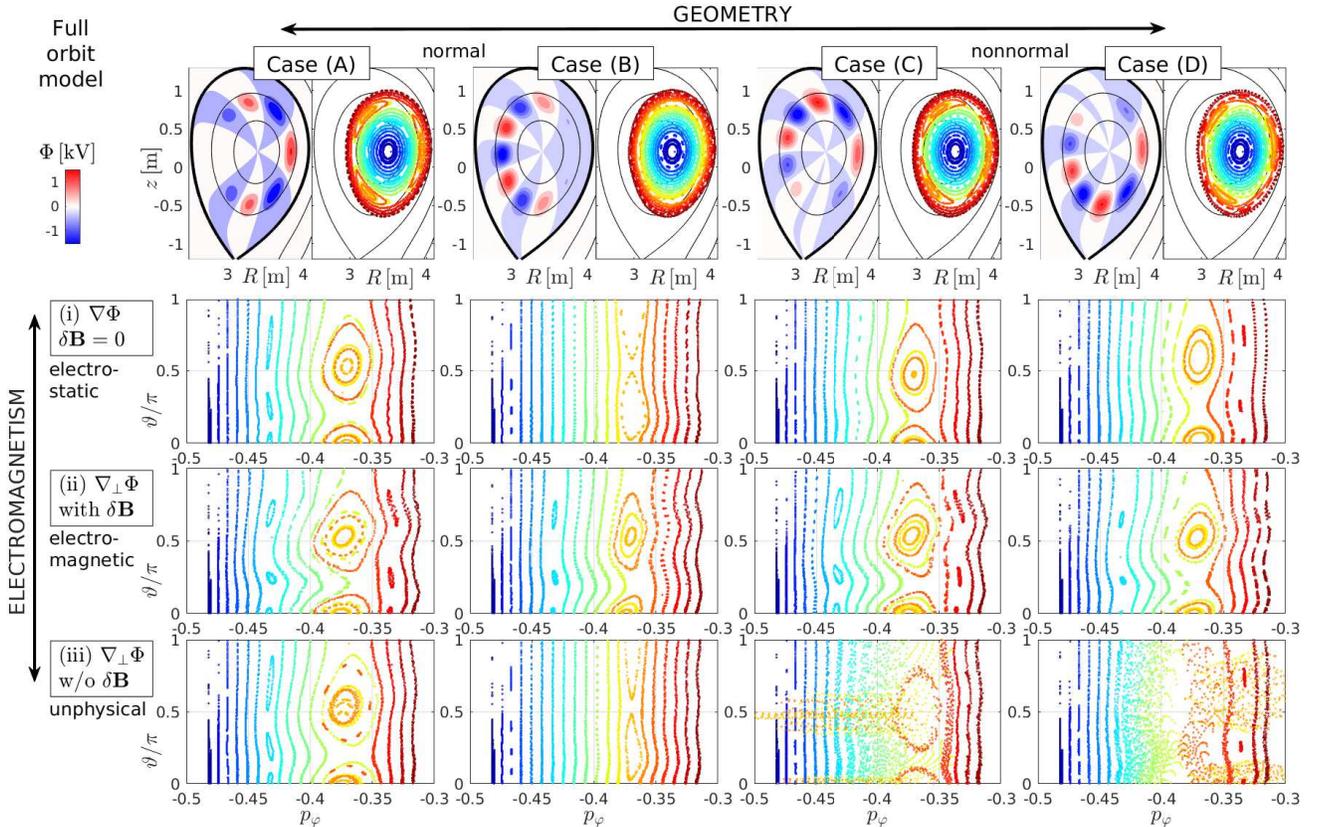}
\caption{Overview of results from full orbit simulations. The top row shows the mode structures of the electric potential $\Phi$ for cases (A)--(D), arranged column-wise. Next to each mode structure is the respective Poincar\'{e} plot for scenario (i) in the $(R,z)$ plane. The three rows below show the respective Poincar\'{e} plots in polar coordinates $(p_\varphi,\vartheta)$ for scenarios (i)--(iii) defined by Eqs.~(\protect\ref{eq:scenario_1})--(\protect\ref{eq:scenario_3}), arranged row-wise. Note that only the range $0 \leq \vartheta \leq \pi$ is shown, since the structures in the other half of the domain ($\pi\leq \vartheta \leq 2\pi$) look similar (cf.~Fig.~\protect\ref{fig:04_cfo_overview_C}(d) above). The particles were followed for about $3\,{\rm ms}$ and the time traces of $\hat{\E}'$ and $\hat{\mu}$ for the unphysical scenario (iii) are shown in Figs.~\protect\ref{fig:06_cfo_k_ABCD} and \protect\ref{fig:07_cfo_3d_CD}(c,d) below.}
\label{fig:05_cfo_overview_ABCD}%
\end{figure*}

\begin{figure*}[tbp]
\centering
\includegraphics[width=0.96\textwidth]{\figures/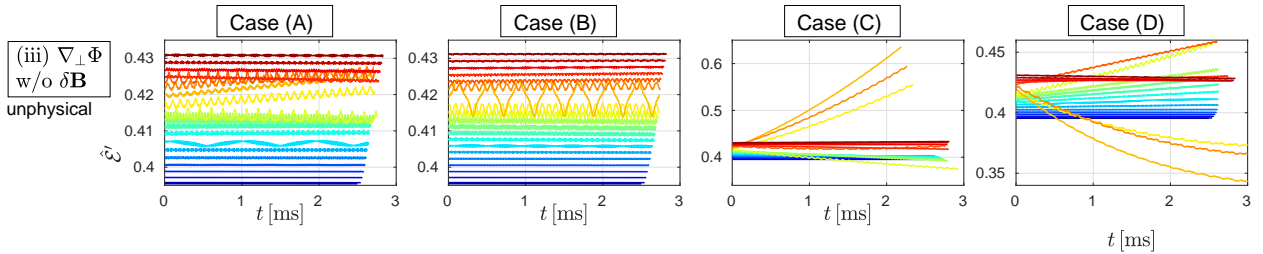}
\caption{Time traces of the rotating frame energy $\hat{\E}'(t)$ for the unphysical scenario (iii) shown in the bottom row of Fig.~\protect\ref{fig:05_cfo_overview_ABCD} above.}
\label{fig:06_cfo_k_ABCD}%
\end{figure*}

\begin{figure}[tbp]
\centering\vspace{-0.05cm}
\includegraphics[width=0.48\textwidth]{\figures/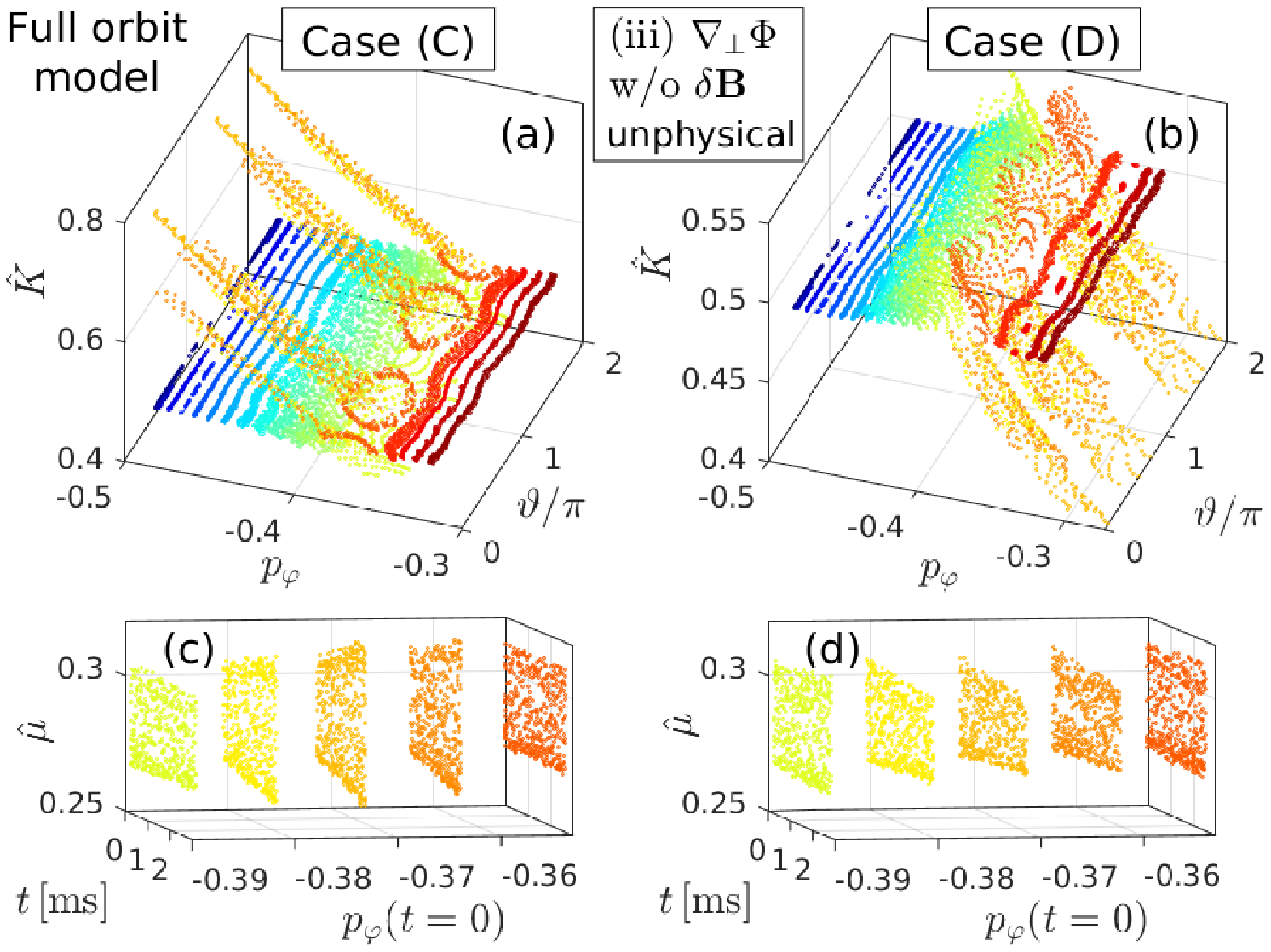}
\includegraphics[width=0.48\textwidth]{\figures/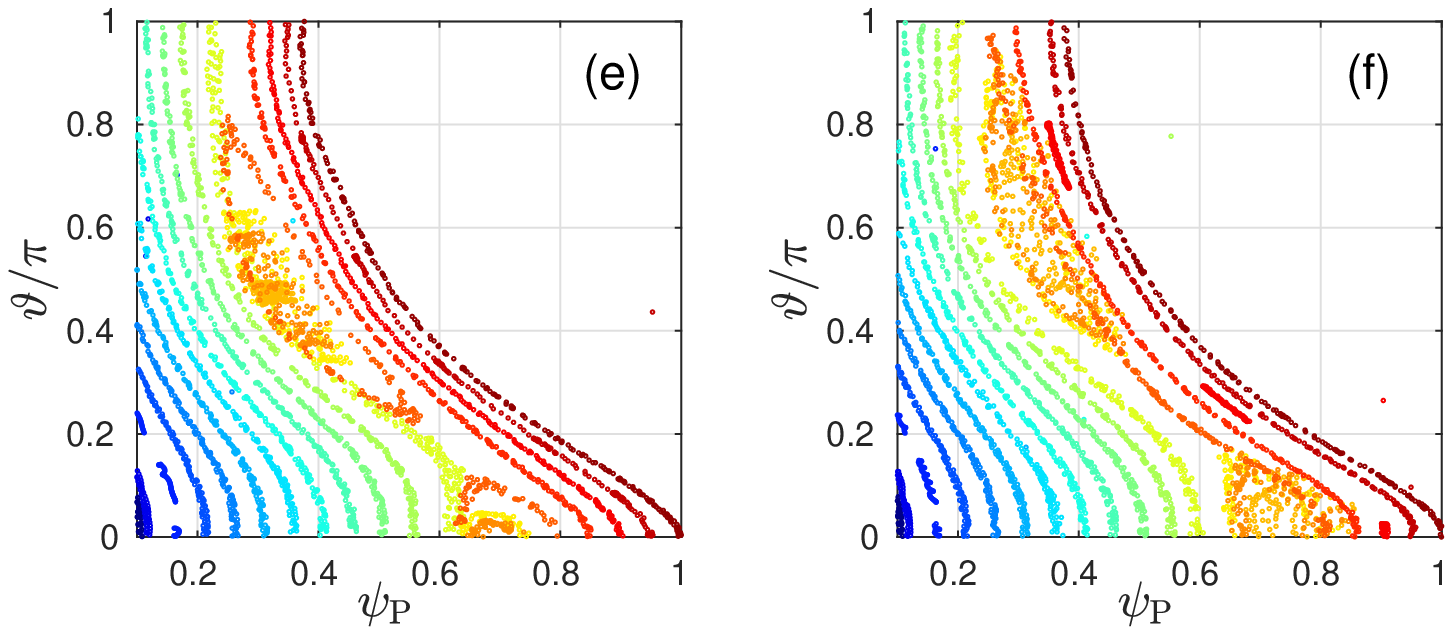}
\caption{Panels (a) and (b) show 3-D plots of the data for the unphysical scenario (iii) of cases (C) and (D) from Fig.~\protect\ref{fig:05_cfo_overview_ABCD} above, with kinetic energy $\hat{K}(t)$ along the vertical axis. Panels (c) and (d) show the time traces of the magnetic moment $\hat{\mu}(t)$ for five simulation particles around the dominant resonance at $p_\varphi \approx -0.37$. Panels (e) and (f) show plots of the Poincar\'{e} data in magnetic coordinates $(\psi_{\rm P},\vartheta)$.}
\label{fig:07_cfo_3d_CD}%
\end{figure}

A 3-D view of the situation is shown in Fig.~\ref{fig:07_cfo_3d_CD}(a,b), which combines the data from Figs.~\ref{fig:05_cfo_overview_ABCD}(C-iii,D-iii) and Fig.~\ref{fig:06_cfo_k_ABCD}(C,D). In addition, Fig.~\ref{fig:07_cfo_3d_CD}(c,d) shows time traces of the magnetic moment for particles near the dominant resonance. The secular acceleration occurs primarily in $v_\parallel = \hat{\bm b}\cdot{\bm v}$ ($\approx u$ in the GC model), so that $\hat{\mu} \propto v_\perp^2$ (and thus the gryoradius) remains roughly conserved on average. Only the magnitude of its variation $\Delta\hat{\mu}$ around the mean changes somewhat. Moreover, panels (e) and (f) of Fig.~\ref{fig:07_cfo_3d_CD} show that there is relatively little secular displacement along $\psi_{\rm P}$ (and, thus, in configuration space), so the large secular drift in $p_\varphi(t)$ is primarily due to acceleration in $v_\varphi$, which is similar to $v_\parallel$ in a tokamak.

Interestingly, the unphysical secular acceleration is relatively small in cases (A) and (B) with normal mode structures. One has to look closely at the Poincar\'{e} plots in panels (A-iii) and (B-iii) of Fig.~\ref{fig:05_cfo_overview_ABCD} to see the deviations from conservative motion. Only the orbits closest to the island center can be seen to exhibit a suspicious drift in $p_\varphi$, and the time traces of their rotating frame energy $\hat{\E}'(t)$ in Fig.~\ref{fig:06_cfo_k_ABCD}(A) have to be traced for at least a millisecond to see some appreciable acceleration. Nevertheless, the violation of rotating frame energy conservation is already evident on shorter time scales as $\hat{\E}'(t)$ in Fig.~\ref{fig:06_cfo_k_ABCD}(A,B) oscillates at what appears to be the nonlinear bounce frequency.

\begin{figure}[tbp]
\centering
\includegraphics[width=0.48\textwidth]{\figures/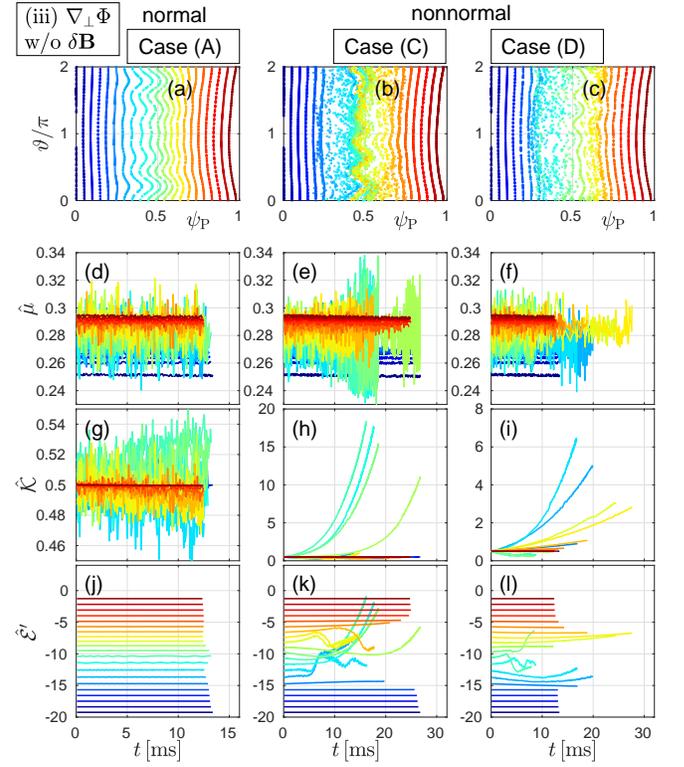}
\caption{Motion of far-off-resonant co-passing $1\,{\rm keV}$ deuterons in the unphysical scenario (iii) simulated with the full orbit model. Results for cases (A), (C) and (D) are arranged column-wise. Panels (a--c) show Poincar\'{e} plots in the $(\psi_{\rm P},\vartheta)$ plane. Panels (d--f), (g--h) and (j--l) show, respectively, the time traces of the magnetic moment $\hat{\mu}(t)$, kinetic energy $\hat{K}(t)$, and rotating frame energy $\hat{\E}'(t)$.}
\label{fig:08_cfo_1keV_ACD}%
\end{figure}

The secular acceleration in the unphysical scenario (iii) is not limited to particles inside or near a primary resonance. Figure~\ref{fig:08_cfo_1keV_ACD} shows similar behavior in the case of far-off-resonant deuterons with an energy of only $K_0 = 1\,{\rm keV}$. Again, the nonconservative character of scenario (iii) remains largely hidden in the normal mode case (A), while it causes strong secular acceleration in the nonnormal mode cases (C) and (D).

Compared to the $K_0 = 400\,{\rm keV}$ orbits analyzed in Fig.~\ref{fig:07_cfo_3d_CD}, the relative increase in the kinetic energy $\Delta K / K_0$ is larger in Fig.~\ref{fig:08_cfo_1keV_ACD} for $K_0 = 1\,{\rm keV}$ because we kept the large value of the wave's potential energy $Ze\Phi$ while reducing $K_0$. The large potential energy $Ze\Phi$ of the wave field in our setup also exceeds the magnitude of the fluctuations $\Delta\mu B_0$ of the magnetic moment for $1\,{\rm keV}$ deuterons. This can be clearly seen in the time traces of $\hat{\mu}(t)$ in Fig.~\ref{fig:08_cfo_1keV_ACD}(d--f), which fluctuate by a large amount only near the mode's peak, namely in the domain of green and yellow contours. In contrast $\hat{\mu}$ is essentially constant for particles that are located outside the mode (dark blue and dark red contours), because the magnetic field varies very little across their small gyroorbit with a diameter of only $2\rho_{\rm g} \approx 2\rho_0 v_\perp/v_0 \approx 8\,{\rm mm}$, unlike in Fig.~\ref{fig:04_cfo_overview_C}(b).

\begin{figure*}[tbp]
\centering
\includegraphics[width=0.96\textwidth]{\figures/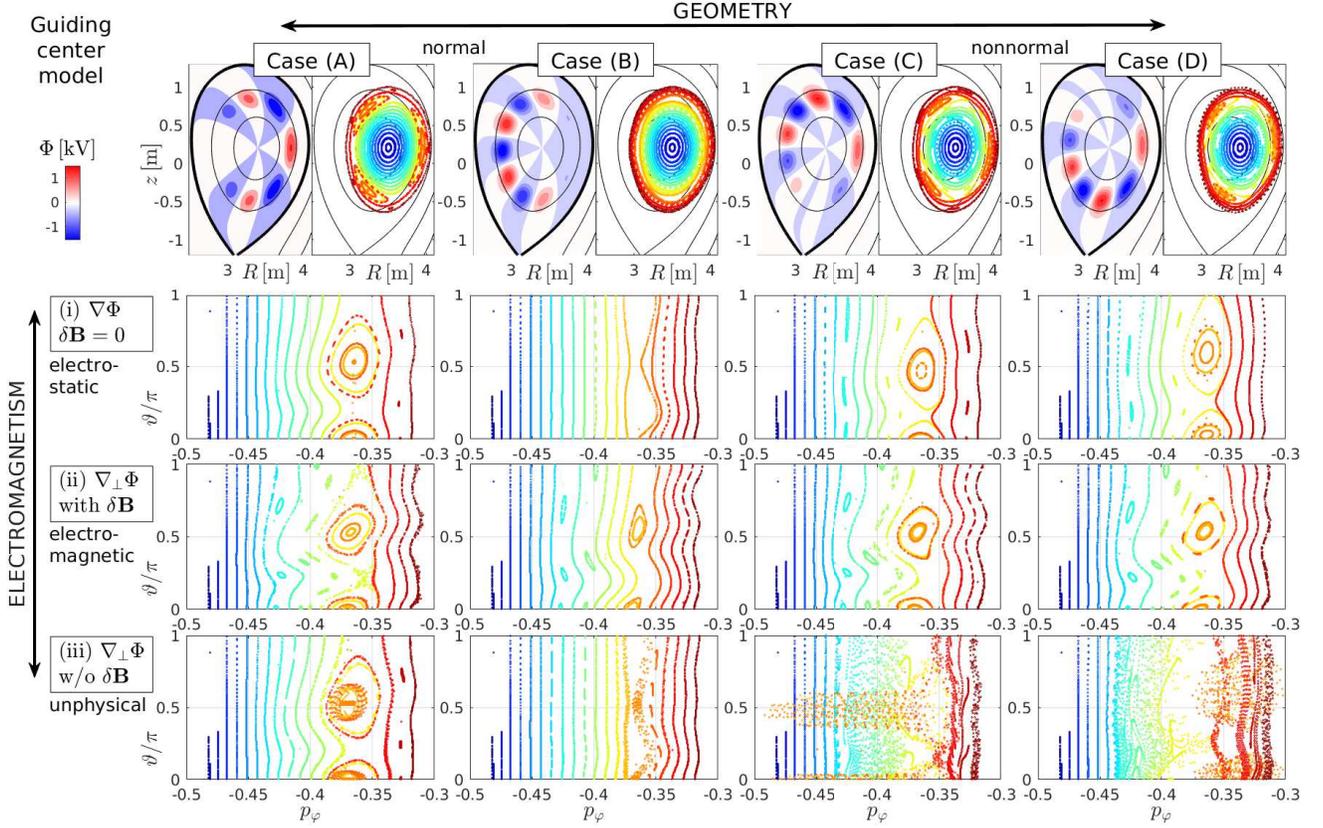}
\caption{Overview of results from GC orbit simulations of co-passing $400\,{\rm keV}$ deuterons. The top row shows the mode structures of the electric potential $\Phi$ for cases (A)--(D), arranged column-wise. Next to each mode structure is the respective Poincar\'{e} plot for scenario (i) in the $(R,z)$ plane. The three rows below show the respective Poincar\'{e} plots in polar coordinates $(p_\varphi,\vartheta)$ for scenarios (i)--(iii) defined by Eqs.~(\protect\ref{eq:scenario_1})--(\protect\ref{eq:scenario_3}), arranged row-wise. Note that only the range $0 \leq \vartheta \leq \pi$ is shown, since the structures in the other half of the domain ($\pi\leq \vartheta \leq 2\pi$) look similar (cf.~Fig.~\protect\ref{fig:04_cfo_overview_C}(d) above). The particles were followed for about $3\,{\rm ms}$ and the time traces of the rotating frame energy $\hat{\E}'$ are shown in Fig.~\protect\ref{fig:10_gco_k_ABCD} below.}
\label{fig:09_gco_overview_ABCD}%
\end{figure*}

\begin{figure*}[tbp]
\centering
\includegraphics[width=0.96\textwidth]{\figures/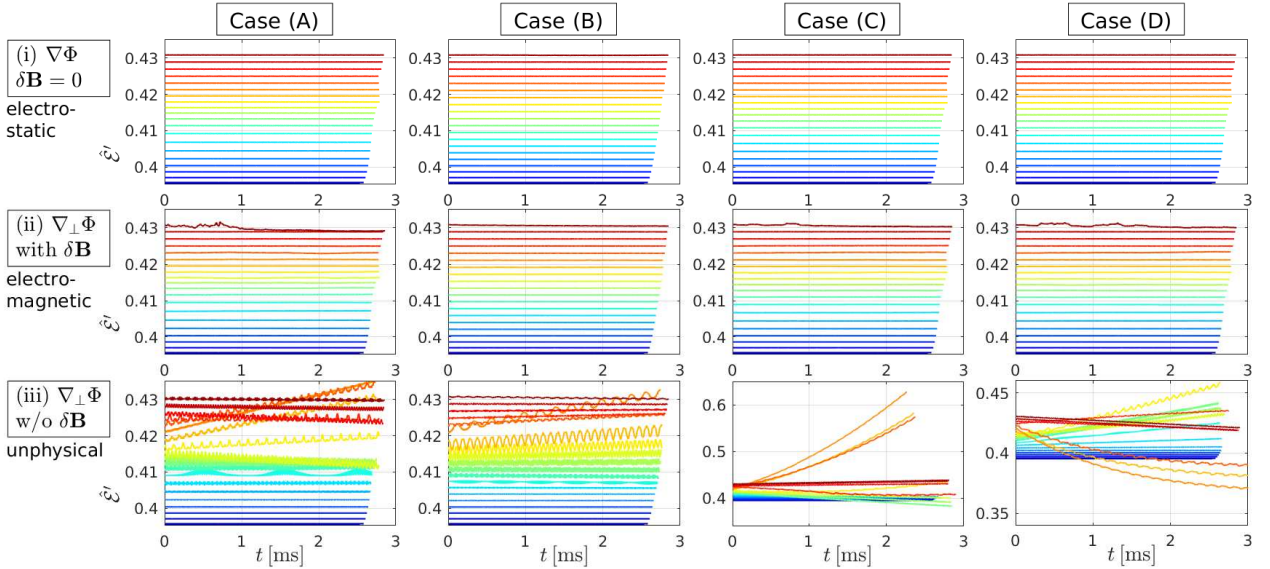}
\caption{Time traces of the rotating frame energy $\hat{\E}'(t)$ for the three scenario (i)--(iii) shown in Fig.~\protect\ref{fig:09_gco_overview_ABCD} above.}
\label{fig:10_gco_k_ABCD}%
\end{figure*}

The main features of the full orbit simulations that we discussed so far are reproduced by simulations using the GC model described in Appendix~\ref{apdx:gc}. The GC model adopted here is based on Hamiltonian theory and includes small terms such as $\partial_t\hat{\bm b}$ and $\nablab|{\bm v}_{\rm E}^2| = \nablab(|{\bm E}^2_\perp|/B^2)$ that are sometimes omitted in other codes. Although formally of higher order, the term $\partial_t\hat{\bm b}$ is essential for this GC model to be conservative. Meanwhile, all terms containing $|{\bm v}_{\rm E}|^2$ may be safely ignored.

The results are summarized in Figs.~\ref{fig:09_gco_overview_ABCD} and \ref{fig:10_gco_k_ABCD}. Overall, the GC simulations reproduce the results of the full orbit simulations, except for a small radial shift of the resonance, which we will discuss in more detail in Sections~\ref{sec:results_ga} and \ref{sec:results_res} below. By definition, the magnetic moment is conserved exactly in the GC model, so we do not show plots of $\hat{\mu}(t)$. Instead, Fig.~\ref{fig:10_gco_k_ABCD} shows the time traces of the rotating frame energy $\hat{\E}'(t)$ for the four cases (A)--(D) in all scenarios (i)--(iii).

For the physical scenarios (i) and (ii), Fig.~\ref{fig:10_gco_k_ABCD} shows excellent conservation of $\hat{\E}'$. The only readily visible exception is the outermost orbit (dark red) that passes near the plasma boundary and experiences the largest amount of numerical noise that is present in our model of $\delta{\bm B}$ (see Fig.~\ref{fig:03_rotE_converg}), and whose effect we will inspect in the following Section~\ref{sec:results_noise}.

In the unphysical scenario (iii), we observe some differences in the rates of secular acceleration in the full orbit and GC models. Especially the normal mode cases (A) and (B) exhibit more rapid acceleration in the GC simulation in Fig.~\ref{fig:10_gco_k_ABCD}(A-iii,B-iii) than in the full orbit simulation in Fig.~\ref{fig:06_cfo_k_ABCD}(A,B). The cause for this difference is not clear,\footnote{Note that scenario (iii) does not contain $\delta{\bm B}$, so the noise-like artifacts in Fig.~\protect\ref{fig:03_rotE_converg} do not enter the gradients of ${\bm B}$. The acceleration seen in Fig.~\ref{fig:10_gco_k_ABCD}(A-iii,B-iii) does not depend on the time step, and the full orbit results in Fig.~\ref{fig:06_cfo_k_ABCD}(A,B) are independent of the numerical method and coordinates used (MLF or RK4, in Cartesian or cylinder coordinates).}
and the particular trend observed here may be accidental. One possible reason is that the resonance location and effective field amplitude differ in the full orbit and GC simulations, as will be discussed in Sections~\ref{sec:results_ga} and \ref{sec:results_res} below (Fig.~\ref{fig:14_comparison_cfo-gco-ga}).

\begin{figure*}[tbp]
\centering\vspace{-0.3cm}
\includegraphics[width=0.96\textwidth]{\figures/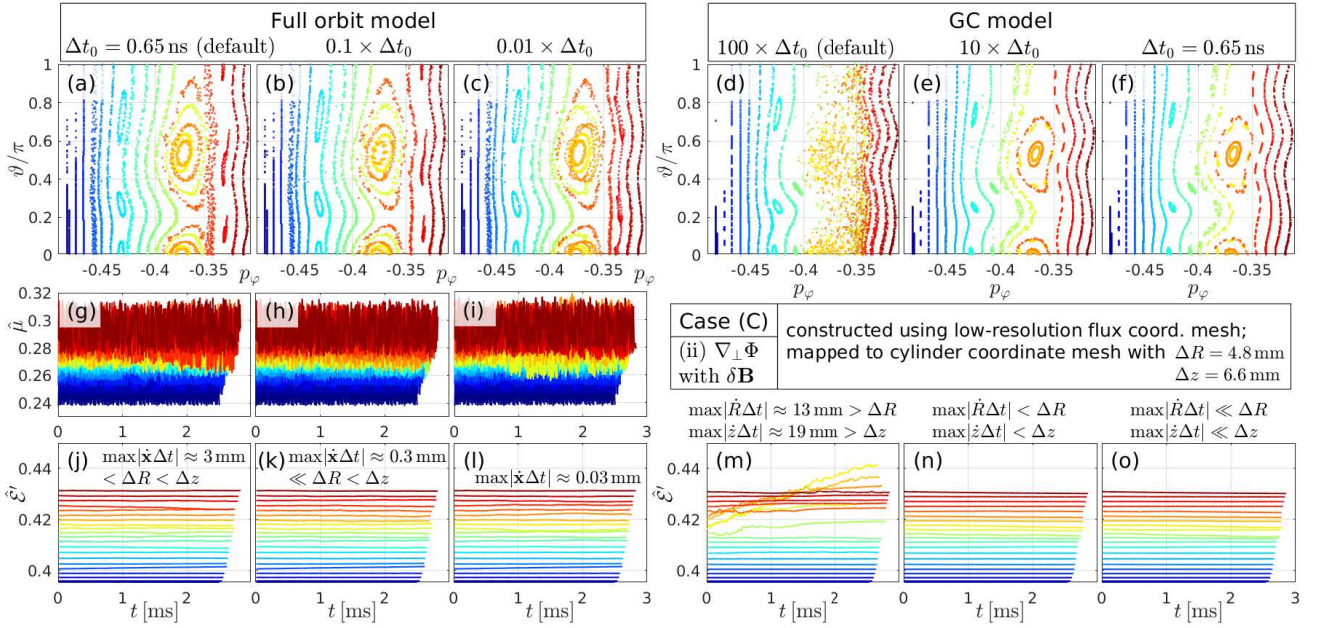}
\caption{Comparison between the results for case (C) obtained with the full orbit model (left) and the GC model (right) using different time steps sizes $\Delta t$. In contrast to all other simulations discussed in this paper, the perturbations ${\bm E}$ and $\delta{\bm B}$ for the electromagnetic scenario (ii) were constructed here with a low-resolution mesh in the flux-coordinates $(\psi_{\rm P},\vartheta)$, causing visible interpolation artifacts in $\delta{\bm B}$ as shown in Fig.~\protect\ref{fig:03_rotE_converg}(b) above. We followed co-passing $400\,{\rm keV}$ deuterons. Panels (a--f) show Poincar\'{e} plots of the orbit topology in the upper half-plane ($0 \leq \vartheta \leq \pi$). Panels (g--i) show time traces of the magnetic moment $\hat{\mu}(t)$ for the full orbit simulations run with $\Delta t = \Delta t_0 \times (1,\, 0.1,\, 0.01)$, where $\Delta t_0 = 0.65\,{\rm ns}$ is our default step size for full orbits. Panels (j--o) show time traces of the rotating frame energy $\hat{\E}'(t)$ for all cases.}
\label{fig:11_cfo-gco_low-res_C}%
\end{figure*}

\subsection{Effect of noise-like interpolation artifacts}
\label{sec:results_noise}

The GC simulations whose results were shown in Figs.~\ref{fig:09_gco_overview_ABCD} and \ref{fig:10_gco_k_ABCD} above were run with $100$ times larger time steps than the default time step
\begin{equation}
\Delta t_0 \approx 0.65\,{\rm ns} \approx 3\times 10^{-5}\,\tfrac{2\pi}{\omega} \approx 0.006\,\tfrac{2\pi}{\Omega_{\rm g0}},
\end{equation}

\noindent that was used in our full orbit simulations with $\Omega_{\rm g0} \approx 2\pi\times 9\,{\rm MHz}$. The corresponding steps the GCs take in the poloidal plane reach ${\rm max}|\dot{R}\Delta t| \approx 13\,{\rm mm}$ and ${\rm max}|\dot{z}\Delta t| \approx 19\,{\rm mm}$. These steps are sufficiently small to satisfy the Courant-Friedrichs-Lewy (CFL) condition with respect to the characteristic scale length of the modes shown in the top row of Fig.~\ref{fig:09_gco_overview_ABCD} (except near the axis, where the poloidal wavelength becomes vanishingly small, but so does the amplitude in that domain).

The situation changes when the fields contain noise or noise-like artifacts like those in Fig.~\ref{fig:03_rotE_converg}, whose scale length is the same as the grid spacing, in our case $\Delta R = 4.8\,{\rm mm}$ and $\Delta z = 6.6\,{\rm mm}$. With the default GC time step $\Delta t = 100\times \Delta t_0$, the CFL condition is violated for the noisy component of the mode structure, since ${\rm max}|\dot{R}\Delta t| > \Delta R$ and ${\rm max}|\dot{z}\Delta t| > \Delta z$. The consequences can be seen in panels (d) and (m) of Fig.~\ref{fig:11_cfo-gco_low-res_C}: KAM surfaces are destroyed and the simulation particles are subject to numerical heating.

The results in Fig.~\ref{fig:11_cfo-gco_low-res_C} were obtained with fluctuating fields of poorer quality than in Figs.~\ref{fig:09_gco_overview_ABCD}(C-ii) and \ref{fig:10_gco_k_ABCD}(C-ii). Instead of $\delta{\bm B}$ in Fig.~\ref{fig:03_rotE_converg}(a), we used the noisier version in Fig.~\ref{fig:03_rotE_converg}(b) that we obtained through the use of a flux coordinate mesh with lower resolution when modeling the wave field for case (C) in the electromagnetic scenario (ii). As was already mentioned in Section~\ref{sec:model_transform} above, the noise in Fig.~\ref{fig:03_rotE_converg}(b) satisfies $\partial_t\delta{\bm B} = -\nablab\times{\bm E}$ and $\nablab\cdot\delta{\bm B} = 0$ with high accuracy, so the signal is physical, but it has an unrealistic (and undesirable) fine structure.

Panels (e) and (n) in Fig.~\ref{fig:11_cfo-gco_low-res_C} show that the KAM surfaces and conservation of $\hat{\E}'$ for GCs are recovered when the time step is reduced to $10\times\Delta t_0$, so that ${\rm max}|\dot{R}\Delta t| < \Delta R$ and ${\rm max}|\dot{z}\Delta t| < \Delta z$. Further reduction of the time step to $\Delta t = \Delta t_0$ in panels (f) and (o) does not show any further improvement (one may even perceive a slight deterioration).

The left half of Fig.~\ref{fig:11_cfo-gco_low-res_C} shows the full orbit results for $\Delta t = \Delta t_0$, $\Delta t_0/10$ and $ \Delta t_0/100$. The default time step already satisfies the CFL condition since the gyroorbit advances in steps of size ${\rm max}|\dot{\bm x}\Delta t| \approx 3\,{\rm mm} < \Delta R < \Delta z$. Nevertheless, the rotating frame energy $\hat{\E}'$ in Fig.~\ref{fig:11_cfo-gco_low-res_C}(j) shows a small amount of drift, and the Poincar\'{e} plot in Fig.~\ref{fig:11_cfo-gco_low-res_C}(a) looks significantly more diffuse than that in Fig.~\ref{fig:05_cfo_overview_ABCD}(C-ii). Moreover, Fig.~\ref{fig:11_cfo-gco_low-res_C}(g) shows some variation in the bounds between which the magnetic moment $\hat{\mu}$ oscillates. Clearly, the noise-like artifacts affect the accuracy of the gyroorbit calculation. The results for $\Delta t_0/10$ in panels (b,h,k) of Fig.~\ref{fig:11_cfo-gco_low-res_C} look somewhat better, but seem to deteriorate again in panels (c,i,l) when the time step is further reduced to $\Delta t_0/100$.

Further tests (not shown) using the normal mode case (A) produced results similar to those shown in Fig.~\ref{fig:11_cfo-gco_low-res_C} for the nonnormal mode case (C). Apparently, the numerical artifacts in Fig.~\ref{fig:03_rotE_converg}(b) affect simulations with normal and nonnormal modes in a similar way.

Evidently, results of higher quality can be obtained only by reducing the numerical artifacts. Only if the noise-like artifacts are sufficiently small, one can take full advantage of the higher computational speed of the GC model, since the time step is no longer bound by the mesh used to discretize the fields, only by their characteristic scale length.

In contrast, the secular acceleration seen in Figs.~\ref{fig:05_cfo_overview_ABCD}--\ref{fig:10_gco_k_ABCD} for the unphysical scenario (iii) is numerically robust, as it cannot be cured by increasing resolution in time or space. Moreover, it is sensitive to the poloidal structure of the wave field.

\subsection{Role of the $t$-dependent magnetic perturbation}
\label{sec:results_dB}

The GC equations are more complicated than the full orbit equations, but that complexity has the advantage that different physical mechanism can be attributed to certain terms, which can be isolated or manipulated. The GC model can hence be used for systematic numerical experiments with the goal to identify which effect causes which observation. In the following, we manipulate the GC equations to gain a better understanding of the results shown in the Figs.~\ref{fig:05_cfo_overview_ABCD}--\ref{fig:10_gco_k_ABCD} above. We begin by examining the role of $\delta{\bm B}(t)$.

\begin{figure}[tbp]
\centering
\includegraphics[width=0.48\textwidth]{\figures/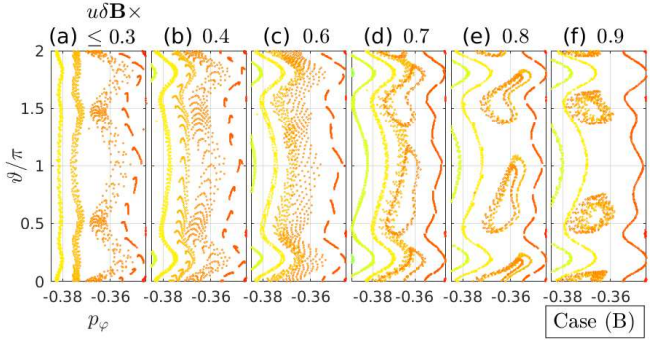}
\caption{Demonstration of the influence of the magnetic perturbation on the poloidal phase of the resonant island in case (B). Beginning from scenario (iii), where we had artificially set $\delta{\bm B} \rightarrow 0$, we scanned the strength of the parallel streaming term $u\delta{\bm B}$ in Eq.~(\protect\ref{eq:dxgc_dt}) by an artificial scaling factor $s$. Panels (a)--(f) show Poincar\'{e} plots of the main resonance for $s = 0.3,0.4,0.6,0.7,0.8$ and $0.9$.}\vspace{-0.15cm}
\label{fig:12_gco_dBu-scan_B}%
\end{figure}

While the two physical scenarios (i) and (ii) in Figs.~\ref{fig:05_cfo_overview_ABCD} and \ref{fig:09_gco_overview_ABCD} are similar in many respects, two notable differences are the amount of distortion of Poincar\'{e} contours outside the main resonance and the poloidal phase of the main resonance; i.e., the position of resonant islands along the poloidal angle $\vartheta$. The parameters in Table~\ref{tab:parm_pol} were chosen such that the elliptic (O-)point of the main resonance is approximately aligned with $\vartheta = 0$ in the electromagnetic scenario (ii), where $\delta{\bm B}(t)$ is present. When applying a purely electrostatic perturbation as in scenarios (i) and (iii), the phase of the resonance is changed only slightly in cases (A), (C) and (D). In contrast, case (B) in Figs.~\ref{fig:05_cfo_overview_ABCD} and \ref{fig:09_gco_overview_ABCD} shows a complete phase flip by $180$ degrees, so that the positions of O- and X-points are interchanged.

The main mechanism via which the magnetic perturbation can influence the phase of the resonant islands is the bending of the magnetic field lines and the resulting redirection of the particle orbits. This effect is captured by the parallel streaming term $u\delta{\bm B}$ in Eq.~(\ref{eq:dxgc_dt}) of the GC model. This term competes with the electric drift ${\bm v}_{\rm E} = {\bm E}\times\hat{\bm b}/B$. In case (A), ${\bm v}_{\rm E}$ dominates on the outboard side of the plasma, where our co-passing deuteron orbits are located. In contrast, the ${\bm v}_{\rm E}$ field in case (B) is very small on the outboard side due to destructive interference, so there the small term $u\delta{\bm B}$ is able to outcompete the electric drift.

This is proven in Fig.~\ref{fig:12_gco_dBu-scan_B} as follows. We started from scenario (iii), where we had artificially set $\delta{\bm B} \rightarrow 0$. The strength of the $u\delta{\bm B}$ term was then varied by an artificial scaling factor $0 \leq s \leq 1$. Obviously, our numerical experiment breaks the Hamiltonian character of the system, but we believe that the effect of the main forces at work can still be seen in a meaningful way. There is no significant effect for $s \leq 0.3$ as one can infer from Fig.~\ref{fig:12_gco_dBu-scan_B}(a), which is still similar to Fig.~\ref{fig:09_gco_overview_ABCD}(B-iii). The effect of $u\delta{\bm B}$ becomes noticeable for $s \approx 0.5$ in Fig.~\ref{fig:12_gco_dBu-scan_B}(b,c), where the island is obscured. The island reappears in distorted form for $s = 0.7$ in panel (d) and becomes increasingly symmetric around the new O-point location $\vartheta = 0$ as the scaling factor $s$ is increased towards unity in panels (e) and (f). This scan shows that, as far as the form of the resonance is concerned, the electric drift ${\bm v}_{\rm E}$ in the electromagnetic scenario (ii) of case (B) is outcompeted by $u\delta{\bm B}$ by about a factor 2 in the region populated by our co-passing $400\,{\rm keV}$ deuterons.

Note that $E_\parallel$ plays no role in the present context: turning the corresponding terms on and off does not affect the phase of the resonant islands, as one can ascertain by comparing panels (B-i) and (B-iii) in Fig.~\ref{fig:09_gco_overview_ABCD}.

\begin{figure}[tbp]
\centering\vspace{-0.3cm}
\includegraphics[width=0.48\textwidth]{\figures/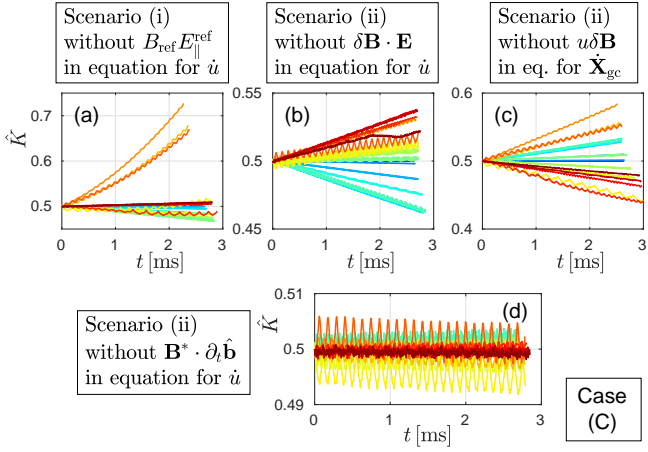}
\caption{Time traces of the kinetic energy $\hat{K}(t)$ showing how the omission of certain terms in the GC equations of motion (\protect\ref{eq:dxgc_du_dt}) cause varied amounts of secular acceleration as seen in Figs.~\protect\ref{fig:06_cfo_k_ABCD} and \protect\ref{fig:10_gco_k_ABCD}. Panel (a) shows the effect of ignoring the parallel electric field $E_\parallel^\eq$ in the electrostatic scenario (i), which is equivalent to scenario (iii) defined in Eq.~(\protect\ref{eq:scenario_3}). Panels (b)--(d) show three examples of terms whose role we have tested in the electromagnetic scenario (ii), where multiple terms contribute, most notably $u\delta{\bm B}$ in panel (c).}
\label{fig:13_gco_tests}%
\end{figure}

\subsection{Identification of secularly accelerating terms}
\label{sec:results_accel}

Let us now use the GC model to investigate how inconsistencies in the poloidal mode structure of the fluctuating fields causes secular acceleration such as that seen in Figs.~\ref{fig:06_cfo_k_ABCD} and \ref{fig:10_gco_k_ABCD}. Equation~(\ref{eq:du_dt}) for the parallel acceleration $\dot{u}$ of a GC can be written
\begin{equation}
B_\parallel^* M \dot{u} \approx -\mu{\bm B}^*\cdot\nablab B + Ze{\bm B}^*\cdot{\bm E} + \text{small terms}.
\label{eq:accel_approx}
\end{equation}

\noindent The first term on the right-hand side is the mirror force and the second term can be decomposed using ${\bm B} = {\bm B}_\eq + \delta{\bm B}$ and ${\bm E} = {\bm E}_\perp^\eq + \hat{\bm b}_\eq E_\parallel^\eq$ as follows:
\begin{align}
&{\bm B}^*\cdot{\bm E} = {\bm B}_\eq^*\cdot{\bm E}_\perp^\eq + B_\parallel^* E_\parallel^\eq + \delta{\bm B}^*\cdot{\bm E} \nonumber
\\
&\approx \underbrace{\rho_\parallel{\bm J}_\eq\cdot{\bm E}_\perp^\eq/\mu_0}\limits_{\rm (o)} + \underbrace{B_\eq E_\parallel^\eq}\limits_{\rm (i)} + \underbrace{\delta{\bm B}\cdot{\bm E}}\limits_{\rm (ii)} + \text{small terms},
\label{eq:accel}
\end{align}

\noindent where $\delta{\bm B}^* \equiv {\bm B}^* - {\bm B}^*_\eq$ and $\mu_0{\bm J}_\eq = \nablab\times{\bm B}_\eq$. We have confirmed numerically that the conservative character of our electrostatic scenario (i) is established by the cancellation of secular accelerations caused by the terms labeled ``(o)'' and ``(i)'' in Eq.~(\ref{eq:accel}). This means that the parallel electric field in $E_\parallel^\eq$ in term ``(i)'' plays an important role and must not be neglected in the absence of magnetic fluctuations. Failure to include term ``(i)'' causes significant secular acceleration as shown in Fig.~\ref{fig:13_gco_tests}(a). This setup is, in fact, equivalent to our unphysical scenario (iii) in Figs.~\ref{fig:06_cfo_k_ABCD} and \ref{fig:10_gco_k_ABCD}(iii), where all $\delta{\bm B}(t)$ terms were ignored.

Secular acceleration occurs not only when all $\delta{\bm B}(t)$ terms (or, equivalently, $E_\parallel^\eq$) are missing. Let us now manipulate the electromagnetic scenario (ii) in order to identify the dominant terms. Figure~\ref{fig:13_gco_tests}(b) shows that the secular acceleration one obtains by neglecting the magnetic fluctuation term $\delta{\bm B}\cdot{\bm E}$ labeled ``(ii)'' in Eq.(\ref{eq:accel}) is significant, but much smaller than in Fig.~\ref{fig:13_gco_tests}(a) that corresponds to letting $\delta{\bm B}(t) \rightarrow 0$ everywhere. Various other terms, including $\mu\delta{\bm B}\cdot\nabla B$ of the mirror force, have a similar degree of importance. The term proportional to $\rho_\parallel{\bm B}^*\cdot(\nablab\times{\bm E}_\perp) = -\rho_\parallel B {\bm B}^*\cdot\partial_t\hat{\bm b}$ in Eq.~(\ref{eq:du_dt_terms}) plays no significant role. Its neglect causes only very benign acceleration as can see in Fig.~\ref{fig:13_gco_tests}(d). It turns out that a significant portion of the secular acceleration seen in scenario (iii) can be attributed to the omission of the perturbed parallel streaming term $u\delta{\bm B}/B_\parallel^*$ in Eq.~(\ref{eq:dxgc_dt}) for the GC velocity $\dot{\bm X}_{\rm gc}$ as Fig.~\ref{fig:13_gco_tests}(c) clearly shows.

In summary, we find that term $\rho_\parallel{\bm J}_\eq\cdot{\bm E}_\perp^\eq/\mu_0$ labeled ``(o)'' of Eq.~(\ref{eq:accel}) causes unphysical secular acceleration unless it is balanced by
\begin{itemize}
\item the parallel electric field $E_\parallel^\eq$ labeled ``(i)'' of Eq.~(\ref{eq:accel}) in the electrostatic scenario (i), and/or
\item the magnetic fluctuations $\delta{\bm B}(t)$ in the electromagnetic scenario (ii), most notably the perturbed parallel streaming term $u\delta{\bm B}$ and, to a lesser degree, terms like $\delta{\bm B}\cdot{\bm E}$ labeled ``(ii)'' in Eq.~(\ref{eq:accel}) and several others.
\end{itemize}

\noindent Secular acceleration is seen not only for fast ions, like our $400\,{\rm keV}$ deuterons whose time traces are shown in Fig.~\ref{fig:13_gco_tests}, but also at low energies, such as the $1\,{\rm keV}$ deuterons that we inspected earlier in Fig.~\ref{fig:08_cfo_1keV_ACD}. This means that the extent of magnetic drifts and the size of the gyroradius does not play a role, and even the existence of efficient low-order resonances is not essential, although they do seem to enhance acceleration.

Instead, the secular acceleration is sensitive with respect to the geometry of the mode structure: nonnormal modes as in case (C) shown in Fig.~\ref{fig:13_gco_tests} yield much stronger acceleration than normal ones. We have also confirmed that the up-down asymmetry of our JT-60U plasma model plays no significant role here. The same sensitivity with respect to mode geometry was seen in a plasma model with shifted circular flux surfaces.

The key factor appears to be the toroidal geometry of the plasma, although this is, in a sense, a trivial fact (or circular argument), since a mode's relation to the nonuniformity of $B$ in toroidal geometry is precisely what distinguishes our nonnormal modes (C,D) from normal modes (A,B). A complete explanation of their different influence on the conservative character of imperfect simulations remains to be found.

\begin{figure*}[tbp]
\centering\vspace{-0.35cm}
\includegraphics[width=0.96\textwidth]{\figures/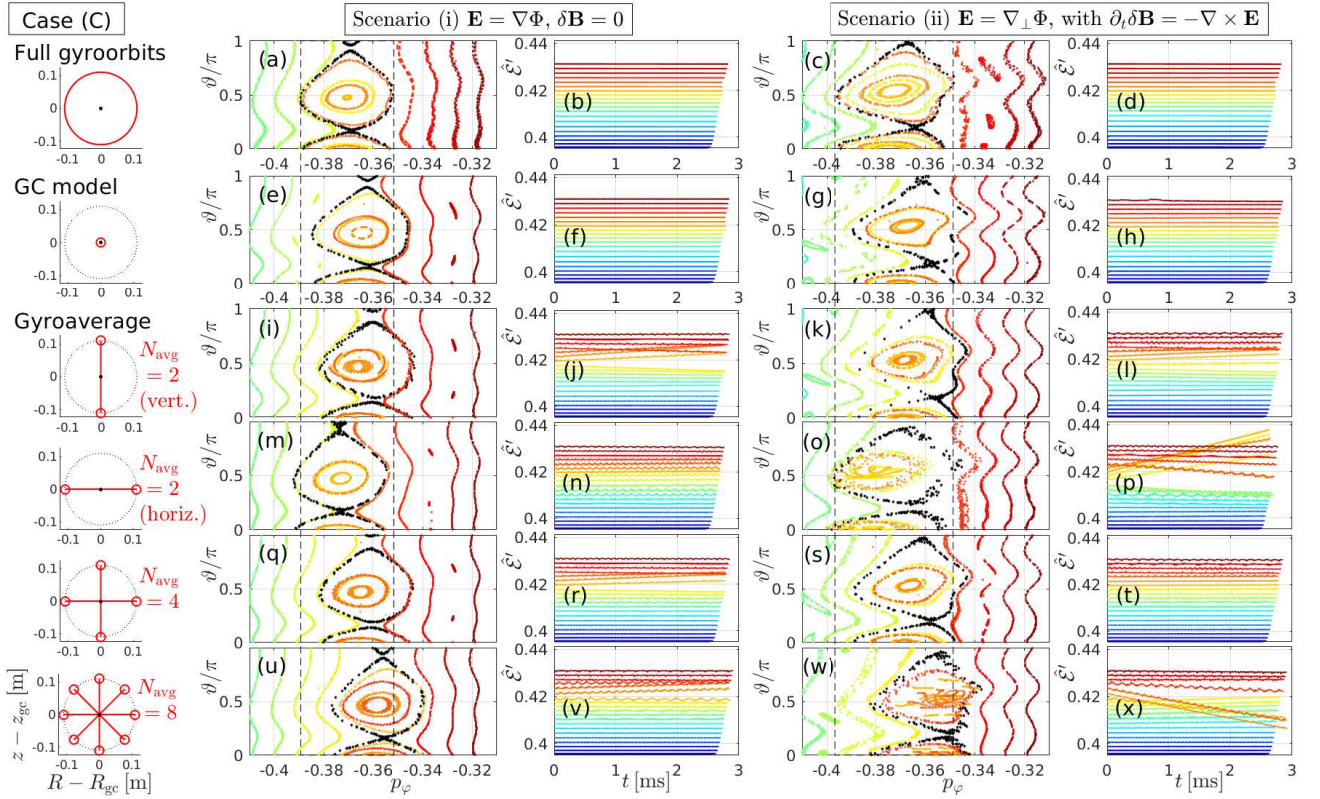}
\caption{Comparison of results of simulations using the full orbit model (a--d), the GC model (e--h), and the GC model with $N$-point gyroaveraging (i--x). We followed co-passing $400\,{\rm keV}$ deuterons, and the nonnormal mode of case (C) was used as a perturbation. Results are presented in terms of Poincar\'{e} plots (columns 2 \& 4) and time traces of the rotating frame energy $\hat{\E}'(t)$ (columns 3 \& 5) for scenarios (i) and (ii) with electrostatic and electromagnetic perturbation, respectively. The first column shows schematically the differences between the three methods for representing particle orbits. In the full orbit case, particles travel along a helix, whose poloidal projection is drawn here as a red circle of radius $\rho_{\rm g} \sim 0.1\,{\rm m}$ around the guiding center (black dot). In the GC model, fields are evaluated at the GC only. Gyroaveraging is done by placing $N_{\rm avg}$ satellite particles (small red circles) on the gyrocircle and taking the average fluctuating electromagnetic field at those locations to push the GC. Here we compare results obtained with $N_{\rm avg} = 2,4,8$ and different satellite positions. As a visual aid, Poincar\'{e} plots are overlaid with a pair of vertical dashed lines that indicate the limits of the resonance's separatrix in the full orbit simulations.}
\label{fig:14_comparison_cfo-gco-ga}\vspace{-0.05cm}
\end{figure*}

\subsection{Gyroaveraging gives wrong trends and breaks the Hamiltonian character of GC motion}
\label{sec:results_ga}

A side-by-side comparison between full orbit and GC simulation results for the nonnormal mode in case (C) is shown in the first two rows of Fig.~\ref{fig:14_comparison_cfo-gco-ga}. The most remarkable difference between the full orbit results in panels (a--d) and the GC results in panels (e--h) is that the resonance in the GC case is shifted outward in radius by $\Delta p_\varphi \approx +0.005$, which corresponds to about $10...20\,{\rm mm}$ in the $(R,z)$ plane. The GC resonances are shifted by the same amount in all cases (A)--(D) and in both scenarios (i) and (ii). This means that $E_\parallel$ does not play a role as it is zero in scenario (ii). In fact, the fluctuation amplitude does not seem to have any significant influence on this resonance shift, so it must be related to the properties of the gyroorbit and how it is represented in the GC model. In this section, we examine how the results are changed if one averages the forces of the fluctuating fields over the gyroradius. This is followed in Section~\ref{sec:results_res} by an inspection of higher-order corrections in the GC equations.

The gyroaveraging procedure is an {\it ad hoc} attempt to advance GCs using the mean force that a physical particle experiences when rapidly gyrating across the electromagnetic field, effectively attenuating the field's nonuniformities. The method employed here was described and tested in Ref.~\onlinecite{Bierwage16c}, and is discussed in some more detail in Appendix~\ref{apdx:gc_ga}. Although not strictly based on Hamiltonian theory, the procedure has been devised with certain constraints in mind. In particular, the averaging is performed only in the poloidal $(R,z)$ plane to preserve the toroidal waveform $\exp(-in\varphi)$. Moreover, only the fluctuating components of the electromagnetic fields are averaged. The purpose of this section is to subject this procedure to further scrutiny under more controlled conditions than those we had used in Ref.~\onlinecite{Bierwage16c}.

The diameter of the gyroradius of our co-passing $400\,{\rm keV}$ deuterons is about $2\rho_0 v_\perp/v_0 = v_\perp/\Omega_{\rm g0} \approx 0.16\,{\rm m} \approx R_0/20$ (on axis). This is comparable to the radial and poloidal size of the ${\bm E}\times{\bm B}$ vortices associated with our $n=2$ modes shown in the upper rows of Fig.~\ref{fig:10_gco_k_ABCD}. We have chosen the nonnormal mode in case (C) and looked at both the electrostatic scenario (i) and the electromagnetic scenario (ii). The results are summarized in panels (i--x) in the lower four rows of Fig.~\ref{fig:14_comparison_cfo-gco-ga}.

We begin our discussion with the electrostatic scenario in the left half of Fig.~\ref{fig:14_comparison_cfo-gco-ga}. Although the use of a mere $N_{\rm avg} = 2$ satellite particles hardly counts as a gyroaverage, the results in panels (i) and (m) highlight the influence of the satellite positions. The location and structure of the resonance varies significantly depending on whether the pair of satellites is aligned vertically (i) or horizontally (m). The case with $N_{\rm avg} = 4$ in panel (q) looks similar to panel (i) with $N_{\rm avg} = 2$ aligned vertically. Finally, $N_{\rm avg} = 8$ in panel (u) gives a resonance shift $\Delta p_\varphi \gtrsim +0.01$ that is even larger than in panel (e) without gyroaveraging, where we had $\Delta p_\varphi \approx +0.005$. Further increase of $N_{\rm avg}$ has only little effect. Test runs with $N_{\rm avg} = 16,19,32$ (not shown) gave essentially the same result as $N_{\rm avg} = 8$, which seems to be `converged', albeit not in a favorable way.

These results show that our gyroaveraging procedure yields the wrong trend. Instead of reducing the difference $\Delta p_\varphi$ between the resonance locations in the full orbit and GC simulations, namely panels (a) and (e), the discrepancy is increased. The reason for this is currently unclear, but possible candidates are the following two features of a true gyroorbit that gyroaveraging does not capture: its helical structure due to the superposition of gyration and parallel streaming, and the variation of the instantaneous value of $\mu$ that was shown in Fig.~\ref{fig:04_cfo_overview_C}(b).

Further tests (not shown here) confirmed that the resonance shift is not affected when the background field ${\bm B}_\eq$ and its gradients are also subject to gyroaveraging. The resonance shift can be reduced or increased if one reduces or increases the averaging radius artificially. Applying a higher-order correction in the angular distribution of the satellite particles as described in Appendix~\ref{apdx:gc_ga} had no effect on the results.

Besides yielding a somewhat inaccurate resonance, another disappointment is that our gyroaveraging procedure can cause unphysical secular acceleration as one can infer from the time traces of the rotating frame energy $\hat{\E}'(t)$ in Fig.~\ref{fig:14_comparison_cfo-gco-ga}, especially in panel (x) for the electromagnetic scenario (ii). Incidentally, $N_{\rm avg} = 4$ shows better conservation than $2$- and $8$-point gyroaveraging; in fact, the results in panels (q)--(t) look similar to those in (e)--(h) that were obtained without gyroaveraging.

In conclusion, our results suggest that $N$-point gyroaveraging is not a recommendable extension of the Hamiltonian GC model. On a slightly more positive note, the secular acceleration in Fig.~\ref{fig:04_cfo_overview_C} seems to be limited to resonantly trapped particles, the separatrix region and near-resonant particles. Orbits located sufficiently far away from a strong resonance seem to be well-behaved. Similar tests (not shown) with $8$-point gyroaveraging around far-off-resonant $1\,{\rm keV}$ deuterons (as in Fig.~\ref{fig:08_cfo_1keV_ACD}) did not reveal any problems in that domain either. Thus, the gyroaveraging procedure may be useful to approximate the mean forces acting on nonresonant particles.

\begin{figure}[tbp]
\centering
\includegraphics[width=0.48\textwidth]{\figures/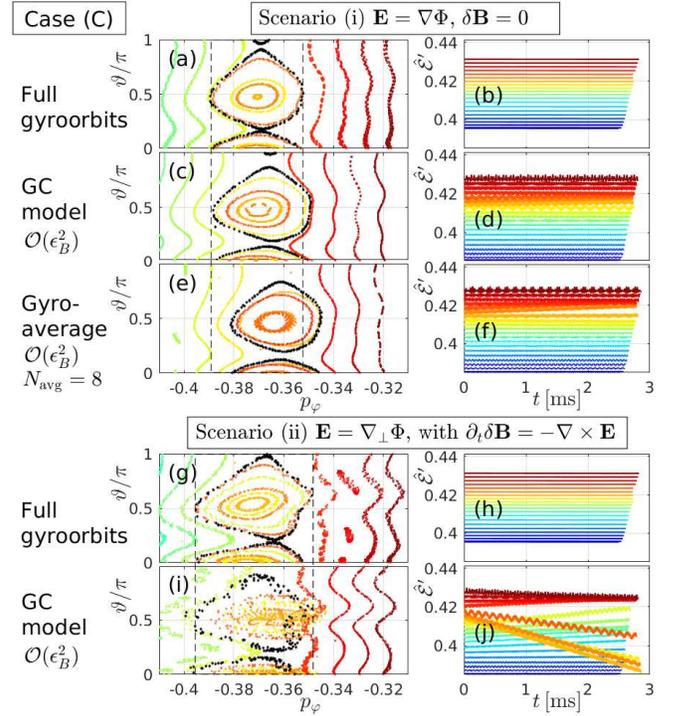}
\caption{Comparison of results from simulations using the full orbit model in panels (a,b,g,h), and the GC model with $\O(\epsilon_B^2)$ corrections given by Eq.~(\protect\ref{eq:gc_corr}) in panels (c--f,i,j). As in Fig.~\protect\ref{fig:14_comparison_cfo-gco-ga}, we followed co-passing $400\,{\rm keV}$ deuterons, and the nonnormal mode of case (C) was used as a perturbation. The left column shows Poincar\'{e} plots of the main resonance with vertical dashed lines indicating the limits of the resonance's separatrix in the full orbit simulation. The right column shows time traces of the rotating frame energy $\hat{\E}'(t)$. The three rows at the top are for the electrostatic scenario (i) and the two rows at the bottom are for the electromagnetic scenario (ii). The GC results in panels (e,f) were obtained with perturbed fields gyroaveraged over $N_{\rm avg}=8$ satellite particles as in the bottom row of Fig.~\protect\ref{fig:14_comparison_cfo-gco-ga}.}
\label{fig:15_gco2-cfo-ga}%
\end{figure}

\subsection{Correction of GC resonance location}
\label{sec:results_res}

The magnitude $|\Delta\hat{\mu}|/\hat{\mu} \sim 10\%$ of the oscillations of the magnetic moment $\hat{\mu}$ in Fig.~\ref{fig:04_cfo_overview_C}(b) suggests that the first-order GC model with respect to the condition $\epsilon_B \equiv \rho_{\rm g}/L_{\rm B} \ll 1$ could be near the limit of its validity, so that $\O(\epsilon_B^2)$ corrections may have noticeable effects. Indeed, we find that the discrepancy between the GC resonance and its full orbit counterpart is significantly reduced if one adds the Ba\~{n}os drift correction to the parallel velocity $u$ and applies the closely related correction to the magnetic moment using the expressions given on p.~718 of Ref.~\onlinecite{Cary09}:
\begin{subequations}
\begin{align}
u \rightarrow u_{\rm corr} &= u + \frac{\mu}{M\Omega_{\rm g}}{\bm B}\cdot\nablab\times\hat{\bm b},
\\
\mu \rightarrow \mu_{\rm corr} &= \mu \left(1 - (B_\parallel^* - B)/B\right).
\end{align}
\label{eq:gc_corr}
\end{subequations}

\noindent The results of making these substitutions in the GC equations of motion (\protect\ref{eq:dxgc_du_dt}) are shown in Fig.~\ref{fig:15_gco2-cfo-ga}. The normalized canonical toroidal angular momentum that is used for the horizontal axes of the Poincar\'{e} plots in the left column of Fig.~\ref{fig:15_gco2-cfo-ga} has also been modified as $p_\varphi({\bm X}_{\rm gc},u) \rightarrow p_\varphi({\bm X}_{\rm gc},u_{\rm corr})$. Similarly, the kinetic energy inside $\E'$ is replaced by
\begin{equation}
K \rightarrow K_{\rm corr} = Mu_{\rm corr}^2/2 + \mu_{\rm corr} B,
\end{equation}

\noindent which causes a reduction of its value by up to $2\%$. The values of $\hat{\E}'$ in the right column of Fig.~\ref{fig:15_gco2-cfo-ga} are reduced by up to $2.6\%$. In fact, instead of applying the above corrections in the GC model, the resonance can also be aligned with that of the full orbit simulation by slightly increasing the initial pitch angle from $\alpha(t=0) = \sin^{-1}(v_\parallel(t=0)/v_0) = 0.25\pi$ to $0.254\pi$, which reduces the magnetic moment by about $2.5\%$.

Note that the corrections in Eq.~(\ref{eq:gc_corr}) are valid only for a stationary magnetic field, as in our electrostatic scenario (i), results for which are shown in the upper three rows of Fig.~\ref{fig:15_gco2-cfo-ga}. The separatrix of the GC resonance in panel (c) now agrees well with that of the full orbit resonance in panel (a). The $8$-point gyroaverage still causes an outward shift of the resonance as on can see in panel (e), and the weak secular acceleration in Fig.~\ref{fig:15_gco2-cfo-ga}(f) is similar to that seen earlier in Fig.~\ref{fig:14_comparison_cfo-gco-ga}(v).

In the electromagnetic scenario (ii), where Eq.~(\ref{eq:gc_corr}) is not valid, these $\O(\epsilon_B^2)$ corrections cause a significant violation of conservation laws in the form of broken KAM surfaces in Fig.~\ref{fig:15_gco2-cfo-ga}(i) and secular acceleration in panel (j). The resonance shift is corrected nevertheless, as panel (i) shows.

Qin \& Davidson\cite{Qin06} treated the case of a time-dependent uniform magnetic field, where an exactly invariant magnetic moment exists. We have not attempted to derive equations of motion for the general nonuniform electromagnetic case from a GC Lagrangian  that is accurate to order $\O(\epsilon_B^2)$. One reason is that gyroaveraging effects are likely to become important when these corrections matter, and we have seen in Figs.~\ref{fig:14_comparison_cfo-gco-ga} and \ref{fig:15_gco2-cfo-ga} that $N$-point gyroaveraging has undesirable side effects. Therefore, instead of constructing corrections, it seems to be more meaningful to use the full orbit model in parameter regimes where the accuracy of the GC theory of order $\O(\epsilon_B)$ no longer suffices. Further pros and cons of full orbit simulations are discussed in Appendix~\ref{apdx:reasons}.

\section{Discussion}
\label{sec:summary}

The purpose of this work was to draw attention to the role of the mode structure as a possible source of inaccuracies and unphysical behavior in numerical simulations of charged particle motion in a toroidally confined plasma. For this purpose, we simulated the motion of deuterons in a tokamak plasma. The simulations were performed with two models: the classical Newton-Lorentz equations governing the dynamics of full gyroorbits, and a guiding center (GC) model. Mathematically, both models represent Hamiltonian systems and it can be shown (see Appendices~\ref{apdx:fo_erot} and \ref{apdx:gc_erot}) that the orbits lie on invariant tori when they are subject to electrostatic or electromagnetic perturbations that consist of a single toroidal mode that has a constant and unique angular phase velocity $\omega/n$. This implies that the mode has a time-independent poloidal mode structure $\widetilde\Phi_{n,\omega}(R,z)$ in the wave's moving frame of reference, where $\varphi' = \varphi + \omega t/n = {\rm const}$. The invariant tori are identified as phase space surfaces on which the rotating frame energy $\E' = K + Ze\Phi + \omega \P_\varphi/n$ is constant. Such a setup was used here as a well-controlled test scenario.

This study was motivated by the realization that the conservation law $\dot{\E}' = {\rm const}$.\ holds for an arbitrary poloidal mode structure only when the fluctuating fields are expressed in terms of potentials $\Phi$ and $\delta{\bm A}$, which satisfy Faraday's law $\partial_t{\bm B} = -\nablab\times{\bm E}$ by definition. In contrast, when the physical fields ${\bm E}$ and $\delta{\bm B}$ are used, Faraday's law must be explicitly enforced in order to preserve the conservative character of Hamiltonian dynamics. This constrains the poloidal mode structure. (Of course, we also require $\nablab\cdot{\bm B} = 0$.)

We have demonstrated that inconsistencies between ${\bm E}$ and $\delta{\bm B}$ that break Faraday's law can cause a significant amount of secular acceleration (and, to a lesser degree, spatial drifts) that violate conservation of phase space density and energy. Such inconsistencies may arise from subtle numerical inaccuracies, approximations, or gross negligence (in our example, we had simultaneously omitted $E_\parallel^\eq$ and $\delta{\bm B}(t)$, which is prohibited by Faraday's law). The resulting unphysical behavior is numerically robust; i.e., it cannot be cured by increasing the numerical resolution in space and time.

We also looked at the effect of noise-like interpolation artifacts that do not violate the laws of electromagnetism on the grid. The spurious heating caused by such artifacts could be alleviated at least partially through the use of smaller time steps. Errors can be enhanced by differentiation operations, in our case ${\bm E} = -\nablab_\perp\Phi$ followed by $\delta{\bm B}(t) = -\partial_t^{-1}\nablab\times{\bm E}$ when modeling the perturbations. Additional differentiation of $\delta{\bm B}$ is needed in the GC model, which can make matters somewhat worse there. The interpolation procedure used here is admittedly not the most sophisticated, but it serves the purpose of demonstrating the existence of possible traps.

Hence, is crucial to provide high-quality fluctuation data for orbit following codes. The effort put into the development and implementation of sophisticated solvers with high accuracy can be for naught when the mode structure of the fluctuations contains errors like noise and physical inconsistencies. In practical situations, where such errors may be inevitable, their influence on the simulation results must be taken into account. The analyses in this work were performed using conventional solvers; namely, explicit 4th-order Runge-Kutta and modified (semi-implicit) leap-frog. We obtained results of acceptable accuracy even when the mode structure contained small imperfections in the form of interpolation artifacts that were not readily visible by eye as in Fig.~\ref{fig:03_rotE_converg}(a).

In the main part of this work, we focused on the quality of the poloidal mode structure and avoided problems along the toroidal direction by using the particle-in-Fourier (PIF) method\cite{Evstatiev13, AmeresPhdThesis, Mitchell19} along the toroidal angle $\varphi$. In principle, the condition $\E' = {\rm const}$.\ is still satisfied by a mode whose toroidal waveform has been discretized using a uniform mesh along $\varphi$, because the mode is allowed to contain higher harmonics of the fundamental angular frequency $\omega_0$ and toroidal mode number $n_0$, as shown in Appendix~\ref{apdx:fo_erot}. However, it is known that the particle-in-cell (PIC) method can cause numerical heating. A benchmark performed in Appendix~\ref{apdx:pif_pic} for our working example with long wavelength modes ($n=2$) showed that the PIC method yields results of similar accuracy as PIF when it is performed with $N_\varphi \gtrsim 16$ grid points (8 points per wavelength) in our electrostatic scenario (i) and $N_\varphi > 32$ in our electromagnetic scenario (ii) with slightly `noisy' $\delta{\bm B}$.

The results of full orbit and GC simulations were similar, except for the fact that the GC resonance was slightly shifted radially relative to the full orbit resonance. Higher-order corrections in the parallel GC velocity $u$ and magnetic moment $\mu$ can largely correct this discrepancy, whereas $N$-point gyroaveraging enhances it. Moreover, the $N$-point gyroaveraging procedure was found to break the conservative character of the system to a significant degree, at least for resonant particles. Therefore, we conclude that this method is not recommendable for cases where resonances play a significant role. When gyroaveraging effects are quantitatively important, one should consider using the full orbit method. Further arguments for and against the use of the full orbit model and discussions pertaining to computational speed are presented in Appendix~\ref{apdx:reasons}.

An interesting observation that we reported in the companion paper\cite{Bierwage22b} was made again here: the geometry of the wave field influences how strongly conservation laws are broken by errors that violate the laws of electromagnetism (here, via the simultaneous neglect of $E_\parallel^\eq$ and $\delta{\bm B}(t)$). The secular acceleration and breaking of KAM surfaces remained relatively small for normal (eigen)modes of the toroidal plasma, but was found to be strongly enhanced when we applied perturbations resembling nonnormal modes. This observation still awaits a full explanation, and some factors were discussed in Section~\ref{sec:results_accel}. In any case, nonnormal modes appear to be useful for testing the conservative character of particle simulations.

\begin{acknowledgments}
One of the authors (A.B.) would like to thank Roscoe B.\ White for providing the Hamiltonian GC code {\tt ORBIT}, which served as a `canonical reference'. The steps taken to eliminate discrepancies between {\tt ORBTOP} and {\tt ORBIT} led to the insights reported in the present paper. A.B.\ is also grateful to Yasushi Todo for providing the code {\tt MEGA} along with valuable advice and support over the years. The development of {\tt ORBTOP} benefited greatly from {\tt MEGA} and from the adoption of some of its algorithms. Helpful discussions with Timur Esirkepov, James Koga, Shuhei Sumida and members of the QST Plasma Theory and Simulation Group are also thankfully acknowledged. The workstation used for the numerical calculations reported here was funded by QST President's Strategic Grant (Creative Research). The work by A.B.\ was partially supported by MEXT as ``Program for Promoting Researches on the Supercomputer Fugaku'' (Exploration of burning plasma confinement physics, JPMXP1020200103).
\end{acknowledgments}

\section*{Data Availability Statement}

The data that support the findings of this study are available from the corresponding author upon reasonable request.


\appendix


\section{Orbit following code {\tt ORBTOP}}
\label{apdx:orbtop}

The simulations and Poincar\'{e} analyses reported in this work were performed using the orbit-following code {\tt ORBTOP} (``Orbit Topology''), which was developed in 2022 using {\tt Julia}.\footnote{{\tt Julia} version 1.8.0 was used in this work.}
The current incarnation of {\tt ORBTOP} offers a few different models, coordinates and discretization schemes, which are outlined in this section.

The perturbation described in Section~\ref{sec:model} above is currently constructed using a {\tt MATLAB} script called \verb#orbtop_prep.m#. This script generates the input data for {\tt ORBTOP}. The reference field is always given as a vector object comprising three real matrices, ${\bm B}_\eq(R,\varphi,z)$, in right-handed cylinder coordinates. Perturbations are composed of either real-valued 3-D matrices in $(R,\varphi,z)$ or complex-valued 2-D matrices in $(R,z)$. Correspondingly, {\tt ORBTOP} performs the mapping from grids to particles using one of the two following methods:
\begin{itemize}
\item a 3-D particle-in-cell (PIC) scheme adopted from the hybrid code {\tt MEGA},\cite{Todo98, Todo05, Todo06} or
\item a mixed PIC-PIF scheme, where the PIC method is used for the poloidal plane $(R,z)$ and the particle-in-Fourier (PIF) method is used for the toroidal direction $\varphi$. Here, we will simply call this option ``PIF''.
\end{itemize}

\noindent In mathematical form, the fields are represented as follows:
\begin{subequations}
\begin{align}
&{\bm E}_{\rm PIC}(R,\varphi,z): \quad \text{3-D PIC method}
\label{eq:input_pic} \\
&\Rightarrow \;\; {\bm E}(R,\varphi,z,t) = {\bm E}_{\rm PIC}(R,\tfrac{\chi(\varphi,t)}{n},z) \nonumber \\
&\Rightarrow \;\; \delta{\bm B}(R,\varphi,z,t) = \tfrac{1}{\omega}[\nablab\times{\bm E}_{\rm PIC}](R,\tfrac{\chi(\varphi,t) + \pi/2}{n},z), \nonumber
\\
&\widetilde{\bm E}_{\rm PIF}(R,z): \quad \text{2-D PIC + 1-D PIF in } \varphi
\label{eq:input_pif} \\
&\Rightarrow \;\; {\bm E}(R,\varphi,z,t) = \Re\left\{\widetilde{\bm E}_{\rm PIF}(R,z)e^{-i\chi(\varphi,t)}\right\} \nonumber \\
&\Rightarrow \;\; \delta{\bm B}(R,\varphi,z,t) = \Re\left\{\tfrac{1}{i\omega}\nablab\times[\widetilde{\bm E}_{\rm PIF}(R,z)e^{-i\chi(\varphi,t)}]\right\}, \nonumber
\end{align}\vspace{-0.4cm}
\label{eq:input}
\end{subequations}

\noindent where the dependence on the toroidal angle and time is captured by the phase
\begin{equation}
\chi(\varphi,t|n,\omega) = n\varphi + \omega t = n\times(\varphi + \omega t/n).
\label{eq:phase_phi_t}
\end{equation}

\noindent Thus, the instantaneous field ${\bm E}$ at angle $\varphi$ and time $t$ corresponds to the input field evaluated at the rotating toroidal angle $\varphi' = \chi/n = \varphi + \omega_n t/n$. The time derivative in Faraday's law shifts the phase of the magnetic fluctuation $\delta{\bm B}(t)$ by $-\pi/2$, which translates into a shift by $+\pi/(2n)$ in the toroidal angle $\varphi$. For multiple modes, each wave would have to be rotated separately relative to the particles.

The current implementation supports only a single toroidally propagating mode with a given toroidal mode number $n$ and angular frequency $\omega$. In that case, the reference field ${\bm B}_\eq$ must be stationary and axisymmetric. Alternatively, {\tt ORBTOP} can follow orbits in an arbitrary 3-D field using the 3-D PIC method, if the field is stationary ($\omega = 0$).

While the input fields are defined in right-handed cylinder coordinates $(R,\varphi,z)$, {\tt ORBTOP} can follow simulation particles in cylinder or Cartesian coordinates. One can choose between a classical full orbit model and a guiding center (GC) orbit model, and two different numerical schemes are available for particle pushing. The following four combinations are currently offered:
\begin{itemize}
\item  Full orbit equations in Cartesian coordinates solved with the modified leap-frog (MLF) scheme.\cite{Hirvioki14}
\item  Full orbit equations in Cartesian coordinates solved with the explicit 4th-order Runge-Kutta (RK4) scheme.
\item  Full orbit equations in cylinder coordinates solved with RK4.
\item  GC equations in cylinder coordinates solved with RK4. Fields can be evaluated at the GC position or via $N$-point gyroaveraging.\cite{Bierwage16c}
\end{itemize}

\noindent Except for the $N$-point gyroaveraged GC model, all models represent Hamiltonian systems. Cartesian coordinates $(v_x,v_y,v_z,x,y,z)$ are canonical (constant Jacobian), while cylinder coordinates $(v_R,v_\varphi,v_z,R,\varphi,z)$ are not. Programming errors were eliminated by comparing the results of different methods against one another. Practically identical results were obtained with MLF and RK4, and in both sets of coordinates once convergence in time step was achieved.

On the millisecond (ms) time scale in our working example, the full orbit solver using the RK4 scheme yielded numerically converged results for time steps $\Delta t \lesssim 0.65\,{\rm ns} \approx 0.006\times2\pi/\Omega_{\rm g0}$, giving about 167 samples per gyration.

Table~\ref{tab:speed} compares the computational performance of different models and solvers For the full orbit model, RK4 is about $20...30\%$ slower than MLF when run with the same default time step $\Delta t_0 = 0.65\,{\rm ns}$. The RK4 solver for the GC model is slower than MLF for full orbits by a factor of $\approx 2$. However, the GC solver can be run with $100$ times larger time steps with only insignificant reduction of accuracy, in which case it runs about $50$ times faster than full orbit MLF at $\Delta t_0$.

The performance penalty of the full orbit model compared to GCs depends on the desired level of accuracy and on physical parameters such as the ratio of gyroradius to magnetic field scale length, $\rho_{\rm g}/L_B$. In our working example based on JT-60U with $400\,{\rm keV}$ deuterons, MLF could be run with somewhat larger time steps $(2...3)\times\Delta t_0$ while maintaining a practically acceptable accuracy of $|\Delta\E'/\E'| \lesssim 0.1\%$ on the millisecond time scale (note that $\Delta\E'$ measures errors in both energy and momentum). In that case, the performance penalty of the full orbit model compared to the GC model may be reduced from a factor $50$ to about $20$. Further enhancement is possible for lower kinetic energies, $K \lesssim 100\,{\rm keV}$.

The model equations and numerical schemes are described in detail in the following Appendices~\ref{apdx:fo}--\ref{apdx:pif_pic}, including a benchmark exercise comparing the results of the PIC and PIF methods. Arguments that we consider when choosing between full orbit and GC models are summarized in Appendix~\ref{apdx:reasons}.

\begin{table}[tbp]
\caption{Comparison of the computational performance of the full orbit model solved with MLF or RK4 schemes, and the GC model solved with RK4 for the working example studied in this paper. The first set of data are the wall times for $10\,{\rm M}$ steps of identical size $\Delta t_0 = 0.65\,{\rm ns}$. The second set of data are the wall times for accumulating $15\times 450$ Poincar\'{e} points, which corresponds to about $40\,{\rm ms}$ of physical time. For the Poincar\'{e} analysis, the GC model was run with $\Delta t = \Delta t_0\times 100$. In all cases, the PIF method was used in the toroidal direction and diagnostics were reduced to a minimum. $^{(*)}$The computational overhead of the GC model was minimized by deleting the code for $|{\bm v}_{\rm E}|^2$ terms and the infrastructure for optional $N$-point gyroaverating. $^{(\dagger)}$Poincar\'{e} analyses using the full orbit model are slowed down by the need to estimate the toroidal angle of the GC, but there remains some unexploited leeway for optimization.}
\begin{ruledtabular}
\begin{tabular}{c|c|c}
Reduced code with & $10^7$ steps @ $\Delta t_0$ & Poincar\'{e} analysis,$^{(\dagger)}$ \\
minima diagnostics & ($t \approx 6.5\,{\rm ms}$) & $6750$ pts.\ ($t \approx 40\,{\rm ms}$) \\
\hline Full orbit MLF & $t_{\rm wall} \approx 15\,{\rm s}$ & $t_{\rm wall} \approx 1050\,{\rm s}$ \\
Full orbit RK4 & $\approx 20\,{\rm s}$ (MLF $\times 1.3$) & $\approx 1300\,{\rm s}$ (MLF $\times 1.2$) \\
GC model$^{(*)}$ RK4 & $\approx 30\,{\rm s}$ (MLF $\times 2$) & $\approx 12\,{\rm s}$ (MLF $\times 1/90$) \\
\end{tabular}\vspace{-0.3cm}
\label{tab:speed}
\end{ruledtabular}
\end{table}

\section{Full orbit model}
\label{apdx:fo}

\subsection{Equations of motion in Cartesian and cylinder coordinates}
\label{apdx:fo_eom}

We consider the nonrelativistic motion of a particle with charge $Ze$ and mass $M$ in a magnetic field ${\bm B}$ and electric field ${\bm E}$. The equations of motion are
\begin{subequations}
\begin{align}
\dot{\bm x} =\,& {\bm v}
\label{eq:eom_x}
\\
M\dot{\bm v} =\,& Ze ({\bm E} + {\bm v}\times{\bm B}).
\label{eq:eom_v}
\end{align}
\end{subequations}

\noindent These empirical equations can be reverse-engineered from the Lagrangian $\L$ and Hamiltonian $\H$ of the form
\begin{equation}
\L = (M{\bm v} + Ze{\bm A})\cdot\dot{\bm x} - \H, \quad \H = Ze\Phi + M{\bm v}^2/2,
\label{eq:cfo_lagrangian}
\end{equation}

\noindent together with the definitions of the potentials $\Phi$ and ${\bm A}$ based on Maxwell's equations. Variation of the Lagrangian gives
\begin{align}
{\rm d}_t(\partial_{\dot{\bm v}}\L) = \partial_{\bm v}\L \quad \Rightarrow \quad &0 = M\dot{\bm x} - M{\bm v}.
\\
{\rm d}_t(\partial_{\dot{\bm x}}\L) = \partial_{\bm x}\L \quad \Rightarrow \quad &M\dot{\bm v} + Ze(\partial_t{\bm A} + \dot{\bm x}\cdot\nablab{\bm A}) \nonumber
\\
&= Ze\left(\nablab\cdot{\bm A}\dot{\bm x} - \nablab\Phi\right),
\end{align}

\noindent where the momentum equation (\ref{eq:eom_v}) is recovered by substituting $\dot{\bm x}\times(\nablab\times{\bm A}) = \nablab\cdot{\bm A}\dot{\bm x} - \dot{\bm x}\cdot\nablab{\bm A}$,  ${\bm B} = \nablab\times{\bm A}$ and ${\bm E} = -\nablab\Phi - \partial_t{\bm A}$.

Several auxiliary variables are also used, for instance, as input parameters and for diagnostics. These include the kinetic energy $K$, the magnetic moment $\mu$, and the velocity components $v_\parallel$ and ${\bm v}_\perp$:
\begin{equation}
K = \frac{M v^2}{2}, \quad \mu = \frac{M v_\perp^2}{2B}, \quad v_\parallel = {\bm v}\cdot\hat{\bm b}, \quad {\bm v}_\perp = {\bm v} - v_\parallel\hat{\bm b}.
\end{equation}

\noindent For normalization, we use characteristic values of the particle velocity $v_0 = \sqrt{2K_0/M}$ and the magnetic field strength $B_0$. In the present work, $K_0 = K_0(t=0)$ is the initial energy and $B_0 = |{\bm B}_\eq|(R_0,z_0)$ is the central value of the axisymmetric reference field, where $(R_0,z_0)$ is the location of the magnetic axis. Replacing time increments ${\rm d}t$ with an equivalent parameter ${\rm d}s = v_0{\rm d}t$ with units of length, the normalized equations of motion become
\begin{subequations}
\begin{align}
\dot{\bm x} =\,& \hat{\bm v} \\
\dot{\hat{\bm v}} =\,& \frac{1}{\rho_0} (\hat{\bm E} + \hat{\bm v}\times\hat{\bm B}),
\end{align}\vspace{-0.2cm}
\label{eq:eom_Cart}
\end{subequations}

\noindent with the characteristic Larmor radius $\rho_0 = v_0/\Omega_{\rm g0}$ and gyrofrequency $\Omega_{\rm g0} = Mv_0/(ZeB_0)$. Energy is normalized by $Mv_0^2$, giving
\begin{equation}
\hat{K} = \frac{\hat{v}^2}{2}, \quad \hat{\mu} = \frac{\hat{v}_\perp^2}{2\hat{B}}.
\end{equation}

In Cartesian coordinates, we have the following six equations of motion:

\vspace{0.2cm}\noindent\fbox{\begin{minipage}{\dimexpr0.49\textwidth-2\fboxsep-2\fboxrule\relax}\vspace{-0.3cm}
\begin{subequations}
\begin{align}
\left[\begin{array}{c} \dot{x} \\ \dot{y} \\ \dot{z} \end{array}\right] =\,& \left[\begin{array}{c} \hat{v}_x \\ \hat{v}_y \\ \hat{v}_z \end{array}\right],
\label{eq:eom_full_cart_x}
\\
\left[\begin{array}{c} \dot{\hat{v}}_x \\ \dot{\hat{v}}_y \\ \dot{\hat{v}}_z \end{array}\right] =\,& \frac{1}{\rho_0} \left[\begin{array}{c} \hat{E}_x + \hat{v}_y\hat{B}_z - \hat{v}_z\hat{B}_y \\ \hat{E}_y + \hat{v}_z\hat{B}_x - \hat{v}_x\hat{B}_z \\ \hat{E}_z + \hat{v}_x\hat{B}_y - \hat{v}_y\hat{B}_x \end{array}\right].
\label{eq:eom_full_cart_v}
\end{align}\vspace{-0.3cm}
\label{eq:eom_full_cart}
\end{subequations}
\end{minipage}}\vspace{0.2cm}

\noindent In order to express these equations in cylinder coordinates $(R,\varphi,z)$, we use the following relations:
\begin{subequations}
\begin{gather}
x = R\cos\varphi, \quad \hat{\bm e}_x = \hat{\bm e}_R\cos\varphi - \hat{\bm e}_\varphi\sin\varphi, \\
y = R\sin\varphi, \quad \hat{\bm e}_y = \hat{\bm e}_R\sin\varphi + \hat{\bm e}_\varphi\cos\varphi, \\
R^2 = x^2 + y^2, \quad \hat{\bm e}_R = \hat{\bm e}_x\cos\varphi + \hat{\bm e}_y\sin\varphi, \\
\varphi = {\rm atan}(y/x), \quad \hat{\bm e}_\varphi = -\hat{\bm e}_x\sin\varphi + \hat{\bm e}_y\cos\varphi,
\end{gather}\vspace{-0.4cm}
\label{eq:cyl_cart}
\end{subequations}

\noindent which imply that
\begin{subequations}
\begin{align}
v_x &= v_R\cos\varphi - v_\varphi\sin\varphi, \\
v_y &= v_R\sin\varphi + v_\varphi\cos\varphi, \\
v_R &= v_x\cos\varphi + v_y\sin\varphi, \\
v_{\hat{\varphi}} &= -v_x\sin\varphi + v_y\cos\varphi.
\end{align}
\end{subequations}

\noindent Here and in the following we denote the toroidal component in physical units is by $v_{\hat{\varphi}}$ to distinguish it from the covariant component $v_\varphi$:
\begin{equation}
v_\varphi = {\bm v}\cdot\partial_\varphi{\bm x}, \quad v_{\hat{\varphi}} = {\bm v}\cdot\partial_\varphi{\bm x}/|\partial_\varphi{\bm x}| = v_\varphi/R.
\end{equation}

\noindent The time derivatives of a particle's position in cylinder coordinates are
\begin{subequations}
\begin{align}
\dot{R} =\,& R^{-1}(x\dot{x} + y\dot{y}) = v_R, \\
\dot{\varphi} =\,& \frac{1}{1 + (y/x)^2}\left(-\frac{y\dot{x}}{x^2} + \frac{\dot{y}}{x}\right) = \frac{v_\varphi}{R}.
\end{align}
\end{subequations}

\noindent The time derivatives of the particle's velocity components are
\begin{subequations}
\begin{align}
\dot{v}_R &= \dot{v}_x\cos\varphi + \dot{v}_y\sin\varphi {\mcb \;+\; \dot{\varphi}(-v_x\sin\varphi + v_y\cos\varphi)} \nonumber \\
&= \frac{1}{\rho_0}\left(E_R + v_{\hat{\varphi}} B_z - v_z B_{\hat{\varphi}}\right) {\mcb \;+\; \frac{v_{\hat{\varphi}}^2}{R}},
\\
\dot{v}_{\hat{\varphi}} &= -\dot{v}_x\sin\varphi + \dot{v}_y\cos\varphi {\mcb \;-\; \dot{\varphi}(v_x\cos\varphi + v_y\sin\varphi)} \nonumber \\
&= \frac{1}{\rho_0}\underbrace{\left(E_{\hat{\varphi}} + v_z B_R - v_R B_z\right)}\limits_{\rho_0\dot{v}_\varphi/R} {\mcb \;-\; \frac{v_{\hat{\varphi}} v_R}{R}}.
\label{eq:eom_vphi}
\end{align}
\end{subequations}

\noindent In summary,

\vspace{0.2cm}\noindent\fbox{\begin{minipage}{\dimexpr0.49\textwidth-2\fboxsep-2\fboxrule\relax}\vspace{-0.3cm}
\begin{subequations}
\begin{align}
\left[\begin{array}{c} \dot{R} \\ \dot{\varphi} \\ \dot{z} \end{array}\right]
=\,& \left[\begin{array}{c} \hat{v}_R \\ \hat{v}_{\hat{\varphi}}/R \\ \hat{v}_z \end{array}\right],
\\
\left[\begin{array}{c} \dot{\hat{v}}_R \\ \dot{\hat{v}}_{\hat{\varphi}} \\ \dot{\hat{v}}_z \end{array}\right]
=\,& \frac{1}{\rho_0} \left[\arraycolsep=0pt\def\arraystretch{1.0}\begin{array}{c}
\hat{E}_R + \hat{v}_{\hat{\varphi}}\hat{B}_z - \hat{v}_z\hat{B}_{\hat{\varphi}} \\
\hat{E}_{\hat{\varphi}} + \hat{v}_z\hat{B}_R - \hat{v}_R\hat{B}_z \\
\hat{E}_z + \hat{v}_R\hat{B}_{\hat{\varphi}} - \hat{v}_{\hat{\varphi}}\hat{B}_R
\end{array}\right]
{\mcb + \frac{1}{\rho_0} \left[\arraycolsep=0pt\def\arraystretch{1.0}\begin{array}{c}
\hat{v}_{\hat{\varphi}}^2\tfrac{\rho_0}{R} \\
-\hat{v}_{\hat{\varphi}} \hat{v}_R\tfrac{\rho_0}{R} \\
0 \end{array}\right]}.
\end{align}\vspace{-0.4cm}
\label{eq:eom_full_cyl}
\end{subequations}
\end{minipage}}\vspace{0.2cm}

\noindent The {\mcbtxt} terms in Eq.~(\ref{eq:eom_full_cyl}) can be interpreted as the centrifugal force ($\hat{v}_{\hat{\varphi}}^2 \rho_0/R$) and the Coriolis force ($-\hat{v}_{\hat{\varphi}}\hat{v}_R \rho_0/R$). If one omits these terms, energy is still conserved but the magnetic mirror effect is lost.

\subsection{Rotating frame energy conservation}
\label{apdx:fo_erot}

The total time derivative of the Hamiltonian
\begin{equation}
\H({\bm x},{\bm v},t) = Ze\Phi({\bm x},t) + M|{\bm v}|^2/2
\end{equation}

\noindent is
\begin{align}
\dot{\H} &= Ze\partial_t\Phi + Ze\dot{\bm x}\cdot\nablab\Phi + M{\bm v}\cdot\dot{\bm v} \nonumber
\\
&= Ze(\partial_t\Phi - {\bm v}\cdot\partial_t{\bm A}) = \partial_t\H  - Ze{\bm v}\cdot\partial_t{\bm A},
\label{eq:dHdt}
\end{align}

\noindent where we substituted the equations of motion (\ref{eq:eom_Cart}) and ${\bm E} = -\nablab\Phi - \partial_t{\bm A}$. Let us now compare this to the total time derivative of the covariant toroidal component of the canonical momentum ${\bm P}_{\rm c} = M{\bm v} + Ze{\bm A}$,
\begin{equation}
{\bm P}_{\rm c}\cdot\partial_\varphi{\bm x} = P_\varphi = R P_{\hat{\varphi}} = M R v_{\hat{\varphi}} + Ze R A_{\hat{\varphi}}({\bm x},t),
\end{equation}

\noindent which becomes
\begin{equation}
\dot{P}_\varphi = M (R \dot{v}_{\hat{\varphi}} + v_R v_{\hat{\varphi}}) + Ze (R\partial_t A_{\hat{\varphi}} + R \dot{\bm x}\cdot\nablab A_{\hat{\varphi}} + v_R A_{\hat{\varphi}}).
\label{eq:dPphidt}
\end{equation}

\noindent Substitution of the equations of motions (\ref{eq:eom_vphi}) and $\dot{\bm x} = {\bm v}$ along with $B_R = R^{-1}\partial_{\hat{\varphi}} A_z - \partial_z A_{\hat{\varphi}}$ and $B_z = R^{-1}\partial_R(R A_{\hat{\varphi}}) - R^{-1}\partial_{\hat{\varphi}}A_R$ gives, after several cancellations,
\begin{equation}
\dot{P}_\varphi = - Ze(\partial_{\hat{\varphi}}\Phi - {\bm v}\cdot\partial_{\hat{\varphi}}{\bm A}) = -\partial_{\hat{\varphi}}\H + Ze{\bm v}\cdot\partial_{\hat{\varphi}}{\bm A}.
\end{equation}

\noindent Consider the case where the fluctuating potentials, $\Phi(t)$ and $\delta{\bm A}(t) = {\bm A} - {\bm A}_\eq$, represent a wave that propagates with a unique and constant angular phase velocity $\omega_0/n_0$ along $\varphi$. In this case, the field's dependence on the toroidal angle $\varphi$ and time $t$ has the form\cite{Bierwage22b}
\begin{equation}
\Phi = \sum_{k=0,1,2...}\frac{1}{2}\widetilde{\Phi}_k(R,z)\exp(\underbrace{-in_k\varphi - i\omega_k t}\limits_{(-in_0\varphi - i\omega_0 t)\kappa_k}) + {\rm c.c.},
\label{eq:pert_erot_const}
\end{equation}

\noindent with a time-independent amplitude and poloidal profile $\widetilde{\Phi}(R,z)$, a constant fundamental frequency $\omega_0$, a single fundamental toroidal mode number $n_0$, and with an arbitrary number of harmonics $(\omega_k,n_k) = (\omega_0\kappa_k,n_0\kappa_k)$ with integer $\kappa_k$. Equations~(\ref{eq:dHdt}) and (\ref{eq:dPphidt}) can then be combined to give
\begin{equation}
n_0\dot{\H} + \omega_0\dot{P}_\varphi = 0 \quad \Rightarrow \quad n_0\E' = n_0\H + \omega_0 P_\varphi = {\rm const}.,
\label{eq:erot_const}
\end{equation}

\noindent so that the rotating frame energy $\E'$ is conserved.

An often-cited reference for this rule is Hsu {\it et al}.\cite{Hsu92} Here and in the companion paper,\cite{Bierwage22b} we generalized it to the case of multi-harmonic waves: Eq.~(\ref{eq:erot_const}) holds for arbitrarily distorted waveforms as long as their Fourier harmonics are phase-locked and satisfy $\omega_k/n_k = \omega_0/n_0$. This is important for numerical simulation codes employing finite-difference grids: although the discretization on a mesh along the toroidal angle $\varphi$ gives rise to deviations from a purely sinusoidal form, this distortion by itself does not cause a violation of rotating frame energy conservation (\ref{eq:erot_const}) when the $\varphi$-grid is uniform. The key requirement is that the angular phase velocity $\omega_0/n_0$ is unique and constant, so that there exists a preferred frame of reference in which the field appears stationary.

\subsection{Discretization in time}
\label{apdx:fo_dt}

The above sets of equations of motion, (\ref{eq:eom_full_cart}) or (\ref{eq:eom_full_cyl}), can be readily solved numerically using a standard explicit 4th-order Runge-Kutta (RK4) algorithm. Each variable ${\bm \xi}_i = \{{\bm v}(t_i),{\bm x}(t_i)\}$ is then advanced by a sum of four increments as\vspace{-0.1cm}
\begin{gather}
\xi_{i+1} = \xi_i + \sum\limits_{k=1}^4 c_k\Delta\xi^k, \\
\label{eq:rk4_sum}
\text{with} \quad c_1=\tfrac{1}{6},\; c_2=\tfrac{1}{3},\; c_3=\tfrac{1}{3},\; c_4=\tfrac{1}{6}, \nonumber
\end{gather}

\noindent where the index $i = 0,1,...$ represents the time step. Writing the respective equation of motion compactly as $\dot{\xi} = F_\xi({\bm \xi}(t),t)$, the increments $\Delta\xi^k$ are computed in four steps $k = 1,...,4$ as follows:
\begin{gather}
\Delta\xi_i^k = F_\xi({\bm \xi}_i^{k-1},t_i)\Delta t \quad \text{for}\quad \xi_i^{k-1} = \xi_i + d_k\Delta\xi_i^{k-1},
\label{eq:rk4_step}
\\
\text{with} \quad d_1=0,\; d_2=\tfrac{1}{2},\; d_3=\tfrac{1}{2},\; d_4=1. \nonumber
\end{gather}

We have also implemented the modified leap-frog (MFL) scheme based on Ref.~\onlinecite{Hirvioki14}, which can be expressed as
\begin{subequations}
\begin{gather}
{\bm x}_{i+\frac{1}{2}} = {\bm x}_{i-\frac{1}{2}} + \Delta s \hat{\bm v}_i, \quad  \text{with} \quad {\bm x}_{0+\frac{1}{2}} = \frac{\Delta s}{2}\hat{\bm v}_0,
\label{eq:mlf_x}
\\
\hat{\bm v}_{i+1} = \hat{\bm v}_i + \frac{\Delta s}{\rho_0} \left(\hat{\bm E}_{i+\frac{1}{2}} + \frac{\hat{\bm v}_{i+1} + \hat{\bm v}_i}{2} \times \hat{\bm B}_{i+\frac{1}{2}}\right),
\label{eq:mlf_v}
\end{gather}\vspace{-0.1cm}
\label{eq:mlf}
\end{subequations}

\noindent where the index $i = 0,1,...$ represents the time step and $\Delta s = v_0\Delta t$ is the normalized time step in units of length (meters). In Cartesian coordinates, Eq.~(\ref{eq:mlf_v}) can be solved straightforwardly as a linear matrix inversion problem for ${\bm v}_{i+1} = \left[\hat{v}_x, \hat{v}_y, \hat{v}_z\right]^{\rm T}_{i+1}$ (where ``T'' denotes the transpose):
\begin{equation}
\tensor{\bm F}^{-1}\cdot\hat{\bm v}_{i+1} = \hat{\bm v}_i + S\left(2 {\bm E}_{i+\frac{1}{2}} + \hat{\bm v}_i\times\hat{\bm B}_{i+\frac{1}{2}}\right)
\end{equation}

\noindent with
\begin{equation}
S = \frac{\Delta s}{2\rho_0}, \quad
\tensor{\bm F}^{-1} = \left[
\begin{array}{ccc}
1 & -S\hat{B}_z & S\hat{B}_y \\
S\hat{B}_z & 1 & -S\hat{B}_x \\
-S\hat{B}_y & S\hat{B}_x & 1
\end{array}\right]_{i+\frac{1}{2}}
\end{equation}

\noindent becomes

\vspace{0.2cm}\noindent\fbox{\begin{minipage}{\dimexpr0.49\textwidth-2\fboxsep-2\fboxrule\relax}\vspace{-0.3cm}
\begin{equation}
\hat{\bm v}_{i+1} = \tensor{\bm F}\cdot\left(\hat{\bm v}_i + 2S{\bm E}_{i+\frac{1}{2}} + S\hat{\bm v}_i\times\hat{\bm B}_{i+\frac{1}{2}}\right)
\end{equation}
\end{minipage}}\vspace{0.2cm}

\noindent with
\begin{subequations}
\begin{align}
\tensor{\bm F} &= \frac{1}{D}\left[
\begin{array}{ccc}
1 & S\hat{B}_z & S^2\hat{B}_x\hat{B}_z \\
S^2\hat{B}_x\hat{B}_y & 1 & S\hat{B}_x \\
S\hat{B}_y & S^2\hat{B}_y\hat{B}_z & 1
\end{array}\right]_{i+\frac{1}{2}}
\\
D &= 1 - S^3\hat{B}_x\hat{B}_y\hat{B}_z.
\end{align}
\end{subequations}

We have not implemented the MLF scheme for cylinder coordinates, because the nonlinear character of the pseudo-forces on the last line of Eq.~(\ref{eq:eom_full_cyl}) complicates the problem. Of course, the fields can be discretized in any coordinates, even when particles are pushed in Cartesian ones. In both cases, RK4 and MLF, the fields ${\bm B}$ and ${\bm E}$ are given on a mesh in cylinder coordinates $(R,\varphi,z)$. In the case of PIC, linear interpolation is used to determine their values ${\bm B}_l = {\bm B}({\bm x}_l)$ and ${\bm E}_l = {\bm E}({\bm x}_l)$ at the location ${\bm x}_l$ of a simulation particle labeled $l$. In the case of PIF, the toroidal dependence is evaluated analytically.

Compared to RK4 scheme, MLF gives better energy conservation (which becomes exact when ${\bm E} = 0$). However, measurements of mixed variable such as $\P_\varphi({\bm x},{\bm v})$ and $\E'({\bm x},{\bm v})$ are less accurate with MLF since positions and velocities are computed at staggered times that lie $\Delta t/2$ apart. Thus, conservation laws involving mixed variables are more easily verified using RK4.

\section{Guiding center model}
\label{apdx:gc}

In Ref.~\onlinecite{Bierwage22b}, we have studied the properties of the Hamiltonian guiding center (GC) model in Boozer coordinates in a canonical and a non-canonical formulation. Here, we revisit the non-canonical Hamiltonian formulation of the GC model in terms of a parallel velocity coordinate $u$ and a coordinate-independent GC position vector ${\bm X}_{\rm gc}$, which can readily be expressed in right-handed cylinder coordinates $(R,\varphi,z)$ as used for the GC module in our code {\tt ORBTOP}.

\subsection{Guiding center phase space Lagrangian}

Our starting point is the GC Lagrangian as reviewed by Cary \& Brizard \cite{Cary09}, with slight changes in notation. See also the Appendix of Ref.~\onlinecite{Bierwage22b}. The ordering parameter $\epsilon \ll 1$ is defined in Eq.~(A2) of Ref.~\onlinecite{Bierwage22b}.

The GC phase space consists of the position vector ${\bm X}_{\rm gc}$, an ignorable gyrophase $\theta$, the parallel velocity variable $u \equiv \dot{\bm X}_{\rm gc}\cdot\hat{\bm b}$, and the magnetic moment $\mu$. At lowest order in the ordering parameter $\epsilon \sim \epsilon_B \equiv \rho_{\rm g}/L_B$, the latter has the form
\begin{equation}
\mu \equiv \frac{M|{\bm w}|^2}{2 B({\bm X}_{\rm gc},t)},
\end{equation}

\noindent where ${\bm w} \equiv {\bm v}_\perp - {\bm v}_{\rm E}$ is the perpendicular velocity of the particle in the local frame moving with the electric drift velocity ${\bm v}_{\rm E} \equiv {\bm E}\times\hat{\bm b}/B$. The GC phase space Lagrangian $\L(\eta,\dot{\eta};t)$ to order $\O(\epsilon)$ with $\eta = \{{\bm X}_{\rm gc}, u, \mu, \theta)$ is
\begin{align}
\L_{\rm gc} =\;& \left(Ze{\bm A}({\bm X}_{\rm gc},t) + M u \hat{\bm b}({\bm X}_{\rm gc},t)\right)\cdot\dot{\bm X}_{\rm gc} + J\dot\theta - \H_{\rm gc},
\label{eq:lgc}
\end{align}

\noindent with $J = \mu B/\Omega_{\rm g}$ and $\Omega_{\rm g} = ZeB/M$. The GC Hamiltonian $\H({\bm X}_{\rm gc},u,\mu;t)$ to order $\O(1)$ is
\begin{align}
\H_{\rm gc} =&\; Mu^2/2 + \mu B({\bm X}_{\rm gc},t) + Ze\Phi({\bm X}_{\rm gc},t)
\label{eq:hgc}
\\
&\;- M|{\bm v}_{\rm E}({\bm X}_{\rm gc},t)|^2/2. \nonumber
\end{align}

\noindent The ponderomotive potential $-M|{\bm v}_{\rm E}|^2/2$ arises from the sum of the electric drift energy $M|{\bm v}_{\rm E}|^2/2$ and a term $-M|{\bm v}_{\rm E}|^2$ arising from the finite-Larmor-radius expansion of $Ze\Phi$ (see Ref.~\onlinecite{Cary09} or Appendix~A.1 of the companion paper\cite{Bierwage22b}).

In contrast to the full-orbit Lagrangian in Eq.~(\ref{eq:cfo_lagrangian}), which has canonical form with three total time derivatives $(\dot{\bm x})$ for a 6-D phase space, the GC Lagrangian in Eq.~(\ref{eq:lgc}) contains four total time derivatives $(\dot{\bm X}_{\rm gc},\dot{\theta})$ for the 6-D phase space $({\bm X}_{\rm gc},u,\mu,\theta)$, or three total time derivatives $\dot{\bm X}_{\rm gc}$ for the reduced 4-D phase space $({\bm X}_{\rm gc},u)$. This means that the factor
\begin{equation}
{\bm P}^{\rm gc} = Ze{\bm A} + M u \hat{\bm b}
\end{equation}

\noindent multiplying $\dot{\bm X}_{\rm gc}$ is not a canonical momentum. Nevertheless, it can be shown (see Appendix~\ref{apdx:gc_erot} below) that its covariant toroidal component
\begin{equation}
\P_\varphi^{\rm gc}  = {\bm P}_{\rm gc}\cdot\partial_\varphi{\bm x} = ZeA_\varphi + Mub_\varphi
\end{equation}

\noindent has all relevant properties of a canonical toroidal angular momentum: namely, it is conserved for GC motion in axisymmetric fields and combines with the Hamiltonian $\H_{\rm gc} = \E_{\rm gc}$ to give the rotating frame energy $\E'_{\rm gc} = \E_{\rm gc} + n\P_\varphi^{\rm gc}/\omega$ that is conserved for perturbations of the form given by Eq.~(\ref{eq:pert_erot_const}).

\subsection{Drifts and parallel acceleration}

Variation of the GC Lagrangian with respect to $u$ gives\footnote{Cary \& Brizard\protect\cite{Cary09} write: ``The $\O(\epsilon)$ terms in the parallel velocity and parallel acceleration, while necessary for keeping the Hamiltonian structure, are not complete. Other $\O(\epsilon)$ terms, such as parallel drifts, would arise if the Hamiltonian [...] were calculated to $\O(\epsilon)$ [...]''. See the discussions in Sec.~III A (p.~702) and at the end of Sec.~III C (p.~707) of Ref.~\protect\onlinecite{Cary09}. These corrections are used in Section~\ref{sec:results_res} of the present paper.}
\begin{equation}
u \equiv \dot{\bm X}_{\rm gc}\cdot\hat{\bm b}({\bm X}_{\rm gc},t) \approx v_\parallel + \O(\epsilon).
\end{equation}

\noindent Variation of the GC Lagrangian with respect to ${\bm X}_{\rm gc}$ gives the Euler-Lagrange equation
\begin{align}
&Ze \dot{u}\hat{\bm b} + Ze\partial_t{\bm A} {\mcb + Mu\partial_t\hat{\bm b}} \nonumber
\\
&= \underbrace{\nablab[\dot{\bm X}_{\rm gc}\cdot(Ze{\bm A} + Mu\hat{\bm b})] - \dot{\bm X}_{\rm gc}\cdot\nablab(Ze{\bm A} + Mu\hat{\bm b})}\limits_{\dot{\bm X}_{\rm gc}\times(Ze\nablab\times{\bm A} + Mu\nablab\times\hat{\bm b}) = Ze\dot{\bm X}_{\rm gc}\times(\nablab\times{\bm A}^*)} \nonumber
\\
&\textcolor{white}{=} -Ze\nablab\Phi - \mu\nablab B {\mcb + M\nablab|{\bm v}_{\rm E}|^2/2}.
\label{eq:gc_euler-lagrange}
\end{align}

\noindent The terms in {\mcbtxt} are sometimes ignored. While $M\nablab|{\bm v}_{\rm E}|^2/2$ is dispensable, $Mu\partial_t\hat{\bm b}$ must be retained to preserve the Hamiltonian character of the model. Note the following identities:
\begin{align}
B\partial_t\hat{\bm b} &= \partial_t{\bm B} - \hat{\bm b}\partial_t B = \partial_t{\bm B} - \hat{\bm b}\hat{\bm b}\cdot\partial_t{\bm B} \nonumber \\
&= (\partial_t{\bm B})_\perp = -(\nablab\times{\bm E})_\perp.
\end{align}

Equation~(\ref{eq:gc_euler-lagrange}) can be written compactly as
\begin{equation}
\dot{u}{\bm B}/\Omega_{\rm g} = {\bm E}^* + \dot{\bm X}_{\rm gc}\times{\bm B}^*,
\label{eq:eom}
\end{equation}

\noindent using the effective scalar and vector potentials
\begin{subequations}
\begin{align}
Ze \Phi^* &\equiv Ze\Phi + \mu B - M|{\bm v}_{\bm E}|^2/2, \\
{\bm A}^* &\equiv {\bm A} + \rho_\parallel{\bm B} = {\bm P}^{\rm gc}/(Ze),
\end{align}\vspace{-0.3cm}
\end{subequations}

\noindent with
\begin{equation}
\Omega_{\rm g} = ZeB/M, \quad \rho_\parallel \equiv u/\Omega_{\rm g}, \quad {\bm v}_{\rm E} = {\bm E}\times\hat{\bm b}/B,
\end{equation}

\noindent and the effective fields
\begin{subequations}
\begin{align}
{\bm B}^* \equiv\;& \nablab\times{\bm A}^* = {\bm B} + \rho_\parallel B \nablab\times\hat{\bm b} \nonumber \\
&\textcolor{white}{\nablab\times{\bm A}^*} = {\bm B} + \rho_\parallel (\nablab\times{\bm B} + \hat{\bm b}\times\nablab B),
\label{eq:bstar} \\
{\bm E}^* \equiv\;& -\nablab\Phi^* - \partial_t{\bm A}^* = {\bm E} + {\bm D},
\label{eq:estar} \\
{\bm D} \equiv\;& - \frac{\mu}{Ze}\nablab B\, {\mcb + \frac{M}{Ze}\frac{\nablab|{\bm v}_{\rm E}|^2}{2} \underbrace{-\rho_\parallel B\partial_t\hat{\bm b}}\limits_{+ \rho_\parallel (\nablab\times{\bm E})_\perp}}.
\label{eq:d}
\end{align}
\end{subequations}

\noindent Equation~(\ref{eq:eom}) readily yields the equation for the GC velocity and its parallel acceleration:

\vspace{0.2cm}\noindent\fbox{\begin{minipage}{\dimexpr0.49\textwidth-2\fboxsep-2\fboxrule\relax}\vspace{-0.3cm}
\begin{subequations}
\begin{align}
\dot{\bm X}_{\rm gc} =\;& u\frac{{\bm B}^*}{B_\parallel^*} + \frac{{\bm E}^*\times\hat{\bm b}}{B_\parallel^*} = {\bm v}_\parallel^* + {\bm v}_{\rm d}^* + {\bm v}_{\rm E}^*,
\label{eq:dxgc_dt}
\\
\dot{u} =\;& \frac{\Omega_{\rm g}}{B} \frac{{\bm B}^*\cdot{\bm E}^*}{B_\parallel^*},
\label{eq:du_dt}
\end{align}\vspace{-0.3cm}
\label{eq:dxgc_du_dt}
\end{subequations}
\end{minipage}}\vspace{0.2cm}

\noindent where $B_\parallel^* \equiv {\bm B}^* \cdot\hat{\bm b}$ is the GC Jacobian (see p.~720 of Ref.~\cite{Cary09}). The velocity components are
\begin{subequations}
\begin{align}
{\bm v}_\parallel^* =\;& u\frac{\bm B}{B_\parallel^*},
\label{eq:vpar}
\\
{\bm v}_{\rm E}^* =\;& \frac{{\bm E}\times\hat{\bm b}}{B_\parallel^*},
\label{eq:vE}
\\
{\bm v}_{\rm d}^* =\;& \frac{\mu}{Ze} \frac{\hat{\bm b}\times\nablab B}{B_\parallel^*} {\mcb\; -\; \frac{M}{Ze}\frac{\hat{\bm b}\times\nablab|{\bm v}_{\rm E}|^2}{2 B_\parallel^*}}
\label{eq:vd}
\\
& + u\rho_\parallel\frac{B \nablab\times\hat{\bm b}}{B_\parallel^*} {\mcb\; +\; \frac{\rho_\parallel}{B_\parallel^*} \underbrace{{\bm B}\times\partial_t\hat{\bm b}}\limits_{\hat{\bm b}\times\partial_t{\bm B}}} \nonumber
\\
=\;& \frac{\hat{\bm b}\times\left(\mu\nablab B {\mcb\; -\; M\nablab|{\bm v}_{\rm E}|^2/2}\right)}{Ze B_\parallel^*} \nonumber
\\
& + \frac{\rho_\parallel}{B^*_\parallel} \left(u \nablab\times{\bm B} + \hat{\bm b}\times({\mcb \partial_t{\bm B}} + u\nablab B)\right), \nonumber
\end{align}
\end{subequations}

\noindent where ${\bm v}_{\rm d}^*$ captures drifts associated with gradients in the electric and magnetic fields. Writing out the equation for the parallel acceleration, we obtain
\begin{equation}
\dot{u} = \frac{{\bm B}^*}{B^*_\parallel} \cdot\left(\Omega_{\rm g}\frac{{\bm E} {\mcb\; +\; \rho_\parallel(\nablab\times{\bm E})_\perp}}{B} - \frac{\mu\nablab B {\mcb\; - \; M\nablab|{\bm v}_{\rm E}|^2/2}}{M}\right)
\label{eq:du_dt_terms}
\end{equation}

\noindent Here, the ponderomotive potential $|{\bm v}_{\rm E}|^2$ can be viewed as a correction to the mirror force and $\nablab B$ drift.

As in Appendix~\ref{apdx:fo_eom} above, we normalize velocities by a characteristic velocity $v_0$ and discretize the time derivatives with increment $\Delta s = v_0\Delta t$, so that $\hat{\bm E} = {\bm E}/v_0$ and $\rho_\parallel = u/\Omega_{\rm g} = \rho_0\hat{u} B_0/B$ with $\rho_0 = v_0/\Omega_{\rm g0}$ and $\Omega_{\rm g0} = ZeB_0/M$. The normalized GC equations of motion are then
\begin{subequations}
\begin{align}
\frac{\Delta{\bm X}_{\rm gc}}{\Delta s} =\;& \hat{u}\frac{{\bm B}^*}{B_\parallel^*} + \frac{(\hat{\bm E} + \hat{\bm D})\times\hat{\bm b}}{B^*_\parallel},
\label{eq:orbtop_dxdt}
\\
\frac{\Delta\hat{u}}{\Delta s} =\;& \frac{{\bm B}^*\cdot(\hat{\bm E} + \hat{\bm D})}{\rho_0 B_0 B^*_\parallel},
\label{eq:orbtop_dudt}
\end{align}\vspace{-0.2cm}
\end{subequations}

\noindent with
\begin{subequations}
\begin{align}
{\bm B}^* =\;& {\bm B} + \rho_0 \hat{u} B_0\nablab\times\hat{\bm b} \nonumber \\
=\;& {\bm B} + \rho_0 \hat{u} B_0(\nablab\times{\bm B} + \hat{\bm b}\times\nablab B)/B,
\\
\frac{\hat{\bm D}}{\rho_0} =\;& -\hat{\mu}\nablab B {\mcb\, +\, B_0\frac{\nablab|\hat{\bm v}_{\rm E}|^2}{2} + \hat{u} \frac{B_0}{B}(\nablab\times\hat{\bm E})_\perp}.
\label{eq:dnorm}
\end{align}\vspace{-0.2cm}
\end{subequations}

\noindent The term proportional to $(\nablab\times{\bm E})_\perp = -(\partial_t\delta{\bm B})_\perp$ in $\hat{\bm D}$ is smaller by a factor $\O(\frac{L_B}{\rho_0}\frac{\omega}{\Omega_{\rm g0}}\frac{\delta B}{B}) \lesssim 10^2\times 10^{-2}\times 10^{-2} \sim 0.01$ than the magnetic drift terms $\hat{u}\rho_\parallel B\nablab\times\hat{\bm b}/\rho_0 - \hat{\mu}\nablab B \sim \O(\tfrac{B}{L_B})\times{\rm max}\{\hat{u}^2,\hat{v}_\perp^2\}$. Its ratio to the electric drift term in Eq.~(\ref{eq:orbtop_dxdt}) is of order $\O(\frac{\hat{u}}{\hat{v}_{\rm E}}\frac{\omega}{\Omega_{\rm g0}}\frac{\delta B}{B}) \lesssim 10^3\times 10^{-2}\times 10^{-2} \sim 0.1$, so the effect may be noticeable. In Eq.~(\ref{eq:orbtop_dudt}) for the parallel acceleration, the contribution of $(\nablab\times{\bm E})_\perp$ becomes $\rho_0\hat{u}\frac{B_0}{B}(\nablab\times{\bm E})_\perp\cdot{\bm B}^* \approx -\rho_0^2 \hat{u}^2 (\partial_t\delta{\bm B})_\perp\cdot\tfrac{{\bm J}_\eq}{\mu_0}$ and is smaller than the electric force ${\bm B}^*\cdot{\bm E} = \rho_\parallel{\bm E}_\perp^\eq\cdot\tfrac{{\bm J}_\eq}{\mu_0} + ...$ in Eq.~(\ref{eq:accel}) by the same factor $\O(\tfrac{\hat{u}}{\hat{v}_{\rm E}} \tfrac{\omega}{\Omega_{\rm g0}} \tfrac{\delta B}{B}) \lesssim 0.1$ as above. Its omission has a relatively small effect as the test in Fig.~\ref{fig:13_gco_tests}(d) showed.

The contribution of the ponderomotive potential $M|{\bm v}_{\rm E}|^2/2$ tends to become noticeable only for $|{\bm v}_{\rm E}|/v \gtrsim 0.1$, which translates to $K = Mv^2/2 \lesssim 10\,{\rm eV}$ for deuterons in our setup, where ${\rm max}|{\bm v}_{\rm E}| \approx 1.5\times 10^4\,{\rm m/s}$.

Note that all terms containing $|{\bm v}_{\rm E}|^2$ can be omitted without breaking the Hamiltonian character of the present GC model, because their neglect is equivalent to omitting the term $|{\bm v}_{\rm E}|^2/2$ in the original Hamiltonian (\ref{eq:hgc}). This is not the case for the term proportional to $\rho_\parallel(\partial_t{\bm B})_\perp = -\rho_\parallel(\nablab\times{\bm E})_\perp$. Although this term is formally of higher order, its omission would break the Hamiltonian character of the system as demonstrated in Fig.~\ref{fig:13_gco_tests}(d) of the main text. Approximations can be made only at the stage of constructing $\L_{\rm gc}$ and $\H_{\rm gc}$. Once these are fixed, all terms --- no matter how small --- must be retained in the resulting equations of motion if the Hamiltonian character of the model is to be preserved.

\subsection{Evaluation of the perturbed field gradient $\nablab B$}

The gradient of the magnetic field strength
\begin{equation}
B = |{\bm B}_\eq + \delta{\bm B}| = \sqrt{B_\eq^2 + 2 {\bf B}_\eq\cdot\delta {\bf B} + \delta B^2}
\end{equation}

\noindent can be written
\begin{align}
B\nabla B  =& \tfrac{1}{2}\nablab B^2 = (\nablab{\bm B})\cdot{\bm B} \nonumber \\
=& \sum\limits_j \left( (\nablab B_{\eq,\hat{j}}) B_{\hat{j}} + (\nablab \delta B_{\hat{j}}) B_{\hat{j}}\right),
\end{align}\vspace{-0.2cm}

\noindent where the index $j$ represents the coordinates $(R,\varphi,z)$ and the hatted $\hat{j}$ symbolizes a physical component, such as $B_{\hat{\varphi}} = {\bm B}\cdot\hat{\bm e}_{\hat{\varphi}} = B_\varphi/|\partial_\varphi{\bm x}| = B_\varphi/R$.

When using the PIC or PIF scheme, we have, respectively,
\begin{subequations}
\begin{align}
\delta B_{\hat{j}}(R,\varphi,z,t) =& \frac{1}{\omega}\left[\nablab\times{\bm E}_{\rm PIC}(R,\tfrac{\chi(\varphi,t)}{n},z)\right]_{\hat{j}},
\\
\delta B_{\hat{j}}(R,\varphi,z,t) =& \Re\left\{\frac{1}{i\omega}\left[\nablab\times\left(\widetilde{\bm E}_{\rm PIF}(R,z) e^{-i\chi(\varphi,t)}\right)\right]_{\hat{j}} \right\}.
\end{align}\vspace{-0.2cm}
\end{subequations}

\noindent with
\begin{equation}
\chi(\varphi,t|n,\omega) = n\varphi + \omega t = n\times(\varphi + \omega t/n).
\label{eq:ve_phase}
\end{equation}

\noindent The formulas for $\nablab\delta B_{\hat{j}}(R,\varphi,z,t)$ are the same, except for an additional gradient operator $\nablab$ in front of $[...]_{\hat{j}}$.

\subsection{Evaluation of the ponderomotive potential $|{\bm v}_{\rm E}|^2$}

The ponderomotive potential does not interfere with any other terms in the present GC model, so it is permissible to omit it or make approximations without affecting the conservative properties of the model. Here, we choose to linearize it by ignoring the magnetic fluctuations:
\begin{equation}
|{\bm v}_{\rm E}|^2 \rightarrow |{\bm v}_{\rm E,lin}|^2 \quad \text{with} \quad {\bm v}_{\rm E,lin} = {\bm E}\times{\bm B}_\eq/B_\eq^2.
\end{equation}

\noindent Given inputs ${\bm E}_{\rm PIC}(R,\varphi,z)$ or $\widetilde{\bm E}_{\rm PIF}(R,z)$ as described in Appendix~\ref{apdx:orbtop}, the term $\nablab|{\bm v}_{\rm E}|^2/2$ is computed as follows.

When using the 3-D PIC method, we define
\begin{equation}
V(R,\varphi,z) = \left|\frac{{\bm E}_{\rm PIC}(R,\varphi,z)\times\hat{\bm b}_\eq(R,\varphi,z)}{B_\eq(R,z)}\right|
\end{equation}

\noindent and simply compute
\begin{equation}
\tfrac{1}{2}\nablab|{\bm v}_{\rm E,lin}|^2(R,\varphi,z,t) = V(R,\tfrac{\chi(\varphi,t)}{n},z)\nablab V(R,\tfrac{\chi(\varphi,t)}{n},z).
\end{equation}

When using the PIF method along $\varphi$, we let
\begin{subequations}
\begin{align}
{\bm V}_{\rm r} &= \Re\left\{\frac{\widetilde{\bm E}_{\rm PIF}(R,z)\times\hat{\bm b}_\eq(R,\varphi,z)}{B_\eq(R,z)}\right\},
\\
{\bm V}_{\rm i} &= \Im\left\{\frac{\widetilde{\bm E}_{\rm PIF}(R,z)\times\hat{\bm b}_\eq(R,\varphi,z)}{B_\eq(R,z)}\right\},
\end{align}
\end{subequations}

\noindent where the subscripts ``r' and ``i'' indicate the real and imaginary components. The physical electric drift is then
\begin{align}
{\bm v}_{\rm E}(R,\varphi,z,t) =&\; \Re\left\{({\bm V}_{\rm r} + i{\bm V}_{\rm i})(\cos\chi - i\sin\chi)\right\} \nonumber \\
=&\; {\bm V}_{\rm r}\cos\chi + {\bm V}_{\rm i}\sin\chi,
\label{eq:ve_rz}
\end{align}

\noindent and its intensity field is
\begin{equation}
|{\bm v}_{\rm E}|^2 = |{\bm V}_{\rm r}|^2\cos^2\chi + |{\bm V}_{\rm i}|^2\sin^2\chi + 2{\bm V}_{\rm r}\cdot{\bm V}_{\rm i}\sin\chi\cos\chi.
\end{equation}

\noindent The ponderomotive force expressed in terms of the scalar fields $V_{\rm r}(R,z) = |{\bm V}_{\rm r}|$, $V_{\rm i}(R,z) = |{\bm V}_{\rm i}|$, $W(R,z) = {\bm V}_{\rm r}\cdot{\bm V}_{\rm i}$ and their gradients becomes
\begin{align}
\tfrac{1}{2}\nablab&|{\bm v}_{\rm E}|^2 =
\label{eq:pond_cyl}
\\
& V_{\rm r}\nablab V_{\rm r} \cos^2\chi + V_{\rm i}\nablab V_{\rm i} \sin^2\chi + \nablab W\sin\chi\cos\chi \nonumber
\\
+ &\frac{n}{R}\left[(V_{\rm i}^2 - V_{\rm r}^2)\sin\chi\cos\chi + W(\cos^2\chi - \sin^2\chi)\right]\hat{\bm e}_{\hat{\varphi}}, \nonumber
\end{align}

\noindent where we used $\nablab\chi = \hat{\bm e}_{\hat\varphi} n/R$. Note that the first line of Eq.~(\ref{eq:pond_cyl}) contains only $\hat{\bm e}_R$ and $\hat{\bm e}_z$ components since $V_{\rm r}$, $V_{\rm i}$ and $W$ are functions of $(R,z)$ only.

\subsection{Rotating frame energy conservation}
\label{apdx:gc_erot}

Using the effective fields and potentials ${\bm E}^* = -\nablab\Phi^* - \partial_t{\bm A}^*$ and ${\bm B}^* = \nablab\times{\bm A}^*$, the total time derivatives of the total energy
\begin{equation}
\E_{\rm gc} = \H_{\rm gc} = M u^2/2 + Ze\Phi^*({\bm X}_{\rm gc},t),
\end{equation}

\noindent and canonical toroidal angular momentum
\begin{equation}
\P_\varphi^{\rm gc} = {\bm P}^{\rm gc}\cdot\partial_\varphi{\bm x} = Ze{\bm A}^*\cdot\partial_\varphi{\bm x} = ZeR A^*_{\hat{\varphi}}(u,{\bm X}_{\rm gc},t),
\end{equation}

\noindent with $A^*_\varphi = R A^*_{\hat{\varphi}}$, can be written as
\begin{subequations}
\begin{align}
\dot{\E}_{\rm gc} =&\; Mu\dot{u} + Ze\partial_t\Phi^* + Ze\dot{\bm X}_{\rm gc}\cdot\nablab\Phi^*
\\
\dot{\P}_\varphi^{\rm gc} =&\; Ze\partial_t A_\varphi^* + ZeR\dot{\bm X}_{\rm gc}\cdot\nablab A^*_{\hat{\varphi}} + Ze \dot{R} A^*_{\hat{\varphi}} + Ze \dot{u} \partial_u A^*_\varphi \nonumber \\
=&\; Ze\partial_t A_\varphi^* - Ze R \hat{\bm e}_\varphi\cdot(\dot{\bm X}_{\rm gc}\times{\bm B}^*) + Ze \dot{\bm X}_{\rm gc}\cdot(\partial_\varphi{\bm A}^*) \nonumber \\
& + M\dot{u}b_\varphi,
\end{align}
\label{eq:gc_erot1}\vspace{-0.4cm}
\end{subequations}

\noindent where we have used $R\dot{\bm X}_{\rm gc}\cdot\nablab A^*_{\hat{\varphi}} + v_R A^*_{\hat{\varphi}} = v_R\partial_R(R A^*_{\hat{\varphi}}) + R v_z\partial_z A^*_{\hat{\varphi}} + v_{\hat{\varphi}}\partial_\varphi A^*_{\hat{\varphi}}$, $\partial_R(R A^*_{\hat{\varphi}}) = R B_z^* + \partial_\varphi A^*_R$ and $R\partial_z A^*_{\hat{\varphi}} = -R B_R + \partial_{\hat{\varphi}} A_z$. Substituting the equations of motion (\ref{eq:dxgc_du_dt}) and (\ref{eq:eom}), we obtain (after a few more cancellations)
\begin{subequations}
\begin{align}
\frac{B^*_\parallel\dot{\E}_{\rm gc}}{Ze} =&\; u{\bm E}^*\cdot{\bm B}^* + B_\parallel^*\partial_t\Phi^* \nonumber \\
&+ u{\bm B}^*\cdot\nablab\Phi^* + \nablab\Phi^*\cdot({\bm E}^*\times\hat{\bm b}) \nonumber \\
=&\; B_\parallel^*\partial_t\Phi^* - u{\bm B}^*\cdot\partial_t{\bm A}^* - \partial_t{\bm A}^*\cdot(\hat{\bm b}\times\nablab\Phi^*),
\label{eq:gc_dedt}
\\
\frac{B^*_\parallel\dot{\P}_\varphi^{\rm gc}}{Ze} =&\; B_\parallel^*\partial_t A_\varphi^* - B_\parallel^* R \hat{\bm e}_\varphi\cdot{\bm E}^* \nonumber \\
&+ (u{\bm B}^* + {\bm E}^*\times\hat{\bm b})\cdot\partial_\varphi{\bm A}^* \nonumber \\
=& -B_\parallel^*\partial_\varphi\Phi^* + u{\bm B}^*\cdot\partial_\varphi{\bm A}^* + \partial_\varphi{\bm A}^*\cdot(\hat{\bm b}\times\nablab\Phi^*).
\label{eq:gc_dpdt}
\end{align}
\end{subequations}

\noindent Equation~(\ref{eq:gc_dedt}) implies that the total energy $\E_{\rm gc}$ of the particle is conserved for time-independent fields, and Eq.~(\ref{eq:gc_dpdt}) implies conservation of $\P_\varphi^{\rm gc}$ for axisymmetric fields. Together, these two equations imply that
\begin{equation}
n\dot{\E}_{\rm gc} = -\omega\dot{\P}_\varphi^{\rm gc},
\label{eq:gc_erot_dt0}
\end{equation}

\noindent and, thus, conservation of the rotating frame energy
\begin{equation}
\E'_{\rm gc} = \E_{\rm gc} + \tfrac{\omega}{n}\P_\varphi^{\rm gc} = {\rm const}.,
\label{eq:gc_erot}
\end{equation}

\noindent for fields whose $t$- and $\varphi$-dependence has the form given by Eq.~(\ref{eq:pert_erot_const}).

Note that Eq.~(\ref{eq:gc_erot_dt0}) holds in spite of the fact that some terms depend nonlinearly on the fluctuating fields. This is because the associated higher harmonics of $\omega$ and $n$ all have the same angular phase velocity $\omega/n$, so they preserve the form of Eq.~(\ref{eq:pert_erot_const}). How this translates into the conservation law (\ref{eq:gc_erot_dt0}) was explicitly shown for $|{\bm v}_{\rm E}|^2$ in Eqs.~(A47)--(A50) of the companion paper.\cite{Bierwage22b} The same holds for factors like $B = \sqrt{B_\eq^2 + 2 {\bf B}_\eq\cdot\delta {\bf B} + \delta B^2}$, whose derivatives satisfy
\begin{equation}
n\partial_t B = \omega \partial_\varphi B = -in\omega({\bm B}_\eq + \delta{\bm B})\cdot\delta{\bm B}/B.
\label{eq:dB_dt_dphi}
\end{equation}

\subsection{$N$-point gyroaveraging}
\label{apdx:gc_ga}

The GC model in {\tt ORBTOP} offers the possibility to perform $N$-point gyroaveraging of the fluctuating fields ${\bm E}(t)$ and $\delta{\bm B}(t)$. This means that the time-dependent forces acting on the GC are not evaluated at the GC position ${\bm X}_{\rm gc}$, but are averaged over the positions of $N_{\rm avg}$ satellite particles that have been placed on a surrounding circle of radius
\begin{equation}
\rho_{\rm gc} \equiv \rho_{\rm g}({\bm X}_{\rm gc}) = \frac{\sqrt{2\mu B({\bm X}_{\rm gc})/M}}{\Omega_{\rm g}({\bm X}_{\rm gc})} \approx \frac{v_\perp}{\Omega_{\rm g}}.
\end{equation}

\noindent Examples with $N_{\rm avg} = 2,4,8$ were shown schematically in the left column of Fig.~\ref{fig:14_comparison_cfo-gco-ga} in Section~\ref{sec:results_ga}.

Gyroaveraging is performed instantaneously at each time step. The averaging circle is taken to lie in the $(R,z)$ plane, irrespective of the exact direction of the local magnetic field vector ${\bm B}({\bm X}_{\rm gc})$. Besides simplifying the implementation, this choice is also motivated by the desire to preserve the Hamiltonian character of the system by leaving the toroidal gradients unaltered and averaging only over the nonuniformities of the poloidal structure of the fluctuating fields. However, the results reported in Section~\ref{sec:results_ga} indicate that, in spite of averaging only in the poloidal plane, this procedure is not entirely successful. Conservation laws are found to be broken on the millisecond time scale, especially when the fluctuations have a magnetic component. Motivated by the arguments presented in the introduction (Section~\ref{sec:intro}), it may be worthwhile checking whether the procedure works better in codes that use potentials $\Phi$ and $\delta{\bm A}$ instead of the physical fields ${\bm E}$ and $\delta{\bm B}$ that we have chosen in this study.

Out of curiosity, we have also performed simulations where gyroaveraging was applied not only to the perturbations ${\bm E}$ and $\delta{\bm B}$ but also to the background field ${\bm B}_\eq$. In the absence of perturbations, the orbits remained on invariant surfaces. In the presence of perturbations, the unphysical secular acceleration was enhanced compared to the results in Fig.~\ref{fig:14_comparison_cfo-gco-ga}, where gyroaveraging was applied only to the fluctuations.

\begin{figure*}[tbp]
\centering
\includegraphics[width=0.96\textwidth]{\figures/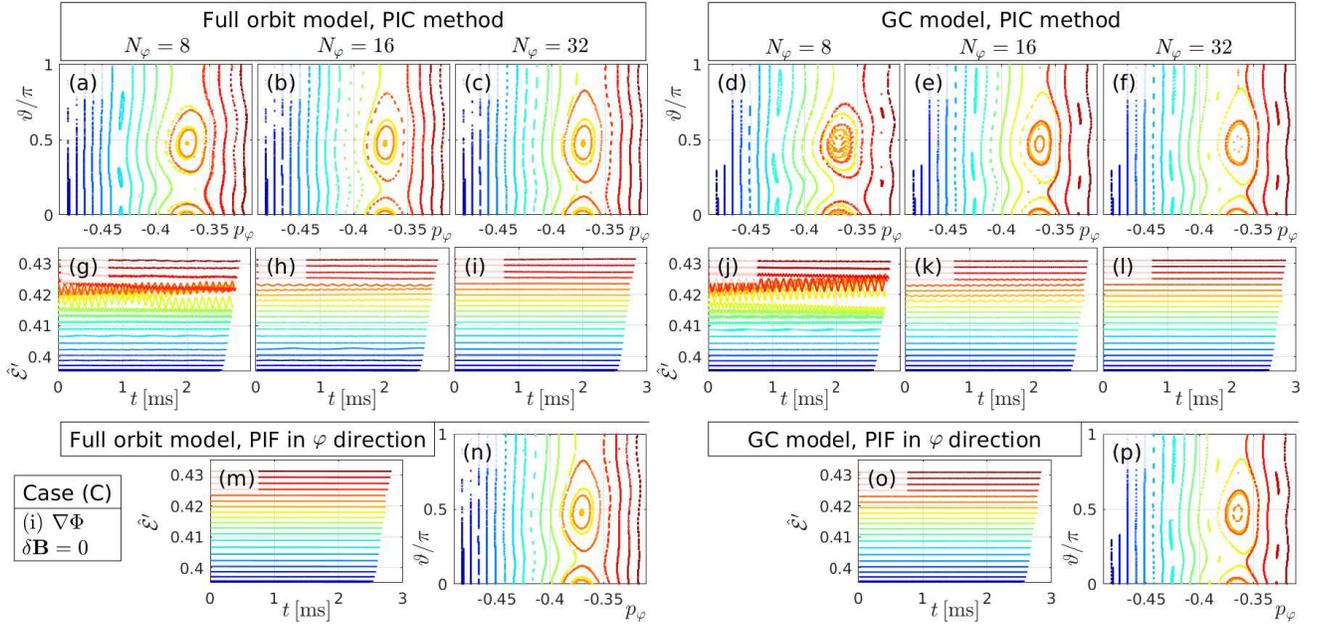}
\caption{Benchmark of the 3-D PIC method (top) versus results of the mixed PIC-PIF method (bottom) for our electrostatic scenario (i) defined in Eq.~(\protect\ref{eq:scenario_1}). The left half show results of the full orbit model and the right half for the GC model. Our PIC scheme uses uniformly spaced toroidal meshes with cell size $\Delta\varphi = 2\pi/N_\varphi$, and results are shown for $N_\varphi = 8,16,32$. Panels (a--f), (n) and (p) show Poincar\'{e} plots in the upper half-plane ($0\leq \vartheta\leq \pi$). Panels (g--l), (m) and (o) show the tine traces of the rotating frame energy $\hat{\E}'(t)$. Theses results are for the nonnormal mode case (C).}
\label{fig:16_cfo-gc_pif-pic_C-ES}%
\end{figure*}

\begin{figure*}[tbp]
\centering
\includegraphics[width=0.96\textwidth]{\figures/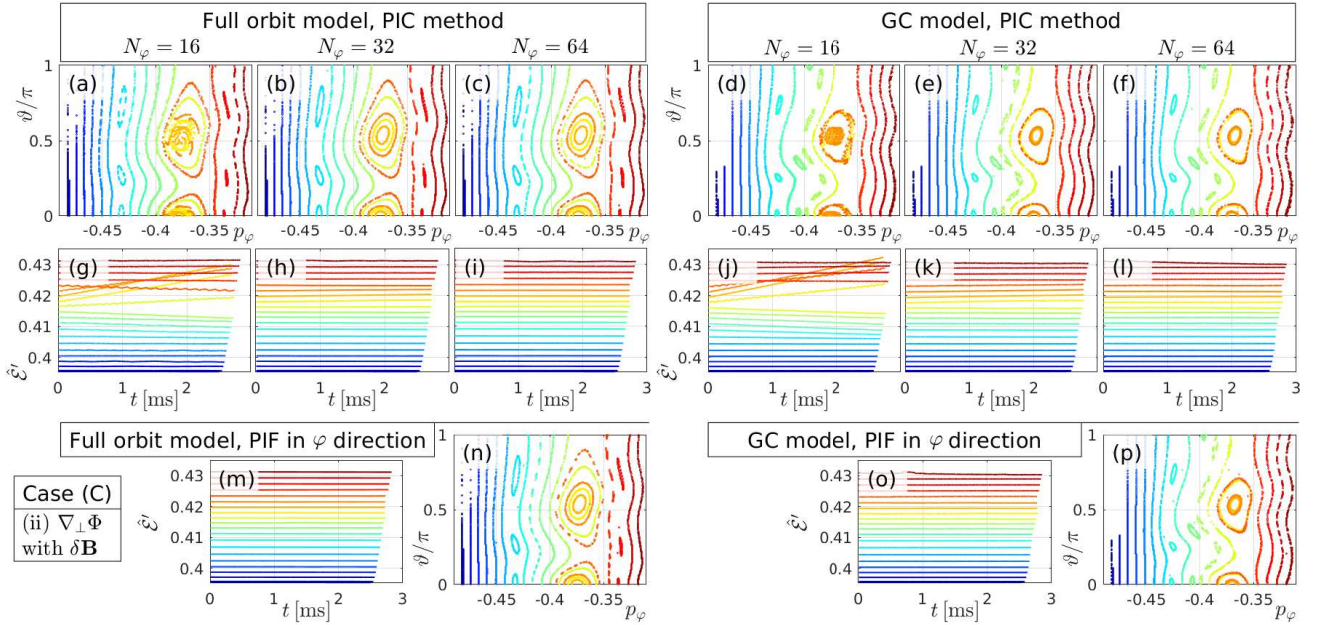}
\caption{Benchmark of the 3-D PIC method (top) versus results of the mixed PIC-PIF method (bottom) for our electromagnetic scenario (ii) defined in Eq.~(\protect\ref{eq:scenario_2}). Arranged like Fig.~\protect\ref{fig:16_cfo-gc_pif-pic_C-ES}, except for the larger values of $N_\varphi = 16,32,64$.}
\label{fig:17_cfo-gc_pif-pic_C-EM}%
\end{figure*}

Another test we performed was to distribute the satellite particles with uniform time increments $\Delta t = 2\pi/(N_{\rm avg} \Omega_{\rm g})$ instead of uniform increments $\Delta\theta = 2\pi/N_{\rm avg}$ in the gyrophase $\theta$. This was done as follows. The increments of the geometric gyroangle $\theta$ for uniform time increments ${\rm d}t$ are given by
\begin{equation}
{\rm d}\theta(t) = \Omega_{\rm g}(\theta){\rm d}t = \frac{\Omega_{\rm g}(\theta)}{\Omega_{\rm gc}}{\rm d}\theta_{\rm g} \approx \frac{R_{\rm gc}}{R_{\rm g}(\theta)}{\rm d}\theta_{\rm g},
\end{equation}

\noindent where ${\rm d}\theta_{\rm g} \equiv {\rm d}t\Omega_{\rm gc}$ is a time-like angle and $\Omega_{\rm gc} \equiv \Omega_{\rm g}({\bm X}_{\rm gc})$. Letting $R_{\rm g}(\theta) \approx R_{\rm gc} + \rho_{\rm gc}\cos\theta$, integration gives
\begin{equation}
\theta_{\rm g}(\theta) = \theta + \frac{\rho_{\rm gc}}{R_{\rm gc}}\sin\theta.
\end{equation}

\noindent Using piecewise cubic hermite interpolating polynomial (PCHIP) interpolation, we inverted this equation numerically to give nonuniformly distributed gyrosatellite samples at $\theta(\theta_{\rm g})$ on a uniform $\theta_{\rm g}$ grid. The results were effectively identical to those obtained by simply placing the satellites uniformly along $\theta$. During the simulated $3\,{\rm ms}$ interval, nearly the same amount of secular acceleration occurred, and the resonance also remained shifted by the same amount as in Fig.~\ref{fig:14_comparison_cfo-gco-ga}. Although this is not unexpected, because $\rho_{\rm gc}/R_{\rm gc} \sim 0.07\,{\rm m}/3.5\,{\rm m} \sim 0.02$ is very small, we wanted to be sure that there is no noticeable effect in the long run. Our test shows that this small correction has no significant effect on the conservative character (or lack thereof) on the multi-millisecond scale in our working example.

\begin{table*}[tbp]\vspace{-0.55cm}
\caption{Pros and cons of full orbit model compared to GC models. An important aspect is particle noise that arises in some applications (e.g., PIC-MHD hybrid codes) and acquires a higher dimensionality. As discussed in the text, it is less the noise level that counts but more the overall effect that noise has on the results. That effect depends on processes like spatio-temporal averaging, and it remains to be shown whether and how much the full orbit model enhances the effect (it could be case-dependent). The outcome has direct impact on the performance of the full orbit model compared to a GC model since it determines the number of particles needed to achieve a certain level of accuracy.}
\begin{ruledtabular}
\begin{tabular}{p{1.5cm}|p{8.5cm}|p{7.3cm}}
Full orbit \newline model has: & Simpler equations & More physics \& is closer to `first principles' \\
\hline Pros: & $\oplus$ Simpler code $\rightarrow$ Less susceptible to errors \& bugs \newline $\oplus$ Can use more efficient solvers (e.g., modified leap-frog) \newline {\color{white}$\oplus$} $\rightarrow$ Speed-up and higher accuracy \newline $\oplus$ Fewer operations $\rightarrow$ Speed-up & $\oplus$ Higher degree of quantitative validity \newline {\color{white}$\oplus$} (e.g., helical orbit shape, correct gyroaveraging) \newline $\oplus$ Broader applicability (e.g., RF heating) \newline $\oplus$ More intuitive \& realistic collision operators \\
\hline Cons: & $\ominus$ Processes like free streaming and drifts are not explicit \newline {\color{white}$\ominus$} $\rightarrow$ Less transparent and less freedom to manipulate terms \newline {\color{white}$\ominus$ $\rightarrow$} in exploratory numerical experiments & $\ominus$ Needs shorter time steps $\Delta t$ \newline {\color{white}$\ominus$} $\rightarrow$ Slow-down of pure orbit-following codes \newline {\color{white}$\ominus$ $\rightarrow$} (but not necessarily PIC-MHD hybrid codes) \newline $\ominus$ 6-D noise instead of 4-D $\rightarrow$ Need more particles? \\
\end{tabular}
\label{tab:cfo-gc}
\end{ruledtabular}
\end{table*}

\section{Toroidal waveform representation: PIF vs.\ PIC}
\label{apdx:pif_pic}

In our working example, where we follow deuterons in the presence of a prescribed wave field with toroidal mode number $n=2$ and frequency $\nu \approx 50\,{\rm kHz}$, the 3-D PIC method requires at least $N_\varphi \approx 16$ grid points (8 per wavelength) in the toroidal direction in order to yield an accuracy comparable to the mixed PIC-PIF method on the millisecond time scale. This is demonstrated in Fig.~\ref{fig:16_cfo-gc_pif-pic_C-ES} for our electostatic scenario (i) defined in Eq.~(\ref{eq:scenario_1}). In the electromagnetic scenario (ii), results for which are shown in Fig.~\ref{fig:17_cfo-gc_pif-pic_C-EM}, the PIC method needs at least $N_\varphi \approx 32$ grid points to obtain acceptable conservation properties, with good KAM surfaces and nearly constant $\E'$.

For lower values of $N_\varphi$, panels (g) and (j) in both figures show that $\E'$ exhibits significant oscillations and drifts. The shape of the resonant island in panels (a--f) is also somewhat affected by insufficient resolution. For the same value of $N_\varphi$, the full orbit model appears to perform slightly better (in terms of conservation) than the GC model, but the difference is small.

Results for normal modes tend to be more accurate than for nonnormal modes. The results in Figs.~\ref{fig:16_cfo-gc_pif-pic_C-ES} and \ref{fig:17_cfo-gc_pif-pic_C-EM} were obtained with the nonnormal modes of our case (C), which may be considered a worst-case scenario. For the normal mode case (A), we obtained results of reasonable accuracy in the electromagntic scenario (ii) when using only $N_\varphi = 16$ grid points.

\section{Pros and cons of full orbit and GC models}
\label{apdx:reasons}

In this work, we have made several comparisons between results of the full orbit and GC models. One purpose was to gather data that can help us and other researchers to make informed decisions when choosing a model for an application. Table~\ref{tab:cfo-gc} shows a (most likely incomplete) summary of advantages and disadvantages that the full orbit model has compared to GC models. In the following paragraphs, we discuss some of those items in more detail, including also information gathered in this study. The important aspect of computational speed (Table~\ref{tab:speed} in Appendix~\ref{apdx:orbtop}) permeates the entire discussion and will be addressed in dependence of the application.

\bigskip

\paragraph{Realism \& physics content.} From the examples listed in Table~\ref{tab:cfo-gc}, the advantages of being closer to what is thought to be `first principles' are probably evident, but a few words shall be said with regard to gyroaveraging. Although GC models can be equipped with a gyroaveraging procedure, we found that this method breaks the Hamiltonian character of the system (Section~\ref{sec:results_ga}) and enhances small discrepancies in location of resonances that may otherwise be cured by higher-order corrections in $u$ and $\mu$ (Section~\ref{sec:results_res}). This may affect the simulation results at least quantitatively. Moreover, a gyroaveraging procedure adds computational overhead to the GC model, thus diminishing one of its main advantages: computational speed. This problem becomes particularly severe when gyroradii are large, as in the case of energetic alpha particles born at $3.5\,{\rm MeV}$ during deuterium-tritium fusion reactions. When a hybrid simulation is highly parallelized on distributed memory platforms using the Message Passing Interface (MPI), it may require complicated MPI communication algorithms to transfer satellite particle data across multiple MPI domain boundaries. On some supercomputers, such kinds of communications diminish performance drastically. (Multi-domain crossings may also happen in a hybrid code whose full orbit solver uses an implicit scheme and is run with too large a time step. This should, of course, be avoided.)

Therefore, when gyroaveraging effects are important, one should carefully consider whether to use a poor imitation thereof in the GC model (violating conservation laws, adding complexity, and reducing performance) or to simply simulate the full gyromotion. We will continue this line of thought below in a paragraph titled ``Gyro- and noise-averaging''.

\smallskip

\paragraph{Simplicity \& physics transparency.} The complexity of the GC model has one advantage that the full orbit model cannot match: In the GC equations, one can identify terms responsible for different physical processes, such as free streaming along magnetic field lines, parallel acceleration, the mirror force and magnetic drifts. These terms can be manipulated in numerical experiments as we have done in Sections~\ref{sec:results_dB}--\ref{sec:results_res} of the main text, and the results can be interpreted in an intuitive way.

On the other hand, the simpler equations of the full orbit model do not only give simpler code but are also linked to accuracy and performance in several ways. For instance, unlike the GC model, a full orbit simulation does not require the computation of derivatives of the magnetic field ${\bm B}$. This can be expected to reduce the effect of interpolation artifacts or PIC noise on the results. In addition, the simpler full orbit equations require fewer operations to be performed per time step and permit to choose more efficient algorithms. In particular, the structure of the Newton-Lorentz equations is such that it can be easily implemented in the modified leapfrog (MLF) scheme as described in Appendix~\ref{apdx:fo_dt}. This requires the inversion of a dense 3-by-3 matrix and possibly a few transformations to and from Cartesian coordinates, which adds a little to the computational cost. As was shown in Table~\ref{tab:speed} of Appendix~\ref{apdx:orbtop}, each 4th-order Runge-Kutta (RK4) step of advancing the complicated GC equations (even with precomputed field gradients) took $50\%$ longer than a step in the full orbit solver. That factor of $1.5$ rises to $2$ when using MLF for full gyroorbits.

\smallskip

\paragraph{Application in pure orbit-following codes.} In a pure orbit-following code such as {\tt ORBTOP} that was used in the present work, fluctuations are usually prescribed and field gradients for the GC model can then be pre-computed. Our GC simulations using the RK4 scheme were performed with $100$ times larger time steps ($\Delta t_{\rm GC} = 65\,{\rm ns}$) than the full orbit simulations ($\Delta t_0 = 0.65\,{\rm ns}$). According to Table~\ref{tab:speed} of Appendix~\ref{apdx:orbtop}, the GC simulation using RK4 was nearly $70$ times faster than the full orbit simulation using RK4. The speed-up factor was reduced to $50$ when compared to the full orbit solver employing the MLF scheme with the same time step as RK4. However, the time step of the MLF scheme may be safely increased by a factor $2$ or more without loosing too much 
accuracy in $\E'$ (combined energy and momentum), so the performance penalty of the full orbit model could be reduced to a factor of $20$. This is still a large value, so we expect that the GC model has a robust advantage in performance for pure orbit-following codes, whenever gyroaveraging and other finite-Larmor-radius effects can be ignored.

\smallskip

\paragraph{Application in PIC-MHD hybrid codes.} The GC model of our orbit-following code {\tt ORBTOP} was mostly adopted from the hybrid code {\tt MEGA}.\cite{Todo98, Todo05, Todo06} Using the explicit RK4 scheme, {\tt MEGA} solves full MHD equations that include fast waves and require short time steps to satisfy the CFL condition. In recent simulations of beam-driven large-amplitude Alfv\'{e}n modes in JT-60U\cite{Bierwage18} and alpha particle transport during so-called `sawtooth crashes' in JET,\cite{Bierwage22a} {\tt MEGA} was run with time steps on the order of $\Delta t_{\rm MHD} \approx 1\,{\rm ns}$. Even smaller time steps were needed for convergence tests performed with finer spatial meshes. Thus, the full MHD solver already runs with the same size of time steps as required for the simulation of gyrating deuterons or alpha particles. If the simulation particles are pushed with the same time step, one may then na\"{i}vely expect that the replacement of the GC model with the full orbit model would not have any detrimental effect on the computational speed. In fact, the data in Table~\ref{tab:speed} of Appendix~\ref{apdx:orbtop} suggest that the full orbit model would run faster than the GC model by a factor $1.5$ with RK4, and the speed-up is raised to a factor $2$ with MLF. The net speed-up may increase even further if one takes into account that the field  gradients needed for the GC model cannot be precomputed but must be updated at each step. However, in practice, the situation is not so simple. Firstly, GC simulation may perform so-called sub-cycling. Secondly, there exists the complex problem of noise. Let us discuss these points in some detail.

\smallskip

\paragraph{Sub-cycling of GC motion in hybrid codes.} Our hybrid simulations such as those reported in Refs.~\onlinecite{Bierwage18,Bierwage22a} usually make use of subcycling, where the particles are not advanced for a certain number $N_{\rm sub}$ of MHD time steps. Typically we use $N_{\rm sub} = 4$ or $8$ for fast ions such as fusion-born $3.5\,{\rm MeV}$ alpha particles in JET\cite{Bierwage22a} or $400\,{\rm keV}$ deuteron beams in JT-60U.\cite{Bierwage18} These values of $N_{\rm sub}$ are much smaller than the factor of $100$ by which the GC time step was increased in pure orbit-following simulations. The choice of a relatively small $N_{\rm sub} \lesssim \O(10)$ in hybrid simulations with fast ions ensures that simulation particles do not usually cross more than one spatial cell in one step. One reason is that PIC and hybrid simulations rely on temporal averaging of noise, so one should avoid freezing the PIC noise for too long. Another reason is that PIC noise creates a fine structure in the field, so that the CFL condition for particle pushing is not given by the wavelength of the physical waves in the system but by the grid-scale noise as we have experienced in Section~\ref{sec:results_noise} of the main text.

Based on the considerations made so far, the full orbit solver using the MLF scheme can be expected to run at a speed comparable to and no more than a factor $2...4$ slower than that of the GC solver that makes use of subcycling ($N_{\rm sub} \approx 4...8$) when run alongside a full MHD solver in a hybrid code.

\smallskip

\paragraph{Gyro- and noise-averaging.} Gyroaveraging effects are often important in hybrid simulations, so their GC modules are sometimes run with $N$-point gyroaveraging.\cite{Bierwage18, Bierwage16c} One may thus expect that the use of the full orbit model can give us a further gain in speed and higher accuracy since the $N$-point gyroaveraging procedure is eliminated. The realism and the conservative character of the simulation are certainly enhanced. However, the impact on overall accuracy and speed is not so straightforward to foresee because it requires an evaluation of PIC noise and its effect on the simulation results.

For instance, experience shows that the noise level of a GC simulation with $4$-point gyroaveraging is comparable to that of a pure GC simulation with a $4$ times larger number of simulation particles per spatial cell in the poloidal $(R,z)$ plane.\footnote{This rule-of-thumb holds when we apply Fourier filtering along the toroidal direction, so that the effect of noise is essentially limited to two dimensions, $R$ and $z$.}
Our full orbit simulations were run with time steps that take about $167$ samples per gyration and even if the time step is increased by a factor $3$, there are still dozens of samples being taken during each gyration; much more than the number of satellites we typically use for gyroaveraging in a GC simulation. From this, one may na\"{i}vely expect a significant reduction of noise effects on time scales much longer than one gyroperiod. However, the actual situation may not be so simple.

Unlike white noise, the phenomenon we call ``PIC noise'' is not entirely random. It arises from the interaction of regular particle motion with regular grids, so it can be expected to constitute a structured albeit complex signal, which undergoes a certain amount of chaotization trough interactions with (noisy) waves in a hybrid simulation. While $N$-point averaging truly smoothes the field's landscape at each step, a gyrating simulation particle influences the field during its gyration and may interact with its own noise --- especially when strong resonances are at play, where relatively large portions of phase space form coherent structures that step in sync with waves in the electric or electromagnetic field. Signal-noise correlations may arise, with consequences that are difficult to foresee.

\smallskip

\paragraph{Dimensionality of noise.} The problem we have just discussed is connected with the noise arising from the variable gyrophase $\theta$. That branch of noise does not exist in the GC model, so a hybrid simulation with full gyroorbits can be thought of as being noisy in an extra dimension. Of course, the moments of particle distributions that enter MHD equations are always 3-D. However, the underlying particle-based representation of phase space density is, in principle, scaleless and of high dimensionality. The PIC method maps this complex signal to a space with a few dimensions. From the viewpoint of wave-particle interactions manifested by phase space structures, the GC model has {\it dynamic noise} in four dimensions: three in space, plus kinetic energy $K$. In addition, there is {\it static noise} in the fifth dimension: the fixed magnetic moment $\mu$. On the one hand, the fact that $\mu$ becomes dynamic in the full orbit model could be an advantage in terms of noise averaging --- at least if one is interested in low-frequency waves, which is where a comparison between GC and full orbit models is meaningful. On the other hand, there is an extra dimension of noise in a full orbit simulation that manifests itself via the rapid oscillation of the five coordinates $\{x,y,z, E, \mu\}$ on the gyration time scale. These oscillations can couple to and exchange energy with fast waves, thus, raising the level of fast wave fluctuations above that of mere MHD activity.

From the point of view of low-frequency phenomena, such as shear Alfv\'{e}n or sound waves, sufficiently complex high-frequency fluctuations may have the appearance of noise. Thus, when measuring the PIC noise level in a hybrid simulation that uses a full orbit model for its kinetic component, one should distinguish the (potentially noisy-looking) physical fast wave signal from numerical PIC noise. Of course, this is easier said than done, since the fast waves will be partially noise-driven. Anyhow, care is required for a fair comparison between the computational performance and resource requirements of GC and full gyroorbit models in a hybrid simulation.

\smallskip

\paragraph{Summary.} A fair and conclusive comparison between the computational performance and accuracy of the full orbit and GC models inside a hybrid code should be done under conditions where the effect of PIC noise is comparable. This is left for future work, noting that the instantaneous PIC noise level does not necessarily reflect the actual impact that the noise has on the results. It also depends on how well the noise's influence is diminished by spatio-temporal averaging on the scales of the waves of interest. When examining low-frequency MHD phenomena, it may be meaningful to isolate PIC noise from complex signals associated with fast waves.

If one leaves aside the elusive effects of PIC noise, which should be monitored on a case-by-case basis, one may conclude that the full orbit model appears to be a practically viable alternative to the GC model when used as part of a hybrid code that solves full MHD equations, which require nanosecond-scale time steps due to the presence of fast waves.

For pure orbit-following codes, the GC model maintains an advantage in speed by a factor $20$ or higher. Moreover, the GC model offers a higher degree of transparency in physics-oriented analyses. It facilitates controlled numerical experiments that can help to identify the mechanisms responsible for a certain observation made in a simulation.



\end{document}